\documentclass[12pt]{article}
\pdfoutput=1
\usepackage{amssymb}
\usepackage{amsmath}
\usepackage{bm} 
\usepackage{graphicx}
\usepackage{color}
\usepackage{xcolor}
\usepackage{dsfont}
\usepackage{cancel}
\usepackage{comment}
\usepackage{hyperref}
\usepackage{enumerate}
\usepackage{setspace}
\definecolor{nicered}{rgb}{0.7,0.1,0.1}
\definecolor{nicegreen}{rgb}{0.1,0.5,0.1}
\hypersetup{colorlinks,citecolor= nicegreen,linkcolor= nicered}
\allowdisplaybreaks
\nopagebreak
\begin{document}
\begin{titlepage}
 \vspace*{-1.9cm}
  \begin{flushright}
ULB-TH/14-03
 \end{flushright}
  \newcommand{\AddrLiege}{{\sl \small IFPA, Dep. AGO, Universit\'e de
      Li\`ege, Bat B5,\\ \small \sl Sart Tilman B-4000 Liege 1,
      Belgium}} 
  \newcommand{\AddrULB}{{\sl \small Service de Physique
      Th\'eorique, Universit\'e Libre de Bruxelles\\
      \sl \small Bld du Triomphe, CP225, 1050 Brussels Belgium}}
\vspace*{.8cm}
 
\begin{center}
  \textbf{\large Scalar triplet flavored leptogenesis: a systematic approach}\\[10mm]
  D. Aristizabal Sierra$^{a,}$\footnote{e-mail addresses: {\tt daristizabal@ulg.ac.be}},
  Mika\"el Dhen$^{b,}$\footnote{e-mail address: {\tt mikadhen@ulb.ac.be}},
  Thomas Hambye$^{b,}$\footnote{e-mail address: {\tt thambye@ulb.ac.be}}
  \vspace{.5cm}\\
  $^a$\AddrLiege.\vspace{0.4cm}\\
  $^b$\AddrULB.\vspace{0.4cm}\\
\end{center}
\vspace*{0cm}

\begin{abstract}
  Type-II seesaw is a simple scenario in which Majorana neutrino
  masses are generated by the exchange of a heavy scalar electroweak
  triplet. When endowed with additional heavy fields, such as
  right-handed neutrinos or extra triplets, it also provides a
  compelling framework for baryogenesis via leptogenesis.  We derive
  in this context the full network of Boltzmann equations for studying
  leptogenesis in the flavored regime.  To this end we determine the
  relations which hold among the chemical potentials of the various
  particle species in the thermal bath.  This takes into account the
  standard model Yukawa interactions of both leptons and quarks as
  well as sphaleron processes which, depending on the temperature, may
  be classified as faster or slower than the Universe Hubble
  expansion.  We find that when leptogenesis is enabled by the
  presence of an extra triplet, lepton flavor effects allow the
  production of the $B-L$ asymmetry through lepton number conserving
  CP asymmetries. This scenario becomes dominant as soon as the
  triplets couple more to leptons than to standard model scalar
  doublets.  In this case, the way the $B-L$ asymmetry is created
  through flavor effects is novel:
  instead of invoking the effect of $L$-violating inverse decays
  faster than the Hubble rate, it involves the effect of $L$-violating
  decays slower than the Hubble rate.  We also analyze the
  more general situation where lepton number violating CP asymmetries
  are present and actively participate in the generation of the $B-L$
  asymmetry, pointing out that as long as $L$-violating triplet decays
  are still in thermal equilibrium when the triplet gauge scattering
  processes decouple, flavor effects can be striking, allowing to
  avoid all washout suppression effects from seesaw interactions.  In
  this case the amount of $B-L$ asymmetry produced is limited only by
  a universal gauge suppression effect, which nevertheless goes away
  for large triplet decay rates.
  
    \end{abstract}
\end{titlepage}
\setcounter{footnote}{0}

\section{Introduction}
\label{sec:intro}
Non-vanishing neutrino masses
\cite{Tortola:2012te,GonzalezGarcia:2012sz,Fogli:2012ua}, and the
cosmic asymmetry between baryons and anti-baryons
\cite{Hinshaw:2012aka,Ade:2013zuv}, constitute two well-established
experimental facts which have particularly well demonstrated that
physical degrees of freedom beyond the standard model (SM) must be at
work at certain unknown energy scale.  Although a large number of
scenarios capable of accounting for these experimental facts exist,
arguably the tree-level seesaw models, type-I \cite{seesaw}, type-II
\cite{Schechter:1980gr} and type-III \cite{Foot:1988aq}, are of
special interest: they constitute the simplest frameworks, are well
theoretically motivated and, through the leptogenesis mechanism,
provide a common explanation for both puzzles (for reviews see
\cite{Davidson:2008bu,Fong:2013wr} and \cite{Hambye:2012fh}).

The type-II leptogenesis scenario
\cite{Hambye:2012fh,Ma:1998dx,Hambye:2000ui,Hambye:2003ka,Hambye:2005tk},
in which the baryon asymmetry is generated from the decay of one or
several scalar triplets, is more intricate than the standard scenario
based in the type-I seesaw.  First of all, while leptogenesis in
type-I seesaw is driven by right-handed (RH) neutrinos which do not
couple to gauge bosons, in type-II seesaw (as well as in type-III
seesaw) the states which dynamically generate the $B-L$ asymmetry do
have electroweak interactions. Since at high temperatures gauge
reactions are much more faster than the Universe Hubble expansion, one
may be tempted to believe that gauge couplings constitute a
non-circumventable obstacle which unavoidably imply the inviability of
leptogenesis in these scenarios. This however is not the case
\cite{Hambye:2012fh,Hambye:2000ui,Hambye:2005tk,Hambye:2003rt,
  AristizabalSierra:2010mv,AristizabalSierra:2011ab}. Once
the temperature of the heat bath reaches the mass of the decaying
triplet, gauge reactions---being doubly Boltzmann suppressed---rapidly
decouple and the dynamics becomes dominated by Yukawa reactions which
then operate to a large extent as in the type-I seesaw case.
Secondly, while a single scalar triplet suffices for fitting neutrino
masses and mixing angles, leptogenesis requires extra degrees of
freedom (e.g. in the form of extra heavier triplets or RH
neutrinos). Thirdly, since the scalar triplet is not a self-conjugated
particle as a RH neutrino is, a scalar triplet/anti-triplet asymmetry
develops \cite{Hambye:2005tk,Hambye:2012fh}, thus calling for an
additional Boltzmann equation accounting for the new asymmetry
populating the heat bath. As a result, while the SM scalar asymmetry
in the type-I case is fully determined by the evolution of the $B-L$
asymmetry, here it is determined in addition by the evolution of the
triplet scalar asymmetry.

 Certainly one of the main differences between
  type-I and type-II seesaws resides on the feasibility of on-shell
  collider production of the seesaw states. At LHC scalar triplet
  production proceeds mainly via gauge boson exchange, with a
  cross-section which depending on the triplet mass can be as large as
  $\sim 1\,\text{pb}^{-1}$
  \cite{Franceschini:2008pz,delAguila:2008cj}. Subsequent decay of the
  scalar triplet, in particular to the dilepton channel, combined with
  possible displaced vertices may eventually allow the reconstruction
  of the Lagrangian parameters, as has been shown in
  \cite{Franceschini:2008pz}. Production, however, requires the scalar
  triplet to be below $\sim 1\,\text{TeV}$, mass values for which
  producing a baryon asymmetry consistent with data is not possible
  due to late $B-L$ production and electroweak sphaleron decoupling
  \cite{Hambye:2005tk,Strumia:2008cf}, a result which even in the flavored regime
  remains valid (see a similar discussion for leptogenesis in the
  type-III seesaw framework in \cite{AristizabalSierra:2010mv}). 

In this paper we aim to study the generation of the $B-L$ asymmetry
arising from the CP violating and out-of-equilibrium decays of a
scalar triplet, taking into account in a systematic way any relevant
effect that a SM interaction could have at a given temperature. This
includes the flavor effects of the charged lepton Yukawa
couplings\footnote{The role played by lepton flavor effects in
  production as well as the evolution of the flavored $B/3-L_i$
  asymmetries have been partially considered in
  Ref.~\cite{Felipe:2013kk}.} and the ``spectator'' effects of the
quark Yukawa couplings (in particular the role of the top Yukawa
reaction) and the sphalerons processes. To this end, we will first
derive the full network of flavored Boltzmann equations and then will
consider the redistribution of the $B/3-L_i$ asymmetries in the heat
bath, which in turn requires considering the conservation laws and
chemical equilibrium conditions implied by slow and fast reactions.

With these tools at hand, and in order to illustrate how does scalar
triplet flavored leptogenesis works, we will analyze two scenarios.
($i$) A scenario where the extra degrees of freedom correspond to
additional scalar triplets, with the lepton number conserving CP
flavored asymmetries naturally dominating the generation of the $B-L$
asymmetry (purely flavored leptogenesis (PFL) scenario
\cite{AristizabalSierra:2007ur,AristizabalSierra:2009bh,
  AristizabalSierra:2009mq,GonzalezGarcia:2009qd}); ($ii$) general
triplet leptogenesis models involving lepton number violating CP
asymmetries stemming from the presence of any seesaw state heavier
than the decaying scalar triplet (RHNs, fermion or extra scalar
electroweak triplets). 

The paper is organized as follows. In Sec.~\ref{sec:generalities} we
fix our notation, discuss tree-level triplet decays, neutrino mass
generation and the different CP asymmetries. In
Sec.~\ref{sec:kinetic-eqs} we derive the network of flavored and
unflavored Boltzmann equations, discuss chemical equilibration and
analytical solutions to the flavored Boltzmann equations. In
Secs.~\ref{sec:PFL-scenarios} and \ref{sec:single-triplet-scenarios}
we study scenarios ($i$) and ($ii$). Finally in Sec.~\ref{sec:concl}
we present our conclusions. In Appx.~\ref{ThermoDef} we present useful
formul{\ae}.

\section{Generalities}
\label{sec:generalities}
As regards the CP asymmetries, the details of scalar triplet
leptogenesis strongly depend on the extra beyond SM degrees of
freedom.  As already pointed out, here we aim to analyze two generic
scenarios: ($i$) models featuring several scalar triplets, or in other
words extended pure type-II seesaw models, focusing on the cases where
the $B-L$ asymmetry production is dominated by the lepton conserving
CP asymmetries; ($ii$) models involving a scalar triplet (minimal
type-II seesaw) and heavier seesaw states, more specifically focusing
on the effects of the lepton number violating asymmetries. The latter
are particularly relevant for models where the generation of a $B-L$
asymmetry becomes possible due to the interplay between type-I and
type-II seesaws, scenarios arising in many well-motivated gauge
extensions of the SM.

\subsection{Interactions and tree-level triplet decays}
\label{sec:int-decays}
The new interactions induced by extending both the scalar and fermion
sectors of the SM with $n_\Delta$ scalar $SU(2)$ triplets
($\Delta_\alpha$) and $n_R$ RH neutrinos ($N_\alpha$) can be written,
in the basis in which the RH neutrino Majorana and charged lepton mass
matrices are diagonal, according to
\begin{align}
  \label{eq:Lagrangian-type-I}
  {\cal L}^{(I)}&= i \overline{N_\alpha}\cancel \partial N_\alpha
  -\overline{N_\alpha} \lambda_{\alpha i}\tilde{\phi}^\dagger
  \ell_{L_i} - \frac{1}{2}\overline{N_\alpha}M_{N_{\alpha\alpha}} C
  \bar N_\alpha^T + \mbox{H.c.}
  \\
  \label{eq:Lagrangian-type-II}
  {\cal L}^{(II)}&= \left(D_\mu \vec{\Delta}
    _\alpha\right)^\dagger\left(D^\mu \vec{\Delta}_\alpha\right)
  -\vec{\Delta}_\alpha^\dagger m^2_{\Delta_\alpha} \vec{\Delta}_\alpha
  \nonumber\\
  & + \ell_{L_i}^T C \,i\tau_2\,Y_\alpha^{ij}
  \left(\frac{\vec{\tau}\cdot \vec{\Delta}_\alpha}
    {\sqrt{2}}\right)\ell_{L_j}+ \mu_{\Delta_\alpha}
  \tilde{\phi}^\dagger \left(\frac{\vec{\tau}\cdot
      \vec{\Delta}_\alpha} {\sqrt{2}}\right)^\dagger \phi +
  \mbox{H.c.}\,,
\end{align}
with $\ell_L^T=(\nu_L, e_L)$ and $\phi^T=(\phi^+, (v + h^0 +
i\phi_3^0)/\sqrt{2})$ the leptons and scalar boson $SU(2)$ doublets,
$\tilde{\phi}=i\tau_2 \phi^*$, $\vec{\tau}^T=(\tau_1,\tau_2,\tau_3)$
(with $\tau_i$ the $2\times 2$ Pauli matrices) and the scalar
$\Delta_\alpha$ triplets given in the $SU(2)$ fundamental
representation i.e.
$\Delta_\alpha=\left(\Delta^1_\alpha,\Delta^2_\alpha,\Delta^3_\alpha\right)$.
Here $Y_\alpha$ and $\lambda$ are $3\times 3$ and $n_R \times 3$
Yukawa matrices in flavor space and $C$ is the charge conjugation
matrix. Throughout the text we will be denoting lepton flavors $e,
\mu, \tau$ with Latin indices $i,j,k\dots$ while RH neutrinos and
scalar triplets with Greek labels $\alpha,\beta,\dots$. The covariant
derivative in (\ref{eq:Lagrangian-type-II}) reads
\begin{equation}
  \label{eq:covariant-derivative-scalars}
  D_\mu=\partial_\mu -ig \vec{T}\cdot \vec{W}_\mu -ig^\prime B_\mu\ ,
 \end{equation}
 where $\vec{T}$ are the dimension three representations of the
 $SU(2)$ generators.  In our notation the $SU(2)$ components of the
 fundamental scalar triplet representation have not all well defined
 electric charges, electric charge eigenstates are instead given by
\begin{equation}
  \label{eq:scalar-triplet}
  \boldsymbol{\Delta_\alpha}\equiv
  \frac{  \vec{\tau}\cdot \vec{\Delta}_\alpha} {\sqrt{2}}=
  \begin{pmatrix}
    \frac{\Delta^+_\alpha}{\sqrt{2}}& \Delta^{++}_\alpha\\
    \Delta^{0}_\alpha & -\frac{\Delta^+_\alpha}{\sqrt{2}}
  \end{pmatrix}\,,
\end{equation}
with the different components reading as
\begin{equation}
  \Delta^{0}_\alpha=\frac{1}{\sqrt{2}}
  \left(
    \Delta^1_\alpha+i\Delta^2_\alpha
  \right)\ ,
  \qquad
  \Delta^{+}_\alpha=\Delta^3_\alpha\ ,
  \qquad
  \Delta^{++}_\alpha\equiv\frac{1}{\sqrt{2}}
  \left(
    \Delta^1_\alpha-i\Delta^2_\alpha
  \right)\ .
\end{equation}
 In a general setup as the one determined by
  Eqs. (\ref{eq:Lagrangian-type-I}) and (\ref{eq:Lagrangian-type-II})
  the number of independent parameters, determined by the Yukawa
  coupling and mass matrices, is given by $4n_R$ moduli and $3(n_R-1)$
  CP phases in the type-I sector, while by $8 n_\Delta$ moduli and
  $3(2n_\Delta-1)$ CP phases in the type-II sector. 

The scalar interactions in (\ref{eq:Lagrangian-type-II}) induce
non-vanishing triplet vacuum expectation values which can be
calculated from the minimization of the scalar potential:
$\langle\Delta_\alpha\rangle=v_{\Delta_\alpha}\simeq
\mu_{\Delta_\alpha}v^2/2m_{\Delta_\alpha}^2$.

Both Lagrangians in (\ref{eq:Lagrangian-type-I}) and
(\ref{eq:Lagrangian-type-II}), involving lepton number violating
sources (from the coexistence of $\lambda$ and $M_N$ and of $Y$ and
$\mu_\Delta$), induce tree-level light neutrino Majorana masses
through the standard type-I (assuming $v\,\lambda\cdot M_N^{-1}\ll 1$)
and type-II seesaw mechanisms. The structure of the full neutrino mass
matrix will of course depend on whether a single or both mechanisms
intervene. Since here we will be dealing with scenarios determined by
either the setup of Eq.~(\ref{eq:Lagrangian-type-II}) or an interplay
between (\ref{eq:Lagrangian-type-I}) and
(\ref{eq:Lagrangian-type-II}), in what follows we write the effective
neutrino mass matrix in each case, namely
 \begin{align}
   \label{eq:neutrino-mass-matrix-pure}
   {\cal M}_\nu^{(II)}&=\sum_\alpha \mathcal{M}_{\Delta_\alpha}^\nu=
   \sum_\alpha \mu_{\Delta_\alpha }\frac{v^2 }{m^2_{\Delta_\alpha}}
   Y_\alpha\ ,
   \\
   \label{eq:neutrino-mass-matrix-mixed}
   {\cal M}_\nu^{(I+II)}&=\sum_\alpha {\cal M}^\nu_{\Delta_\alpha} 
   + {\cal M}_N^\nu=
   \sum_\alpha \mu_{\Delta_\alpha }\frac{v^2 }{m^2_{\Delta_\alpha}}Y_\alpha
   -\frac{v^2}{2}\lambda^T M_N^{-1} \lambda\ .
\end{align}
The light neutrino mass spectrum is thus derived from these matrices
by diagonalization through the leptonic mixing matrix
$U=U(\theta_{23})U(\theta_{13},\delta)U(\theta_{12})\hat P$, with
$\theta_{ij}$ being the neutrino mixing angles, $\delta$ the Dirac CP
phase and $\hat P=\mbox{diag}(1,e^{-i\varphi_1},e^{-i\varphi_2})$
containing the Majorana CP phases.

Regardless of the scenario considered, we are interested in the $B-L$
asymmetry generated in triplet decays. Generating a sufficiently large
$B-L$ asymmetry, that after sphaleron reconversion matches the
observed baryon asymmetry, requires certain balance between production
and washout. Production is controlled by the CP violating asymmetry
($\epsilon_{\Delta_\alpha}$) which structure is determined by the
details (interactions) of the corresponding scenario, but which in any
case arises via the interference of the tree-level decay and its
one-loop corrections, as required by the unitarity of the scattering
matrix \cite{Kolb:1979qa}.

Tree-level triplet decays involve leptonic and scalar final
states. The leptonic partial decay widths, depending on the lepton
flavor composition of the final states, involve extra factors of 1/2
which avoid overcounting:
\begin{align}
  \label{eq:leptonic-tree-level-Scalar-decays}
  \Gamma( \Delta _\alpha \to \bar \ell_i\bar \ell_j) =
  \frac{m_{\Delta_\alpha}}{8\pi} |Y_\alpha^{ij}|^2
 \left[1+|Q-1|(1-\delta_{ij}) \right]\ ,
\end{align}
where $Q$ stands for the  electric charges of the different $SU(2)$
triplet components,
$\Delta^Q_\alpha=(\Delta_\alpha^{0},\Delta_\alpha^{+},\Delta_\alpha^{++})$. 
On the other hand, scalar triplet decay modes can be written according to
\begin{equation}
  \label{eq:scalar-tree-level-Scalar-decays}
  \Gamma( \Delta_\alpha \to  \phi\phi) =
  \frac{|\mu_{\Delta_\alpha}|^2}{8\pi m_{\Delta_\alpha}}\ ,
\end{equation}
so the total decay width, after summing over lepton flavors, can be
written as
\begin{equation}
  \label{eq:Gammatot}
  \Gamma^\text{Tot}_{\Delta_\alpha}=\frac{1}{8\pi}
  \frac{m^2_{\Delta_\alpha}\tilde{m}_{\Delta_\alpha}}{v^2}
  \frac{B_{\ell}^\alpha+B_{\phi}^\alpha}{\sqrt{B_{\ell}^\alpha B_{\phi}^\alpha}}\ ,
\end{equation}
where the ``neutrino mass-like'' parameter $ \tilde m_{\Delta_\alpha}$
is defined as
\begin{equation}
  \label{eq:mtilde-Delta}
 \tilde m_{\Delta_\alpha}^2=|\mu_{\Delta_\alpha}|^2
  \frac{v^4}{m_{\Delta_\alpha}^4}\ \mbox{Tr}[Y_\alpha Y^\dagger_\alpha]  \ ,
\end{equation}
with $B_\ell^\alpha$ and $B_\phi^\alpha$ standing for the
$\Delta_\alpha$ triplet decay branching ratios to lepton and scalar  
final states:
\begin{align}
  \label{eq:lepton-and-scalar-BRs}
  B_\ell^\alpha
  &=\sum_{i=e,\mu,\tau}B_{\ell_i}^\alpha
  =\sum_{i,j=e,\mu,\tau}B_{\ell_{ij}}
  =  \sum_{i,j=e,\mu,\tau} 
  \frac{m_{\Delta_\alpha}}{8\pi  \Gamma^\text{Tot}_{\Delta_\alpha}}
  |Y_\alpha^{ij}|^2\ ,
  \nonumber\\
  B_\phi^\alpha&=\frac{|\mu_{\Delta_\alpha}|^2}
  {8\pi m_{\Delta_\alpha}\Gamma^\text{Tot}_{\Delta_\alpha}}\ ,
\end{align}
where of course the relation $B^\alpha_\ell+B^\alpha_\phi=1$ holds.
As can be seen directly from Eqs.~(\ref{eq:Gammatot}) and
(\ref{eq:mtilde-Delta}), for fixed $\tilde m_{\Delta_\alpha}$ and
$m_{\Delta_\alpha}$, $\Gamma^\text{Tot}_{\Delta_\alpha}$ exhibits a
minimum at $B^\alpha_\ell=B^\alpha_\phi=1/2$. Thus, the farther we are
from $B^\alpha_\ell=B^\alpha_\phi=1/2$, the faster the scalar triplet
decays.

\subsection{CP asymmetries in triplet decays}
\label{sec:CP-asymm}
\begin{figure}
\centering
\includegraphics[scale=0.45]{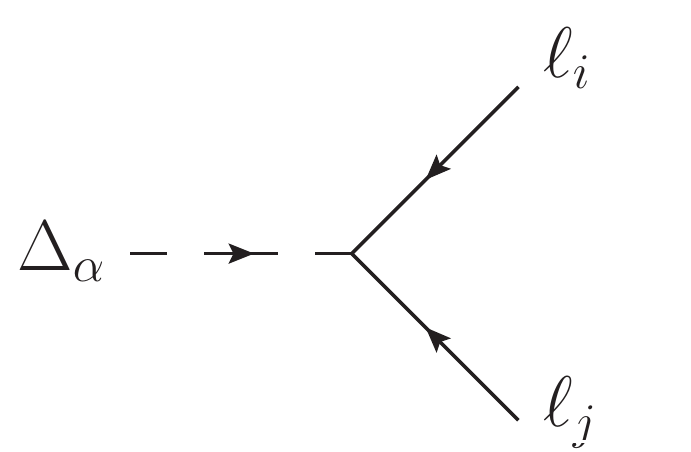}
\includegraphics[scale=0.45]{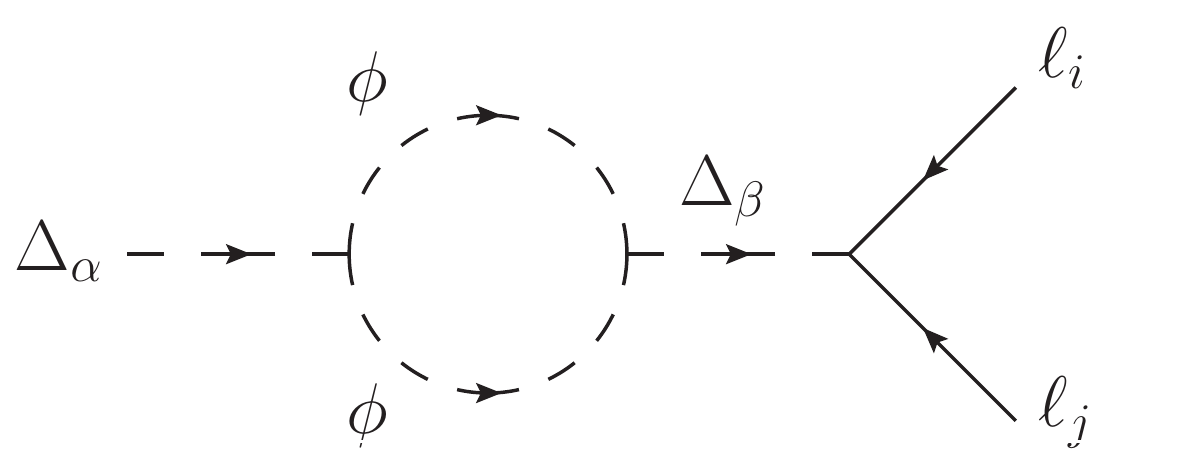}
\includegraphics[scale=0.45]{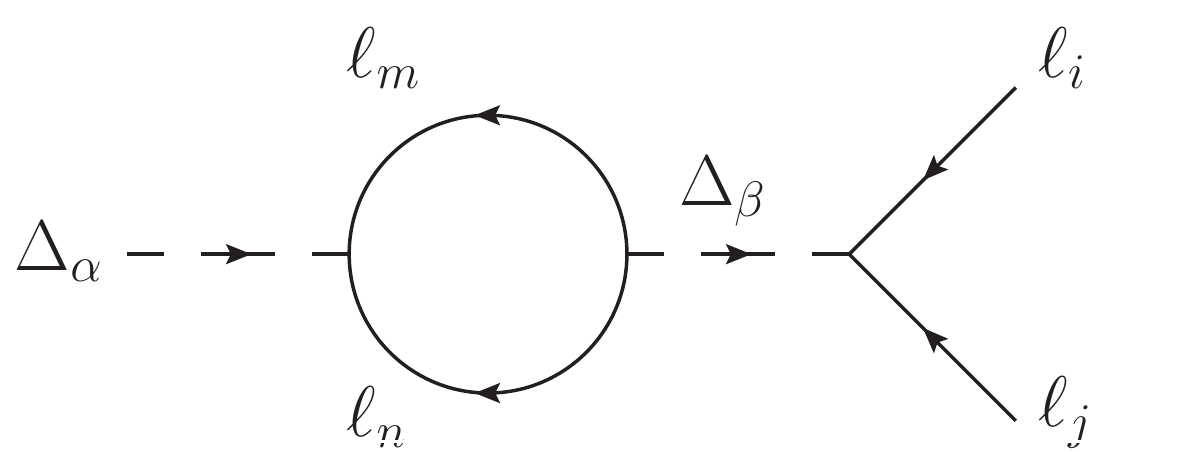}		 
\caption{\it Tree-level and one-loop Feynman diagrams responsible for
  the flavored CP asymmetry $\epsilon^{\ell_i}_{\Delta_\alpha} $ in
  the pure type-II seesaw scenario.}
\label{fig:tree-level-decay}
\end{figure}
As already pointed out, the one-loop corrections to the tree-level
decay depend on the details of the corresponding model. In purely
triplet models, that is to say models entirely determined by the
Lagrangian in (\ref{eq:Lagrangian-type-II}), the corrections to the
leptonic tree-level decay mode involve only wave-function type
corrections \cite{Ma:1998dx}. The CP asymmetry follows from the
interference between the tree-level and wave-function corrections
shown in Fig.~\ref{fig:tree-level-decay}, it therefore consists of two
pieces: a lepton number and flavor violating one (scalar loops) and a
purely flavor violating part (lepton loops). The total flavored CP
asymmetry in $\Delta_\alpha$ decays can then be written as
\begin{equation}
  \label{eq:flavored-CP-asymmetry-type-II-scenario}
  \epsilon^{\ell_i}_{\Delta_\alpha}=
  \epsilon^{\ell_i (\not{L},\not{F})}_{\Delta_\alpha}
  +
  \epsilon^{\ell_i (\not{F})}_{\Delta_\alpha}\ ,
\end{equation}
where the two pieces read
\begin{align}
  \label{eq:CPpuretype2-LNC-1}
  \epsilon^{\ell_i (\not{L},\not{F})}_{\Delta_\alpha}&=
  \frac{1}{2\pi}\sum_{\beta\neq \alpha}
  \frac{\mathbb{I}\mbox{m} 
    \left[
      \left(Y^\dagger_\alpha Y_\beta\right)_{ii}
      \mu^*_{\Delta_\alpha} \mu_{\Delta_\beta}  
    \right]}
  {m^2_{\Delta_\alpha} \mbox{Tr}[Y_\alpha  Y^\dagger_\alpha] 
    + |\mu_{\Delta_\alpha}|^2 
  }\,g(m_{\Delta_\alpha}^2/m_{\Delta_\beta}^2)\ ,
  \\
  \label{eq:CPpuretype2-LNC-2}
  \epsilon^{\ell_i (\not{F})}_{\Delta_\alpha}&=
  \frac{1}{2\pi} \sum_{\beta\neq \alpha} m^2_{\Delta_\alpha}\,
  \frac{\mathbb{I}\mbox{m}
    \left[
      \left(
        Y^\dagger_\alpha Y_\beta
      \right)_{ii}
      \mbox{Tr}[Y_\alpha  Y^\dagger_\beta]    
    \right]}
  {m^2_{\Delta_\alpha} 
    \mbox{Tr}[Y_\alpha  Y^\dagger_\alpha] 
    + |\mu_{\Delta_\alpha}|^2 
  }\,g(m_{\Delta_\alpha}^2/m_{\Delta_\beta}^2)\ ,
\end{align}
with
\begin{equation}
  \label{eq:CP-asymm-function}
  g(x) = \frac{x(1-x)}{(1-x)^2 + x y} 
\end{equation}
and $y=(\Gamma_{\Delta_\beta}^\text{Tot}/m_{\Delta_\beta})^2$. Branco:2011zb 
Note that the CP asymmetry in Eq.~(\ref{eq:CPpuretype2-LNC-1}) is in-line with what has been found in \cite{Branco:2011zb}, and that the one in Eq.~(\ref{eq:CPpuretype2-LNC-2}) is in-line with what has been found in \cite{Felipe:2013kk}. 
This piece, which
we refer to as purely flavored CP violating asymmetry, satisfies the
total lepton number conservation constraint
\begin{equation}
  \label{eq:purely-flavored-CP-asymm}
  \sum_i \epsilon^{\ell_i(\not{F})}_{\Delta_\alpha}=0\ ,
\end{equation}
and so the total CP asymmetry can consequently be written as
\begin{equation}
  \label{eq:total-CP-asymmetry-purely-triplet}
  \epsilon_{\Delta_\alpha}=\sum_{i=e,\mu,\tau}
  \epsilon^{\ell_i}_{\Delta_\alpha}=\sum_{i=e,\mu,\tau}
    \epsilon^{\ell_i (\not{L},\not{F})}_{\Delta_\alpha}\ .
\end{equation}
In terms of triplet decay observables the total flavored asymmetries
can be recasted according to
\begin{align}
  \label{eq:CPpuretype2-bis}
  \epsilon^{\ell_i}_{\Delta_\alpha}=
  -& \frac{1}{2\pi v^2} \sum_{\beta\neq \alpha}
  \frac{m^2_{\Delta_\beta}}{m_{\Delta_\alpha}}
  \frac{\sqrt{B_\ell^\alpha B_\phi^\alpha}}
  {\tilde{m}_{\Delta_\alpha}}\mathbb{I}\mbox{m}
  \left[
    \left(
      \mathcal{M}_{\Delta_\alpha}^{\nu\dagger}\mathcal{M}_{\Delta_\beta}^\nu
    \right)_{ii}
    \left(
      1 + \frac{m_{\Delta_\alpha}}{m_{\Delta_\beta}}
      \frac{\mbox{Tr}[\mathcal{M}_{{\Delta_\alpha}}^\nu
        \mathcal{M}_{{\Delta_\beta}}^{\nu\dagger}]}
      {\tilde{m}_{\Delta_\alpha}\tilde{m}_{\Delta_\beta}}
      \sqrt{\frac{B_\ell^\alpha B_\ell^\beta}
        {B_\phi^\alpha B_\phi^\beta}}
    \right)
  \right]\nonumber\\ 
  &\times g(m_{\Delta_\alpha}^2/m_{\Delta_\beta}^2)\ .
\end{align}
If flavor effects are operative, that is to say if leptogenesis takes
place below $10^{12}$~GeV, the purely flavored CP asymmetry
in~(\ref{eq:CPpuretype2-LNC-2}) will play a role in the generation of
the $B-L$ asymmetry. These asymmetries, conserving total lepton
number, involve only the $Y_\alpha$ Yukawa couplings and not the
lepton number violating parameter $\mu_{\Delta_\alpha}$. Hence, as
also noted in Ref.~\cite{Felipe:2013kk}, they are not necessarily
suppressed by the smallness of the neutrino masses. As can be seen by
comparing (\ref{eq:CPpuretype2-LNC-1}) and
(\ref{eq:CPpuretype2-LNC-2}), when the condition
\begin{equation}
\label{eq:PFLcondition}
\mu_{\Delta_\alpha}^*\mu_{\Delta_\beta}\ll
m_{\Delta_\alpha}^2\mbox{Tr}[Y_\alpha Y_\beta^\dagger]
\end{equation}
is satisfied, the purely flavored CP asymmetry overshadows the lepton
number violating piece, therefore leading to a regime where
leptogenesis is entirely driven by flavor dynamics. In terms of scalar
triplet interactions, this means that a purely flavored scalar triplet
leptogenesis scenario naturally emerges whenever the triplets couple
substantially less to SM scalars than to leptons, $B_\phi^\alpha \ll B_\ell^\alpha
$ for at least one value of $\alpha$.  Note that although
PFL scenarios in type-I seesaw can be defined as well, they differ
significantly from the purely flavored scalar triplet leptogenesis
scenario in that the latter just require suppressed lepton number
violation in a single triplet generation i.e.~suppression of lepton
number breaking interactions in the full Lagrangian is not mandatory,
as can be seen by noting that condition (\ref{eq:PFLcondition}) can be
satisfied even if $\mu_{\Delta_\alpha}/m_{\Delta_\alpha}\ll Y_\alpha$
for a single value of $\alpha$.
\begin{figure}
  \centering
  \includegraphics[scale=0.45]{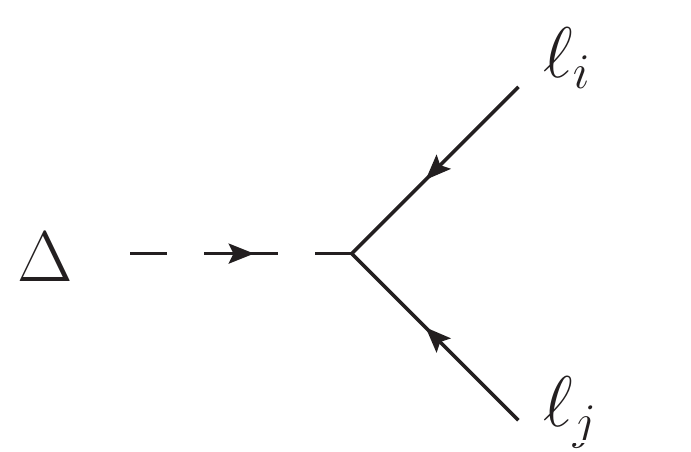}
  \qquad
  \includegraphics[scale=0.45]{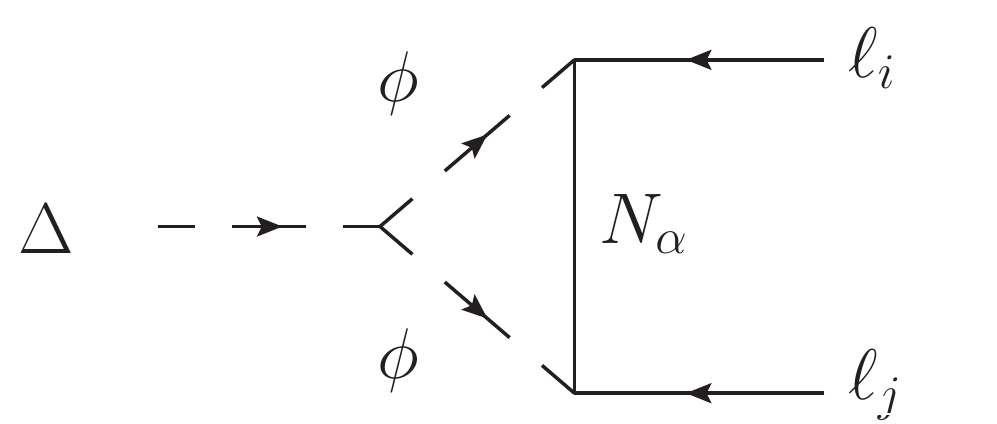}
  \caption{\it Tree-level and one-loop Feynman diagrams accounting for
    the flavored CP asymmetry $\epsilon^{\ell_i}_{\Delta_\alpha}$ in
    scenarios featuring type-I and type-II interplay.}
\label{fig:one-loop-decay-mixed}
\end{figure}

We now turn to the case where the new states beyond the scalar triplet
are RH neutrinos. In these scenarios the tree-level triplet decay
involves only a vertex one-loop correction as shown in
Fig.~\ref{fig:one-loop-decay-mixed}. The interference between the tree
and one-loop level diagrams leads to the following CP asymmetry
\cite{Hambye:2003ka,Hambye:2005tk}:
\begin{equation}
  \label{eq:CPmixedtype1+2}
  \epsilon^{\ell_i}_\Delta= - \frac{1}{4\pi}\sum_{\alpha,j}
  M_{N_\alpha}\frac{\mathbb{I}\mbox{m}
    \left[\mu_\Delta Y^{ij} \lambda^{*}_{\alpha i}\lambda^{*}_{\alpha j}
    \right]}{m^2_\Delta \mbox{Tr}[Y\,Y^\dagger]+|\mu_\Delta|^2}
  \ln\left(1+\frac{m^2_\Delta}{M^2_{N_\alpha}}\right)\ .
\end{equation}
Here the triplet generation index, being superfluous, has been
dropped. In contrast to what has been found in the previous case, the
resulting flavored CP asymmetry violates lepton flavor as well as
lepton number. So, unless a specific (and somehow arbitrary) flavor
alignment is assumed, so that $\sum_i \epsilon_{\Delta}^{\ell_i}=0$,
in these ``hybrid'' schemes PFL scenarios are not definable.

In the hierarchical case, $m_{\Delta}\ll M_{N_\alpha}$, the flavored
CP asymmetry can be recasted in terms of triplet decay observables,
namely
\begin{equation}
  \label{eq:flavored-CP-asymm-type-I-plus-type-II-T-obs}
  \epsilon^{\ell_i}_\Delta=\frac{1}{2\pi}\frac{m_{\Delta}}{v^2}
  \sqrt{B_\ell B_\phi}\frac{\mathbb{I}\mbox{m}
    \left[
      \left(
        \mathcal{M}^\nu_\Delta\mathcal{M}_{N}^{\nu\dagger}
      \right)_{ii}
    \right]}{\tilde{m}_{\Delta}}\ ,
\end{equation}
with ${\cal M}^\nu_\Delta$ and ${\cal M}_N^\nu$ given by
Eqs.~(\ref{eq:neutrino-mass-matrix-pure}) and
(\ref{eq:neutrino-mass-matrix-mixed}).  Note that in
  type-I+type-II scenarios, opposite to the scenario we consider here,
  it is also possible to generate the $B-L$ asymmetry from the decay
  of RH neutrinos via a vertex diagram involving a virtual scalar
  triplet, see in particular
  Refs.~\cite{Hambye:2003ka,Antusch:2004xy,Abada:2008gs}. 
\section{Boltzmann equations}
\label{sec:kinetic-eqs}
In general, the evolution equations of particle asymmetries in the
early Universe couple all particles species and thus involve a large
number of reactions. However, a simplification is possible given that
for specific temperature regimes different reactions have different
timescales. Any reaction occurring in the heat bath will necessarily
fall in one of the following categories\footnote{We thank Enrico Nardi
  for clarifying several aspects of this discussion.}:
\begin{enumerate}[(I)]
\item\label{slow-reactions} Reactions which at a given temperature
  $T_0$ are much slower than the Hubble Universe expansion rate
  $H(T_0)$: $\Gamma_\text{SR}\ll H(T_0)$.
\item\label{fast-reactions} Reactions which at a given temperature
  $T_0$ are much faster than the Hubble Universe expansion rate $H(T_0)$:
  $\Gamma_\text{FR}\gg H(T_0)$.
\item\label{comparable-reactions} Reactions which at a given
  temperature $T_0$ are comparable to the Hubble Universe expansion
  rate $H(T)$: $\Gamma_\text{CR}\sim H(T_0)$.
\end{enumerate}
At $T_0$, reactions falling in category \ref{slow-reactions} basically
have not taken place, so they are of no relevance in the actual
problem. The parameters responsible for such reactions can then be put
to zero at the Lagrangian level, leading to the corresponding early
Universe effective Lagrangian which involves new global symmetries
implying new conservation laws \cite{Fong:2010bv}. In contrast, the
reactions in \ref{fast-reactions} at $T_0$ have occurred so often that
the particles involved attain thermodynamic equilibrium and so are
subject to chemical equilibrium constraints, which enforce relations
among the different particles chemical potentials (the particle
asymmetries)\footnote{For a generic reaction $\sum_i
  X_i\rightleftharpoons \sum_i Y_i$ the chemical equilibrium condition
  read: $\sum_i \mu_{X_i}= \sum_i \mu_{Y_i}$, where $\mu_{X_i}$ and
  $\mu_{Y_i}$ are the chemical potentials of species $X_i$ and
  $Y_i$.}. These chemical equilibrium conditions, when coupled with
the constraints implied by the conservation laws of the early Universe
effective Lagrangian, allow to express the particle asymmetries of all
the species in the thermal bath in terms of quasi-conserved charge
asymmetries, the asymmetries related with charges that are only
(slowly) broken by the reactions in
\ref{comparable-reactions}. Finally, reactions of type
\ref{comparable-reactions} are not fast enough to equilibrate the
distributions of the intervening particles, and so they have to be
accounted for via Boltzmann equations, which dictate the evolution of
the quasi-conserved charge asymmetries and therefore of all the
asymmetries in the heat bath. Note that for reactions of category
\ref{slow-reactions} one has nevertheless to be cautious before
dropping them from the Boltzmann equations. A well-known example,
relevant in some cases for the dark matter abundance, is the freeze-in
regime, i.e.~slow production of dark matter particles from an
out-of-equilibrium process \cite{McDonald:2001vt}.  Further on, in
Sec.~\ref{sec:PFL-scenarios}, we will see that in the PFL scenario a
relatively similar effect, from very slow triplet interactions, can be
relevant or even crucial, i.e. dominates the whole baryogenesis
process (due to the fact that an additional asymmetry, the scalar
triplet asymmetry, populates the heat bath, thus implying an
additional Boltzmann equation).

The comparable-to-the-expansion triplet decays induce a $B-L$
($B/3-L_i$ in the lepton flavored regime) and $\phi$
asymmetries. Although stemming from the same state and occurring at the same
stage, these asymmetries follow different behaviors and so have to be
treated in different ways. Let us discuss this in more detail. In the
absence of triplet interactions, the full Lagrangian possesses a
$U(1)_{B-L}$ symmetry no matter what the value of $T_0$ is.  The $B-L$
charge asymmetry is therefore only affected by slow washouts (in that
sense is a quasi-conserved charge) which implies that it is not
entirely washed away and spreads all over the thermal bath feeding all
the SM particle asymmetries. In the same token, the $\phi$ asymmetry
is partially transferred to asymmetries in SM fermions through Yukawa
interactions (those which at $T_0$ are in thermal
equilibrium). However, while the evolution of the $B-L$ asymmetry is
analyzed with the corresponding Boltzmann equations, the analysis of
the evolution of the $\phi$ asymmetry does not require a Boltzmann
equation: its evolution is determined by the chemical equilibrium
conditions enforced by the reactions that at $T_0$ are in thermal
equilibrium.

The network of Boltzmann equations for scalar triplet leptogenesis, no
matter whether lepton flavor effects are active or not, corresponds to
a system of coupled differential equations accounting for the
temperature evolution of the triplet density
$\Sigma_\alpha=\Delta_\alpha+\Delta_\alpha^\dagger$, the triplet
asymmetry $\Delta_{\Delta_\alpha}=\Delta_\alpha-\Delta_\alpha^\dagger$
and the $B-L$ ($B/3-L_i$ in the lepton flavor regime) charge
asymmetry. The resulting network will of course---and
unavoidably---involve the scalar doublet asymmetry, for which chemical
equilibrium conditions have to be used in order to determine its
dependence with the asymmetries that feed the heat bath
($\Delta_{\Delta_\alpha}$ and $B-L$), as it is done in the standard
leptogenesis case \cite{Buchmuller:2001sr,Nardi:2005hs}. In what
follows we will derive in detail the appropriate set of equations
suitable for tackling the problem. We will closely follow the notation
of \cite{Nardi:2007jp}.

In the hot plasma, triplets are subject to reactions that either tend
to washout the $B-L$ asymmetry or to generate it. Depending on the
interaction inducing the process one can distinguish---at
tree-level---four kind of reactions: pure Yukawa, pure scalar, pure
gauge and Yukawa-scalar reactions. Explicitly, for $\Delta_\alpha$, we
have\footnote{Expressions for all the intervening reaction densities
  can be found in Appx.~\ref{ThermoDef}.}:
\begin{itemize}
\item Yukawa and scalar-induced decay and inverse decays, 
  $\Delta_\alpha\leftrightarrow \bar\ell \bar\ell$  and $\Delta_\alpha \leftrightarrow
  \phi\phi$, described by the reaction densities: 
  $\gamma_{D_\alpha}^\ell\equiv\sum_{i,j}\gamma_{\ell_i
    \ell_j}^{\Delta_\alpha}$ and
  $\gamma_{D_\alpha}^\phi\equiv\gamma_{\phi\phi}^{\Delta_\alpha}$. The
  total decay reaction density thus given by
  $\gamma_{D_\alpha}=\gamma_{D_\alpha}^\ell + \gamma_{D_\alpha}^\phi$.
\item Lepton flavor and lepton number ($\Delta L=2$) violating
  Yukawa-scalar-induced and triplet-mediated $s$ and $t$ channel
  $2\leftrightarrow 2$ scatterings   $\phi \phi \leftrightarrow \bar\ell_i
  \bar\ell_j$ and $\phi\ell_j\leftrightarrow \bar\phi\bar\ell_i$ , which are
  accounted for by the reaction densities
  $\gamma^{\phi\phi}_{\ell_i\ell_j}$ and
  $\gamma^{\phi\ell_j}_{\phi\ell_i}$.
\item Lepton-flavor-violating Yukawa-induced and triplet-mediated $s$
  and $t$ channel $2\leftrightarrow 2$ scatterings: $\ell_n \ell_m
  \leftrightarrow \ell_i \ell_j$ and $\ell_j\ell_m\leftrightarrow
  \ell_i\ell_n$, with reaction densities given by
  $\gamma^{\ell_n\ell_m}_{\ell_i\ell_j}$ and
  $\gamma^{\ell_j\ell_m}_{\ell_i\ell_n}$.
\item Gauge-induced $2\leftrightarrow 2$ scatterings as follows:
  $s$-channel gauge-boson-mediated:
  $\Delta_\alpha\Delta_\alpha\leftrightarrow FF$ ($F$ standing for
  SM fermions), $\Delta_\alpha\Delta_\alpha\leftrightarrow
  \phi\phi$ and $\Delta_\alpha\Delta_\alpha\leftrightarrow VV$ ($V$
  standing for SM gauge bosons); $t$ and $u$ channel
  triplet-mediated: $\Delta_\alpha\Delta_\alpha\leftrightarrow VV$ and
  four-point vertex $\Delta_\alpha\Delta_\alpha\leftrightarrow VV$
  reactions. All together they are characterized by the reaction
  density $\gamma_{A_\alpha}$.
\end{itemize}
Note that if the flavor degrees of freedom were to be neglected, all
the reactions---apart from those in the third item---would still be
present in their unflavored form. The reactions in third item are
therefore inherent to scalar flavored leptogenesis.

All together, these reactions lead to the following network of
flavored classical Boltzmann equations\footnote{This
    network of equations turns out to be consistent and suitable if
    one aims to study the generation of the $B-L$ asymmetry in the
    fully flavored regime, where lepton flavor decoherence is fully
    accomplished. If instead one aims to analyze the problem in
    transition regimes, a treatment based on the density matrix
    formalism will be required, as has been discussed e.g. in
    \cite{Barbieri:1999ma}.}:
\begin{align}
  \label{eq:flavored-BEqs1}
  \dot Y_{\Delta_{\Delta_\alpha}}&=
  -\left[
    \frac{Y_{\Delta_{\Delta_\alpha}}}{Y^\text{Eq}_\Sigma}
    -
    \sum_k
    \left(
    \sum_i B^\alpha_{\ell_i}C^\ell_{ik}
    -
    B^\alpha_\phi C^\phi_k
    \right)\frac{Y_{\Delta_k}}{Y^\text{Eq}_\ell}
  \right]\gamma_{D_\alpha}\ ,
  \\
  \label{eq:flavored-BEqs2}
  \dot Y_{\Sigma_\alpha}&=
  -\left(
    \frac{Y_{\Sigma_\alpha}}{Y^\text{Eq}_\Sigma} - 1
  \right)\gamma_{D_\alpha}
  -2\left[
    \left(\frac{Y_{\Sigma_\alpha}}{Y^\text{Eq}_\Sigma}\right)^2 - 1
  \right]\gamma_{A_\alpha}\ ,
  \\
  \label{eq:flavored-BEqs3}
  \dot Y_{\Delta_{ B/3-L_i}}&=
  -\left(\frac{Y_{\Sigma_\alpha}}{Y^\text{Eq}_\Sigma}-1\right)
  \epsilon^{\ell_i}_{\Delta_\alpha}
  \gamma_{D_\alpha}
  + 2\sum_j
  \left(
    \frac{Y_{\Delta_{\Delta_\alpha}}}{Y^\text{Eq}_\Sigma}
    -\frac{1}{2}\sum_k
    C^\ell_{ijk}\,\frac{Y_{\Delta_k}}{Y^\text{Eq}_\ell}
  \right)B^\alpha_{\ell_{ij}}\gamma_{D_\alpha}
  \nonumber\\
  &-2\sum_{j,k}
  \left(
    C^\phi_k
    +
    \frac{1}{2} C^\ell_{ijk}
  \right)\frac{Y_{\Delta_k}}{Y^\text{Eq}_\ell}
  \left(
    \gamma^{\prime \phi\phi}_{\ell_i\ell_j}
    +
    \gamma^{\phi\ell_j}_{\phi\ell_i}
  \right)
  -\sum_{j,m,n,k}
  C^\ell_{ijmnk}\,
  \frac{Y_{\Delta_k}}{Y^\text{Eq}_\ell}
  \left(
    \gamma^{\prime \ell_n \ell_m}_{\ell_i\ell_j}
    +
    \gamma^{\ell_m \ell_j}_{\ell_i\ell_n}
  \right)\ .
\end{align}
Here we have adopted the following conventions (details can be found
in Appx.~\ref{ThermoDef}). A fraction of the asymmetry generated in
$\ell_i$ is transferred to RH charged leptons, $e_i$, via SM Yukawa
interactions, and so $L_i=2\ell_i+e_i$.  We use the particle number
density-to-entropy ratio defined as $Y_{\Delta_X}=\Delta
n_X/s=(n_X-n_{\bar X})/s$, where $n_X$ ($n_{\bar X}$) is the number
density of species $X$ ($\bar X$) and $s$ is the entropy density.  We
have defined $Y_{\Delta_\Delta}\equiv
Y_{\Delta_{\Delta^0}}=Y_{\Delta_{\Delta^+}}=Y_{\Delta_{\Delta^{++}}}$
and $Y_{\Delta_\phi}\equiv
Y_{\Delta_{\phi^0}}=Y_{\Delta_{\phi^+}}$. The derivative is denoted
according to $\dot Y\equiv s H z_\alpha dY/dz_\alpha$, with $H$ the
expansion rate of the Universe, and as usual
$z_\alpha=m_{\Delta_\alpha}/T$. Primed $s$-channel scattering reaction
densities refer to the rates with resonant intermediate state
subtracted: $\gamma^\prime=\gamma-\gamma^\text{on-shell}$. Finally the
matrices $C^\ell_{ijk}$ and $C^\ell_{ijmnk}$ are defined according to
\begin{align}
  \label{eq:flavor-coupling-matrices}
  C^\ell_{ijk}&=C^\ell_{ik} + C^\ell_{jk}\ ,
  \nonumber\\
  C^\ell_{ijmnk}&=C^\ell_{ik} + C^\ell_{jk} -C^\ell_{mk} - C^\ell_{nk}\ ,
\end{align}
where the $C^\ell$ and $C^\phi$ matrices ({\it asymmetry coupling
  matrices}) relate the asymmetry in lepton and scalar doublets with
the $B/3-L_k$ and triplet asymmetries---the ``fundamental''
asymmetries present in the plasma---according to
\begin{equation}
  \label{eq:lepton-doublets-scalar-doublet-BmL-asymm}
  Y_{\Delta_{\ell_i}}=-\sum_k \,C^\ell_{ik}\,Y_{\Delta_k}
  \qquad
  \mbox{and}
  \qquad
  Y_{\Delta_\phi}=-\sum_k \,C^\phi_k\,Y_{\Delta_k}\ .
\end{equation}
In these relations the asymmetries $Y_{\Delta_k}$ are given by the components
of the asymmetry vector
\begin{equation}
  \label{eq:asymm-vector-full}
  \vec Y_\Delta =
  \begin{pmatrix}
    Y_{\Delta_\Delta}\\
    Y_{\Delta_{B/3-L_k}}
  \end{pmatrix}\ ,
\end{equation}
and the structure of the $C^\ell$ and $C^\phi$ {\it asymmetry coupling
  matrices} becomes determined by the constraints coming from the
global symmetries of the effective Lagrangian and the chemical
equilibrium conditions enforced by those SM reactions
which in the relevant temperature regime (the regime at which the
$B-L$ asymmetry is generated) are faster than the Universe Hubble
expansion rate. The final baryon asymmetry is then given by
\begin{equation}
  \label{eq:final-baryon-asymmetry}
  Y_{\Delta_B}=3\times\frac{12}{37}\sum_i\,Y_{\Delta_{B/3-L_i}}\ ,
\end{equation}
where the factor 3 accounts for the different $SU(2)$ degrees of
freedom of the scalar triplet.

\begin{figure}
  \centering
  \includegraphics[scale=0.61]{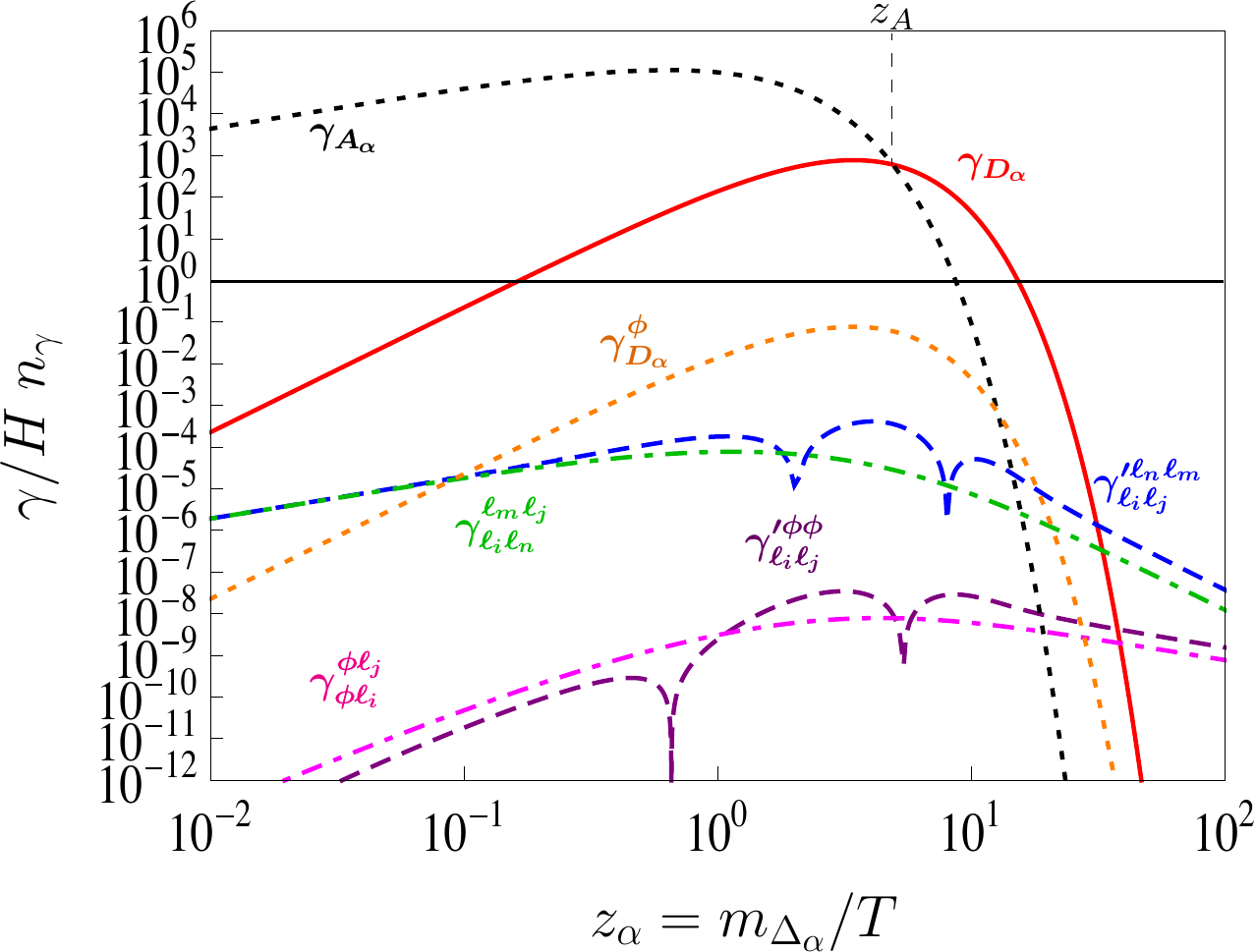}
  \includegraphics[scale=0.61]{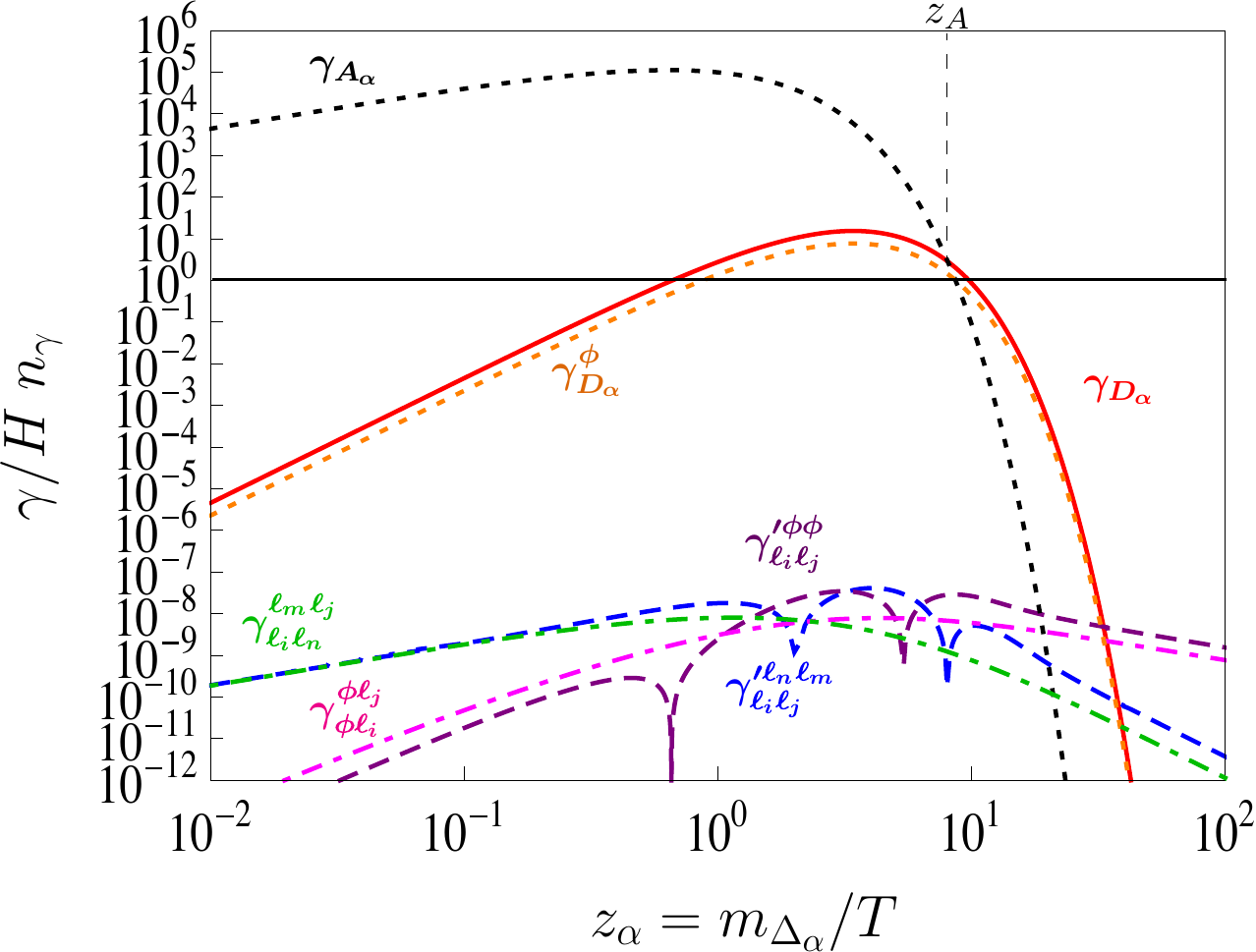}
  \caption{\it Reaction densities for the different processes involved
    in scalar triplet flavored leptogenesis. In the left-hand side
    plot  $B_\phi=10^{-4}$  while in the right-hand side plot
    $B_\phi=B_\ell=0.5$ ($B_\ell=1-B_\phi$). The remaining parameters
    have been fixed according to
     $m_{\Delta_\alpha}=10^9\,\text{GeV}$ and $\tilde
    m_{\Delta_\alpha}= 10^{-2}\,\text{eV}$.} 
  \label{fig:reactionrates}
\end{figure}
Before discussing chemical equilibration, we also write Boltzmann
equations valid in the case where the top Yukawa-related reactions are
either the only fast Yukawa processes ($10^{12}\,\text{GeV}\lesssim
T\lesssim 10^{15}\,\text{GeV}$)  or slow ($T\gtrsim
10^{15}\,\text{GeV}$), or when quantum lepton flavor coherence is
already broken but an alignment in lepton flavor space is fixed (i.e.~$\Delta$ couples to only one flavor combination). We will refer these cases  as ``the one lepton flavor approximation'' (see Appx.~\ref{Cell-one-flavor-limit} for further details).
In addition
to Eq.~(\ref{eq:flavored-BEqs2}) which holds no matter the regime, one
has
\begin{align}
  \label{eq:network-aligned-Triplet}
  \dot Y_{\Delta_{\Delta_\alpha}}&=
  -\left[
    \frac{Y_{\Delta_{\Delta_\alpha}}}{Y^\text{Eq}_\Sigma}
    -
    \sum_k
    \left(
      B^\alpha_\ell C^\ell_k
      -
      B^\alpha_\phi C^\phi_k
    \right)\frac{Y_{\Delta_k}}{Y^\text{Eq}_\ell}
  \right]\gamma_{D_\alpha}\ ,\\
  \label{eq:network-aligned-BmL}
    \dot Y_{\Delta_{B-L}}&=
  -\left(\frac{Y_{\Sigma_\alpha}}{Y^\text{Eq}_\Sigma}-1\right)
  \epsilon_{\Delta_\alpha}
  \gamma_{D_\alpha}
  + 2\left(
    \frac{Y_{\Delta_{\Delta_\alpha}}}{Y^\text{Eq}_\Sigma}
    -\sum_k
    C^\ell_k\,\frac{Y_{\Delta_k}}{Y^\text{Eq}_\ell}
  \right)B^\alpha_\ell\gamma_{D_\alpha}
  \nonumber\\
  &-2\sum_k
  \left(
    C^\phi_k
    +
    C^\ell_k
  \right)\frac{Y_{\Delta_k}}{Y^\text{Eq}_\ell}
  \left(
    \gamma^{\prime \phi\phi}_{\ell\ell}
    +
    \gamma^{\phi\ell}_{\phi\ell}
  \right)\ ,
\end{align}
where in this case the asymmetry vector is reduced to $\vec
Y_{\Delta}^T=(Y_{\Delta_\Delta},Y_{\Delta_{B-L}})$, and so the
relation between the lepton doublet asymmetry and $\vec Y_\Delta$
reads like in (\ref{eq:lepton-doublets-scalar-doublet-BmL-asymm}),
dropping the lepton flavor index. Note that the evolution equations
derived in Ref.~\cite{Hambye:2005tk} match with
(\ref{eq:network-aligned-Triplet})-(\ref{eq:network-aligned-BmL})
provided in the latter all SM Yukawa interactions effects are
neglected, see Eq.~(\ref{eq:hypercharge-neutrality-Hight-T}) below.

A final comment before we proceed with the following section. A quite
accurate calculation of the resulting $B-L$ asymmetry can be done by
considering only decays, inverse decays, gauge induced reactions and
the off-shell pieces of the $s$-channel processes: $\gamma^{\prime\phi
  \phi}_ {\ell_i \ell_j}$ and $\gamma^{\prime\ell_n \ell_m}_{\ell_i
  \ell_j}$, which guarantee that the resulting equations have a
consistent thermodynamic behavior.  This is demonstrated by
Fig.~\ref{fig:reactionrates} where we have plotted the different
reaction densities as a function of $z_\alpha$ by fixing the relevant
parameters according to
 $m_{\Delta_\alpha}=10^9\,\text{GeV}$, $\tilde
m_{\Delta_\alpha}=  10^{-2}$ eV and $B_\phi^\alpha=10^{-4}$
($B_\phi^\alpha=0.5$)  for the plot on the left (right). Thus, from now
on and throughout the numerical calculation we will drop the third and
fourth term in Eq.~(\ref{eq:flavored-BEqs3}). In the network of
unflavored kinetic equations such approximation implies dropping the
third term in Eq.~(\ref{eq:network-aligned-BmL}).
\subsection{Chemical equilibrium conditions}
\label{sec:chemical-equilibrium}
At very high temperatures ($T\gtrsim 10^{15}$~GeV) all SM
reactions are frozen in the sense of item \ref{slow-reactions}. As the
temperature drops, certain reactions (those driven by the largest
couplings first) attain thermal equilibrium which demands kinetic as
well as chemical equilibrium of the corresponding reactions, the
latter in turn enforce constraints among the different chemical
potentials of the intervening particles. Since for a relativistic
species $X$ the particle number density-to-entropy ratio is, at
leading order in $\mu/T$, related with the chemical potential
according to \cite{Harvey:1990qw}:
\begin{equation}
  \label{linkdensitymu1}
  Y_{\Delta_X}= \frac{T^2}{6s}\,g_X\,\mu_X\
  \begin{cases}
    1 \ , &\text{for fermions}
    \\[2mm]
    2\ , &\text{for bosons}\ ,
  \end{cases}
\end{equation}
(with $g_X$ the number of degrees of freedom\footnote{ Using our previous definitions, we have $g_\Delta=g_\phi=g_{Q_i}=g_{u_i}=g_{d_i}=g_{\ell_i}=g_{e_i}=1$.} ) the chemical equilibrium
constraints thus relate the different particle asymmetries of those
species participating in fast reactions. 

In principle, there is a chemical potential (an asymmetry) for each
particle in the thermal bath, which implies that {\it a priori} there
are as many chemical potentials as particles in the plasma:  $61$. 
This
number, however, is largely reduced due to the constraints imposed by
the set of chemical equilibrium conditions and the conservation laws
of the early Universe effective Lagrangian. Depending on the
temperature regime where the $B-L$ asymmetry is generated, the
possible constraints on the chemical potentials are:
\begin{enumerate}[1.]
\item \label{ceqC-1} Chemical potentials for gauge bosons vanish
  $\mu_{W^i}=\mu_{\cal B}=\mu_g=0$, and so the components of the
  electroweak and color multiplets have the same chemical potentials
  \cite{Harvey:1990qw}. This already reduces to $17$ the number of
  independent asymmetries.
\item \label{ceqC-2} Regardless of the temperature regime,
  cosmological hypercharge neutrality must be obeyed, namely
  \begin{equation}
    \label{eq:hyp-neutrality}
    {\cal Y}=
    \sum_i
    \left(
      \mu_{Q_i} + 2 \mu_{u_i} - \mu_{d_i} 
      - \mu_{\ell_i} - \mu_{e_i}
      + 2 \mu_\phi + 6 \mu_\Delta
    \right)=0\ .
  \end{equation}
\item \label{ceqC-3} Non-perturbative QCD instanton and electroweak
  sphaleron reactions---if in thermal equilibrium---enforce the
  following constraints:
  \begin{equation}
    \label{eq:QCD-instanton-EW-sphaleron}
    \sum_i \left(2\mu_{Q_i} - \mu_{u_i} - \mu_{d_i}\right)=0\ ,
    \qquad
    \sum_i \left(3\mu_{Q_i} + \mu_{\ell_i}\right)=0\ .
  \end{equation}
  The temperature at which the QCD instanton reactions attain
  equilibrium has been estimated to be $T\sim 10^{13}$~GeV
  \cite{Moore:1997im,Bento:2003jv} while for electroweak sphaleron
  processes, being controlled by $\alpha_{EW}$ rather than $\alpha_S$,
  it has been found to be about a factor 20 smaller
  \cite{Bento:2003jv}.
\item \label{ceqC-4} Finally Yukawa reactions when being in thermal
  equilibrium lead to the chemical equilibrium constraints:
  \begin{alignat}{2}
    \label{eq:up-type-quarks}
    \mbox{Up-type quarks:}&\qquad &\mu_{u_i}-\mu_{Q_i}-\mu_\phi&=0\ ,\\
    \label{eq:down-type-quarks}
    \mbox{Down-type quarks:}&\qquad &\mu_{d_i}-\mu_{Q_i}+\mu_\phi&=0\ ,\\
    \label{eq:lepton-type-quarks}
    \mbox{Charged leptons:}&\qquad &\mu_{e_i}-\mu_{\ell_i}+\mu_\phi&=0\ .
  \end{alignat}
  Top Yukawa-induced reactions are in thermodynamic equilibrium for
  $T\lesssim 10^{15}$~GeV. Bottom,  charm  and tau Yukawa-induced processes are
  in equilibrium at $T\lesssim 10^{12}$~GeV,  strange and muon  at $T\lesssim 10^{9}$~GeV, and the first
  generation Yukawa-induced processes at $T\lesssim 10^{5}$~GeV \cite{Barbieri:1999ma,Nardi:2006fx,Abada:2006fw}. 
\end{enumerate}
The exact number of non-vanishing chemical potentials as well as the
number of chemical equilibrium conditions are fixed only when a
specific temperature window is settled. Once this is done, the
resulting system of equations is solved in terms of a single set of
variables, which we take to be $\mu_{B/3-L_i}$ and $\mu_\Delta$.  The
solution thus provides the relations between the asymmetries of all
the particles in the heat bath with the independent asymmetries
$\{Y_\Delta\}=\{Y_{\Delta_\Delta},Y_{B/3-L_i}\}$ appearing in the
asymmetry vector in (\ref{eq:asymm-vector-full}).

In what follows we briefly discuss the symmetries of the corresponding
early Universe effective Lagrangian and the relevant chemical
equilibrium conditions in \ref{ceqC-1}-\ref{ceqC-4} which enable us to
calculate the rectangular matrices relating the lepton and scalar
doublet asymmetries with $\{Y_\Delta\}$, as given by
Eqs.~(\ref{eq:lepton-doublets-scalar-doublet-BmL-asymm}). In each
regime, when applicable, we also discuss in
Appx.~\ref{Cell-one-flavor-limit} the one-flavor limit by taking
flavor alignments as in Ref. \cite{Nardi:2005hs} and deriving the
corresponding $C^{\ell,\phi}$ matrices needed in such approximations.
We start by discussing the high temperature regime $T>10^{15}$~GeV,
proceeding subsequently to the  temperature ranges
  $T\subset [10^{12},10^{15}]$~GeV, $[10^{9},10^{12}]$~GeV,  $[10^{5},10^{9}]$~GeV and
  $T<10^{5}$~GeV.  Theses ranges are based on the assumption that all
  SM interactions that approximately enter in thermal equilibrium at a
  similar temperature do it effectively at the same temperature. We
  stress that some of these temperature ``windows'' differ from those
  used in Ref.~\cite{Nardi:2006fx}, in particular in what regards the
  charged lepton Yukawa reaction equilibrium temperatures. They
  however match with those pointed out in
  Ref.~\cite{Barbieri:1999ma}. 
\begin{itemize}
\item {\it None SM reactions in thermal equilibrium,
    $T\gtrsim 10^{15}$~GeV}:\\
  In this regime all SM reactions are slow in the sense of
  \ref{slow-reactions}, and so only triplet-related interactions are
  relevant. A proper treatment of the problem therefore should be done
  with the unflavored kinetic equations in (\ref{eq:flavored-BEqs2}),
  (\ref{eq:network-aligned-Triplet}) and
  (\ref{eq:network-aligned-BmL}), bearing in mind that since in the
  heat bath the triplet is subject only to scalar- and Yukawa-induced
  interactions, in Eq.~(\ref{eq:flavored-BEqs2}) the second term can
  be neglected.  With all SM reactions frozen, only the triplet,
  lepton and scalar doublets develop chemical potentials:
  $\mu_\Delta$, $\mu_\ell$, $\mu_\phi$. These chemical potentials are
  subject only to the hypercharge neutrality constraint in
  \ref{ceqC-2}, which leads to the following $C^\ell$ and $C^\phi$
  matrices:
  \begin{equation}
    \label{eq:hypercharge-neutrality-Hight-T}
    C^\ell=
    \begin{pmatrix}
      0 & 1/2
    \end{pmatrix}\ ,
    \qquad
    C^\phi=
    \begin{pmatrix}
      3 & 1/2
    \end{pmatrix}\ .
  \end{equation}
  The resulting Boltzmann equations match those derived in
  Refs.~\cite{Hambye:2012fh,Hambye:2005tk} since there, regardless of
  the temperature considered, no SM Yukawa interactions effects were
  taken into account, neither from the quarks nor from the charged
  leptons. Strictly speaking, this is an accurate procedure only in
  this regime, $T\gtrsim 10^{15}$~GeV, given that none of the SM
  Yukawa reactions are in thermal equilibrium (from now on, when
  necessary we will refer to this literature-reference-scheme as the
  ``unflavored case''). It is worth stressing that through
  Eq.~(\ref{eq:lepton-doublets-scalar-doublet-BmL-asymm}) the $C^\phi$
  matrix in Eq.~(\ref{eq:hypercharge-neutrality-Hight-T}) leads to the
  sum rule:
  \begin{equation}
  6Y_{\Delta_\Delta} + 2Y_{\Delta_\phi} + Y_{\Delta_{B-L}}=0\,.
  \end{equation}
  This expression is nothing else but the sum rule employed in
  Ref.~\cite{Hambye:2005tk} (taking into account the fact that, in
  this reference, $Y_{\Delta_\Delta}$ and $Y_{\Delta_\phi}$ involve a
  sum on the $SU(2)$ degrees of freedom, so that one must substitute
  $Y_{\Delta_\Delta}\to 3 Y_{\Delta_\Delta}$ and $Y_{\Delta_\phi}\to 2
  Y_{\Delta_\phi}$).  Note that above $10^{15}$~GeV there is
  relatively little time for the reheating to occur before the
  temperature goes below the scalar triplet mass (assuming the
  reheating occurs below the Planck scale).  So, unless the triplet
  Yukawa couplings are such that at $T\sim m_\Delta$ the triplet still
  follows a thermal distribution (strong washout regime), the final
  baryon asymmetry produced will depend on the initial scalar triplet
  number density (initial condition), i.e. will further depend on the
  details of the reheating.  We will not consider these possible
  effects here.  Note also that above $10^{15}$~GeV, for
  $\tilde{m}_\Delta\sim 0.05$~eV one gets non perturbative Yukawa
  couplings if $B_\phi\lesssim 8 \cdot 10^{-3}$.
\item {\it Only top Yukawa-related reactions in thermal equilibrium,
     $T\subset [10^{12},10^{15}]$~GeV }:\\
  Within this temperature regime, apart from top Yukawa-related
  interactions which are fast, all SM Yukawa-induced reactions fall in
  category \ref{slow-reactions}. Accordingly, the correct description
  of the problem is given by the one lepton flavor approximation
  equations in (\ref{eq:flavored-BEqs2}),
  (\ref{eq:network-aligned-Triplet}) and
  (\ref{eq:network-aligned-BmL}).

  The global symmetries of the effective Lagrangian are those of the
  SM kinetic terms broken only by the top Yukawa coupling,
  and so the group of global transformations is:
  \begin{equation}
    \label{eq:Geff-R1}
    G_\text{Eff}=U(1)_Y\times U(1)_B\times U(1)_e \times U(1)_\text{PQ}
    \times SU(3)_d\times SU(3)_e\times SU(2)_Q\times SU(2)_u\ .
  \end{equation}
  The $SU(3)$ factors combined with the exact $U(1)_B$,
  $U(1)_\text{PQ}$ and the absence of Yukawa couplings for all
  SM particles, except the top quark, imply:
  $\mu_{d_i}=\mu_{e_i}=\mu_{u_{1,2}}=\mu_{Q_{1,2}}=\mu_{B}=0$. Taking
  this constraints into account and the relevant chemical
  equilibrium conditions (\ref{eq:hyp-neutrality}) and
  (\ref{eq:up-type-quarks}), the latter written only for the top
  quark, we obtain
  \begin{equation}
    \label{eq:cell-cphi-R1}
    C^\ell=
    \begin{pmatrix}
      0 & 1/2
    \end{pmatrix}\ ,
    \qquad
    C^\phi=
    \begin{pmatrix}
      2 & 1/3
    \end{pmatrix}\ .
  \end{equation}
  Including these effects will enhance the efficiency by about $20\%$
  with respect to the unflavored case, the precise value being of
  course dependent upon the parameter choice.   \item
    {\it QCD instantons, electroweak sphalerons, bottom, charm and tau
      Yukawa-related reactions in
      thermal equilibrium, $T\subset [10^{9},10^{12}]$~GeV}:\\
    In this temperature window the lepton doublets lose their quantum
    coherence due to the tau Yukawa-related interactions being in
    thermal equilibrium \cite{Nardi:2006fx,Abada:2006fw}. On the other
    hand, since electroweak sphaleron reactions are in thermal
    equilibrium, baryon number is no longer conserved, while they
    conserve the individual $B/3-L_i$ charges.  An appropriate study
    of the evolution of the $B-L$ asymmetry should then be done by
    tracking the evolution of the flavored charge asymmetries
    $B/3-L_i$ ($i=a, \tau$, the state $a$ being a coherent
    superposition of $e$ and $\mu$ lepton flavors) with the network of
    Eqs.~(\ref{eq:flavored-BEqs1})-(\ref{eq:flavored-BEqs3}). 

  The QCD instantons reactions break the global $U(1)_\text{PQ}$, the
  bottom and tau Yukawa couplings break the RH down-type quark and
  charged lepton $SU(3)$ flavor multiplet and in addition the tau
  Yukawa coupling also breaks the global $U(1)_e$. The Lagrangian is
  as expected ``less symmetric'', with the group of global
  transformations given by  
    \begin{equation}
    \label{eq:Geff-R3}
    G_\text{Eff}=U(1)_Y\times SU(2)_d
    \times SU(2)_e\times  U(1)_Q \times  U(1)_u\ .
  \end{equation}
  These global symmetries imply:
  $\mu_{u_1}=\mu_{Q_1}=0$ and $\mu_{d_i}=\mu_{e_i}=0$ with $i=1,2$, while the complete set of chemical equilibrium conditions
  correspond to (\ref{eq:hyp-neutrality}) for hypercharge neutrality
  (written so to include the now non-vanishing bottom, charm and tau chemical
  potentials), (\ref{eq:QCD-instanton-EW-sphaleron}) for QCD
  instantons, (\ref{eq:QCD-instanton-EW-sphaleron})  for electroweak sphalerons, and (\ref{eq:up-type-quarks}),
  (\ref{eq:down-type-quarks}) and (\ref{eq:lepton-type-quarks})
  written for top, bottom,charm and tau Yukawa interactions. Due to
  sphaleron reactions, lepton flavor is no longer conserved so that  
  chemical potentials develop in three independent lepton doublets:
  $\ell_\tau$, $\ell_a$ and $\ell_b$. Conservation of the $B/3-L_i$
  charges however provide the constraint $\mu_{B/3-L_b}=0$, which when
  coupled with the corresponding chemical equilibrium conditions  yields the following flavored $C^{\ell,\phi}$ matrices: 
\begin{align}
 \label{eq:C-ell-PFL-used}
 C^\ell=
 \begin{pmatrix}
   - 6/359 & 307/718 &- 18/359 \\
   39/359 &- 21/718 &  117/359
 \end{pmatrix} \ ,
\quad 
C^\phi=
\begin{pmatrix}
  258/359 &  41/359 & 56/359
\end{pmatrix}\ . 
\end{align}

  \item {\it Strange and muon Yukawa interactions in thermal equilibrium, $T\subset [10^5,10^{9}]$~GeV:}\\ 
  As pointed out in Ref.~\cite{Nardi:2006fx,Abada:2006fw}, in this temperature
  regime the lepton doublets completely lose their quantum coherence,
  implying that chemical potentials develop in each orthogonal lepton
  flavor doublet: $\ell_\tau$, $\ell_\mu$ and $\ell_e$. Since the
  second generation Yukawa reactions are no longer of type
  \ref{slow-reactions}, the symmetries of the effective Lagrangian are
  reduced to $U(1)$ factors:
  \begin{equation}
    \label{eq:Geff-R4}
    G_\text{Eff}=U(1)_Y\times U(1)_d\times U(1)_e\times 
    U(1)_Q \times U(1)_u\ .
  \end{equation}
  These constraints imply $\mu_d=\mu_e=\mu_{Q_1}=\mu_u=0$, and when
  combined with the corresponding chemical equilibrium conditions (the
  ones from previous item complemented with (\ref{eq:up-type-quarks}),
  (\ref{eq:down-type-quarks}) and (\ref{eq:lepton-type-quarks}) for
  the charm, strange and muon Yukawa interactions) yield:
  \begin{align}
  \label{second-generation-th-eq}
    C^\ell&=
    \begin{pmatrix}
      -6/179 & 151/358 & -10/179 & -10/179\\
      33/358 & -25/716 & 172/537 & -7/537\\
      33/358 & -25/716 & -7/537  & 172/537
    \end{pmatrix}\ ,
    \\
    C^\phi&=
    \begin{pmatrix}
      123/179 & 37/358 & 26/179 & 26/179
    \end{pmatrix}\ .
  \end{align}
 \item {\it  All SM reactions in thermal equilibrium, $T\lesssim
    10^5$~GeV:} \\
  In this case and until electroweak symmetry breaking, the only
  surviving symmetry is $U(1)_Y$. Due to all SM reactions
  being fast,                               all SM particles develop non-vanishing
  chemical potentials, with the chemical equilibrium conditions given 
  by the full list in items \ref{ceqC-1}-\ref{ceqC-4}. The flavored
  $C^{\ell,\phi}$ rectangular matrices in this regime therefore read:
  \begin{align}
    \label{eq:fully-flavored-Cell}
    C^\ell&=
    \begin{pmatrix}
      9/158 & 221/711 & -16/711 & -16/711\\
      9/158 & -16/711 & 221/711 & -16/711 \\
      9/158 & -16/711 & -16/711 & 221/711
    \end{pmatrix}\ ,
    \\
    \label{eq:fully-flavored-Cphi}
    C^\phi&=
    \begin{pmatrix}
      39/79 & 8/79 & 8/79 & 8/79
    \end{pmatrix}\ .
  \end{align}
\end{itemize}

\subsection{Domain of validity of the various sets of flavored Boltzmann equations}
\label{domain-of-validity}
The temperature ranges discussed in the previous Sec. are determined
from the assumption that lepton flavor decoherence
 happens  as soon as the corresponding lepton Yukawa
interaction rate becomes faster than the Hubble rate, at a temperature
$T\equiv T_h$.  
  Lepton flavor decoherence is a delicate issue which requires a pure
  quantum treatment, which in full generality does not even exist for
  the more widely considered {\it standard} leptogenesis
  picture. Here, in this Sec. rather than providing an exhaustive
  treatment of this issue, we will consider a simplified treatment considering the two most relevant processes: SM lepton Yukawa reactions (given approximately by Eq. (\ref{eq:Yuk-reactions})) and  lepton-related triplet inverse decays, basically along the lines of  Ref. \cite{Blanchet:2006ch}. 

If at the time when a lepton Yukawa interaction rate becomes faster
than the Hubble rate, the triplet inverse decay processes
 $\ell\ell\to\bar\Delta$  are much faster than this
reaction,  the  coherent superposition of leptons
produced from the decay of a scalar triplet will inverse decay before
it has the time to undergo any  red charged  lepton Yukawa
interaction.  In this case it is expected that decoherence is fully
achieved only later when the inverse decay rate, which is Boltzmann
suppressed at low temperatures, gets smaller than the SM lepton Yukawa
rate, at a temperature $T\equiv T_\text{decoh}$. Between $T_h$ and
$T_\text{decoh}$, one lies in an intermediate regime where flavor
effects are suppressed.

\begin{figure}
  \centering 
  \includegraphics[width=7.9cm,height=7.cm]{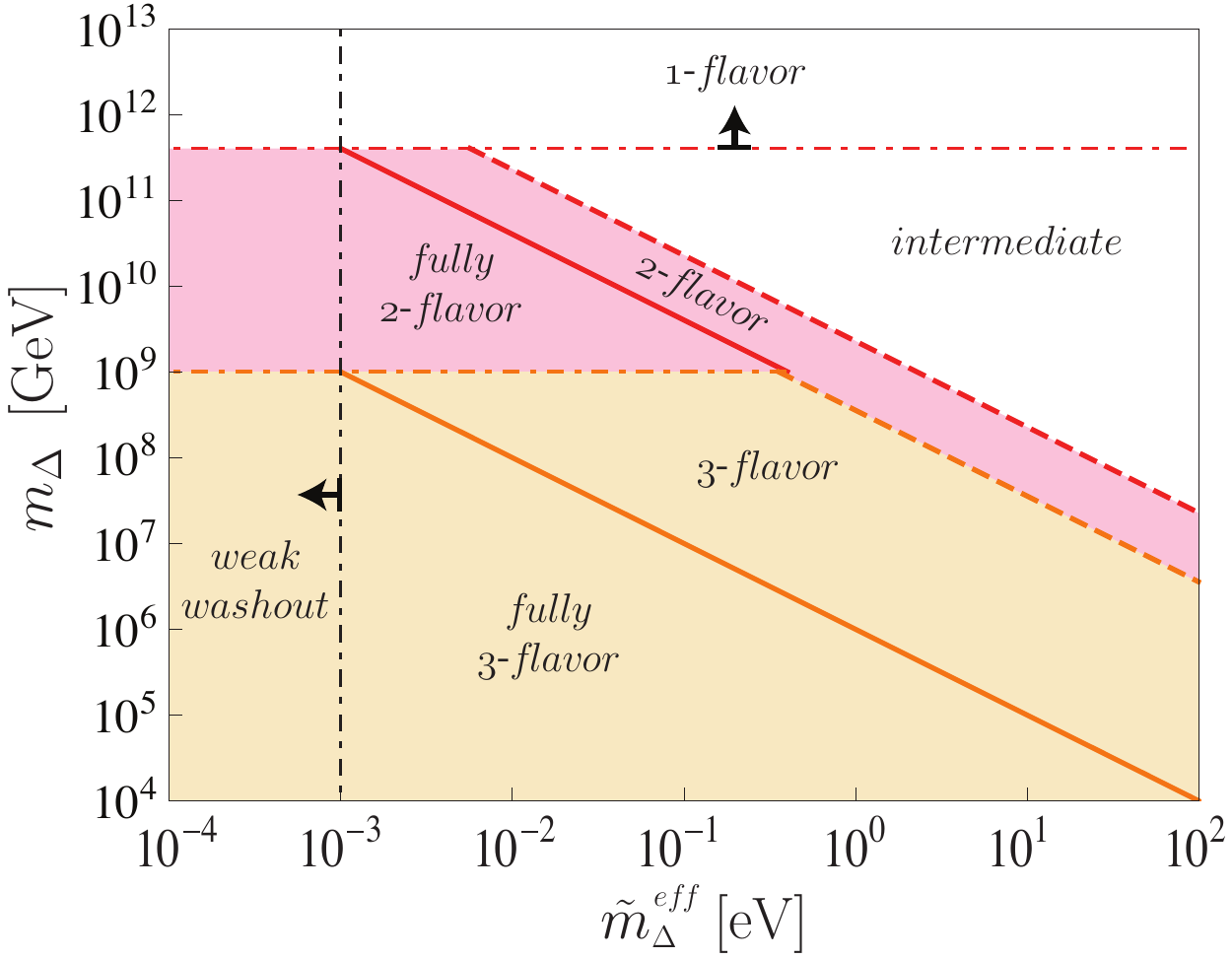}
  \includegraphics[width=7.9cm,height=7.cm]{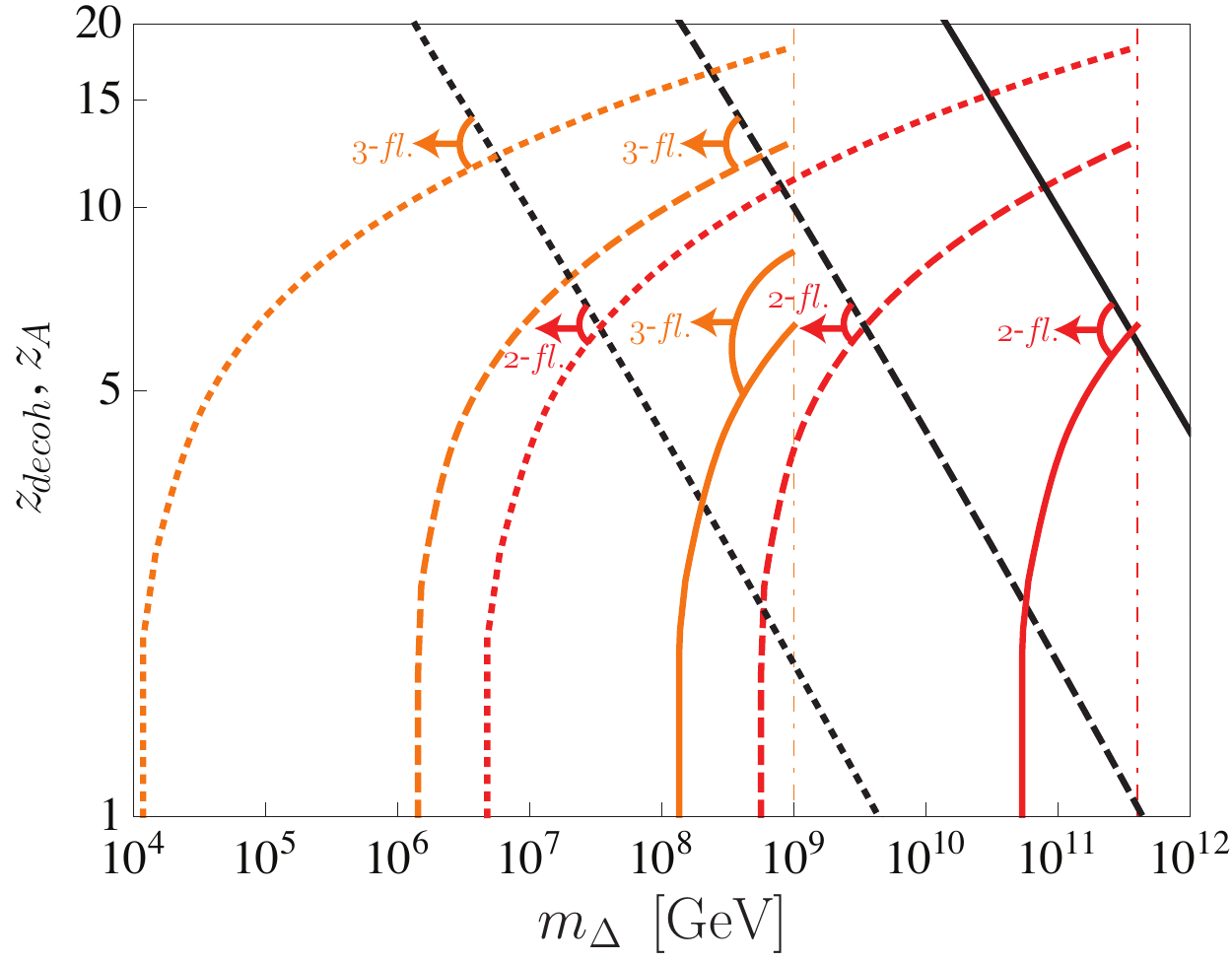}
  \caption{\it \underline{Le}f\underline{t:} regions determining the different flavor regimes
    as a function of $\tilde{m}^{\text{eff}}_\Delta$ and $m_\Delta$. The
    region below the red (orange) solid line is obtained by the
    requirement that the $\tau$($\mu$) Yukawa rate is always faster
    than the $\ell\ell\to\bar\Delta$ inverse decay rate, determining
    the fully 2(3)-flavors regime. The region below the red (orange) dashed
    line is obtained by the requirement that the $\tau$($\mu$) Yukawa
    rate is faster than the $\ell\ell\to\bar\Delta$ inverse decay rate
    for $z\geq z_A$ (see the right plot).  The red (orange) horizontal dot-dashed line
    corresponds to the value of $m_\Delta$ above which the $\tau$($\mu$) Yukawa never reach thermal equilibrium. The vertical
    dot-dashed line corresponds to the value of
    $\tilde{m}^{\text{eff}}_\Delta$ below which inverse decays never
    reach thermal equilibrium.  \underline{Ri}g\underline{ht:} $z^{\tau}_\text{decoh}$ (red),
    $z^{\mu}_\text{decoh}$ (orange) and $z_A$ (black) as a function of
    $m_\Delta$, for $\tilde{m}_\Delta^{\text{eff}} = 0.01$~eV (solid),
    $1$~eV (dashed) and $100$~eV (dotted), where
    $z_\text{decoh}=m_\Delta/T_\text{decoh}$ and
    $z_{A}=m_\Delta/T_{A}$. 
    The vertical red (orange) dot-dashed line corresponds to
    the value of $m_\Delta$ above which the $\tau$($\mu$) Yukawa
    never reach thermal equilibrium. 
     As explained in the text, for $z\leq z^\tau_{decoh}$ ($z> z^\tau_{decoh}$), the 1(2)-flavor Boltzmann equations must be used (and similarly for $z^\mu_{decoh}$ with 2(3)-flavors). On the other hand,  the $z_A$ lines determine  when the use of a simple set of Boltzmann equations for all $z$ gives a reliable result. Since the   asymmetry   is mainly produced when $z>z_A$,   if $z_A>z^\tau_{decoh}$ ($z_A>z^\mu_{decoh}$) it is indeed a good approximation to use a 2(3)-flavor Boltzmann equation set for all $z$.}    
  \label{fig:zdecoh-vs-mdelta}
\end{figure}

The parameters which determine $T_\text{decoh}$ are $m_\Delta$ and the
 inverse leptonic decay effective parameter:
\begin{equation}
    \label{m-tilde-eff}
    \tilde{m}^{\text{eff}}_\Delta \equiv 
    \tilde{m}_\Delta \sqrt{\frac{1-B_\phi}{B_\phi}}\ .
\end{equation} 
 Imposing that the lepton-related triplet inverse
  decays never get faster than a given SM Yukawa reaction at a given
  temperature, one can derive  upper   bounds on the triplet mass as a
  function of $\tilde m^\text{eff}_\Delta$, in the same way it has
  been done in the type-I seesaw case \cite{Blanchet:2006ch}.  
 These bounds are shown in the left-hand side plot in
  Fig.~\ref{fig:zdecoh-vs-mdelta}, with the constraints applying in
  the tau (muon) case displayed in solid red (orange) line, labeled by  ``fully 2(3)-flavor''.
Analytically  the bounds are  given by
 the requirement that 
\begin{equation}
  \label{tau-muon-constraints}
  \Gamma_{f_i}\gtrsim B_\ell \, \Gamma^{\text{Tot}}_{\Delta}
  \frac{ Y^{\text{Eq}}_\Sigma}{Y^{\text{Eq}}_\ell}\qquad 
  (f_i=\tau,\mu)\ ,
\end{equation}
where $B_\ell \, \Gamma^{\text{Tot}}_{\Delta}\propto
\tilde{m}^{\text{eff}}_\Delta $ (see Eq.~(\ref{eq:Gammatot})) and  where the
corresponding SM reaction rates  are  given by $\gamma_{f_i}/n_{f_i}$, with
$\gamma_{f_i}$ approximately given by
Eq.~(\ref{eq:Yuk-reactions}). The constraints in
(\ref{tau-muon-constraints}) then translate into constraints over
$m_\Delta$ and $\tilde m_\Delta^\text{eff}$, and fix the values that
these parameters should have in order to assure that triplet dynamics
takes place in either a ``fully'' two or three flavor regime, namely
\begin{align}
  \label{bound-decoher-tau}
  m_\Delta&\lesssim 4\times \left(\frac{10^{-3}
      \text{eV}}{\tilde{m}^{\text{eff}}_\Delta}\right)
  \times 10^{11}\ \text{GeV}
  &\quad& \text{fully 2-flavor regime;}\\
  \label{bound-decoher-mu}
  m_\Delta&\lesssim 1\times \left(\frac{10^{-3}\text{
        eV}}{\tilde{m}^{\text{eff}}_\Delta}\right)\times 10^{9}\
  \text{GeV}& \quad& \text{fully 3-flavor regime.}
\end{align}
For illustration we take the example of the decoherence effect
associated with the $\tau$ SM Yukawa interaction. If below the
temperature $T=T_h^\tau\simeq 10^{12}$~GeV (at which the $\tau$ Yukawa
rate gets faster than the Hubble rate) the
 $\ell\ell\to\bar\Delta$  inverse decay rate is slower
than this Yukawa rate, one enters in the 2 flavor regime defined by
Eqs.~(\ref{eq:Geff-R3}) already at $T=T_h^\tau$.  For example, for
$B_\phi=0.5$ and $\tilde{m}_\Delta\simeq 10^{-3}$~eV which gives
 $\tilde{m}_\Delta^{\text{eff}}=10^{-3}$~eV, one
trivially satisfies the condition in Eq.~(\ref{bound-decoher-tau}) for
any mass since in this case triplet inverse decays never reach thermal
equilibrium. On the contrary, if
 $\tilde{m}_\Delta^{\text{eff}}\gtrsim 10^{-3}$~eV  the
triplet inverse decay rate is faster than the $\tau$ Yukawa rate down
to a smaller temperature, $T_\text{decoh}^\tau\lesssim T_h^\tau$. In
this case one can use the network of flavored Boltzmann equations in
(\ref{eq:flavored-BEqs1})-(\ref{eq:flavored-BEqs3}), written in the
two-flavor regime, only below $T^\tau_\text{decoh}$.
 Fig.~\ref{fig:zdecoh-vs-mdelta} (right-hand plot),
  which shows the dependence of
  $T_\text{decoh}=m_\Delta/z_\text{decoh}$ (for tau and muon reactions
  displayed in red and orange colors respectively) with the triplet
  mass for several values of $\tilde{m}_\Delta^{\text{eff}}$
  (solid-dashed-dotted: $10^{-2}$-$1$-$10^2$ eV), proves that for large
  values of $\tilde{m}^{\text{eff}}_\Delta$---not satisfying the lower
  limit given by Eq.~(\ref{bound-decoher-tau})---this temperature can
  be far below $10^{12}$~GeV. 

 In more detail, taking   $m_\Delta=10^{10}$ GeV,
$\tilde{m}_\Delta=0.01$~eV and  $B_{\phi}=10^{-4}$,
which gives  $\tilde{m}_\Delta^{\text{eff}}\approx
  1$~eV, one gets $T_\text{decoh}^\tau \simeq
  10^9$~GeV.  Above $T^\tau_\text{decoh}$  one
expects the decoherence effect to be mild,
 so that for $10^9~\text{GeV}\lesssim T \lesssim
  10^{12}~\text{GeV}$   and for this parameter choice 
one should better use a set of Boltzmann equations where the QCD and
electroweak instantons as well as the top, bottom and charm Yukawa
interactions are all in thermal equilibrium but the $\tau$ Yukawa is
still effectively ``off'' for what concerns the $B-L$ asymmetry
production process (even if faster than the Hubble rate). Therefore
within this temperature range,  and for this parameter
  choice, one has still a single lepton flavour Boltzmann equation
with $C^\ell$ and $C^\phi$ matrices which take into account the
effects of all these instantons and $t$, $b$ ,$c$ Yukawa interactions
\begin{equation}
  \label{eq:Geff-tauoff}
  C^\ell= 
  \begin{pmatrix}
    0& 3/10
  \end{pmatrix}
  \qquad
  C^\phi=
  \begin{pmatrix}
    3/4 &  1/8
  \end{pmatrix}
\end{equation} 
In other words in this case, one does not consider the chemical
potential relation associated to the $\tau$ Yukawa interaction even if
the corresponding rate is faster than the Hubble rate. Strictly
speaking this relation holds for an infinitely fast reaction
rate. Here the rate is slower than the inverse decay rate (closely
related to the $B-L$ asymmetry production) and cannot be considered as
infinitely fast.  To sum up, within the $10^{12}-10^{9}$~GeV range,
 strongly
  depending on parameter configurations
  ($m_\Delta$ and $\tilde m_\Delta^\text{eff}$),
  one has two possible sets of $C^\ell$ and $C^\phi$ matrices,
  thus implying that the problem of tracking the evolution of the
  $B-L$ asymmetry is described   either by the kinetic equations given in  Eqs~(\ref{eq:network-aligned-Triplet})-(\ref{eq:network-aligned-BmL}) with    $C^\ell$
  and $C^\phi$ matrices given by Eq.~(\ref{eq:Geff-tauoff}) 
  if $T>T^{\tau}_\text{decoh}$, either   by the kinetic equations given in
  (\ref{eq:flavored-BEqs1})-(\ref{eq:flavored-BEqs3}) with  $C^\ell$
  and $C^\phi$ matrices given by Eq.~(\ref{eq:C-ell-PFL-used})
  whenever $T<T^\tau_\text{decoh}$. 

As for the next temperature range, between $T_h^\mu\simeq10^9$~GeV and
$T_h^e\simeq10^5$~GeV, where both the $s$ and the $\mu$ rate are also
faster than the Hubble rate, there one has three possible regimes:
$(i)$~the one-flavor case, as long as the $\tau$ is ``off'' (if still
it is, which implies that the $\mu$ Yukawa ``off'' too); $(ii)$~the
two-lepton-flavor case when the $\tau$ is ``on'' but the $\mu$ Yukawa
is still ``off''; $(iii)$~the three-flavor case discussed in the
previous Sec. when both the $\tau$ and $\mu$ Yukawas are ``on'', and
for which the $C^\ell$ and $C^\phi$ matrices are given by
Eq.~(\ref{second-generation-th-eq}). Finally below $10^5\,$~GeV where
the up, down and electron Yukawa interaction rates are faster than the
Hubble rate, one has 4 situations depending on which interactions are
``off'': $(i)$~$\tau,\mu,e$ ``off'' (one flavor), $(ii)$~$\mu,e$
``off'' (2 flavors), $(iii)$~$e$ ``off'' (3 flavors) and $(iv)$~all
interactions ``on'' (3 flavors).  Apart from case
  $(iv)$, for which the $C^\ell$ and $C^\phi$ matrices are given in
  Eqs.~(\ref{eq:fully-flavored-Cell}) and
  (\ref{eq:fully-flavored-Cphi}), the corresponding sets of $C^\ell$
and $C^\phi$ matrices for the remaining situations are given in
Appx.~\ref{Cell-other-regime}. The temperature ranges where they hold
are given in the right panel of Fig.~\ref{fig:zdecoh-vs-mdelta}, for $\mu$ and $\tau$ only for the sake of clarity.

For the cases where one would have several sets of Boltzmann equations
to take into account successively as the temperature goes down, one
important remark to be done is that the transition between these
regimes might be non-trivial to treat in a satisfactory way. Just
assuming a step function in temperature from one regime to the next
one could easily constitute a too rough procedure. In the following we
will not consider such kind of cases. In fact in many situations this
question turns out to be of little numerical importance.

In the type-I case, basically this is of no numerical importance if
the inverse decays have never been faster than the lepton Yukawa rate
\cite{Blanchet:2006ch}, a condition which in the type-II case gives
Eq.~(\ref{bound-decoher-tau}).  However, for the type-II case this
condition turns out to be too conservative. The difference comes from
the fact that the scalar triplets, unlike right-handed neutrinos, have
gauge interactions. As explained in detail in
e.g.~Ref.~\cite{Hambye:2012fh,Hambye:2005tk}, see also
section~\ref{sec:single-triplet-scenarios} below, this implies that as
long as the gauge scattering rate is faster than the decay rate,
scalar triplets gauge scatter before they have the time to decay, and
the asymmetry production is highly suppressed.  Only from the
temperature ``$T_A$'' where the gauge scattering rate gets smaller
than the decay rate, a substantial asymmetry can develop itself.  This
means that if $T_\text{decoh}\gtrsim T_A$, all what happens at
$T>T_\text{decoh}$ is anyway irrelevant and one can safely use only
the set of Boltzmann equations where decoherence is assumed. If
instead $T_\text{decoh}\ll T_A$ the asymmetry produced for $T<
T_\text{decoh}$ will be suppressed from the fact that the number of
triplets remaining at $T\sim T_\text{decoh}$ is Boltzmann suppressed.
In this case one expects the unflavored period to dominate the
production of the asymmetry, as the number of triplets still present
at $T\simeq T_A$ is larger.  This means that for the PFL case to be
discussed in Sec. \ref{sec:PFL-scenarios}, where there is no asymmetry
production in the unflavored regime, better $T_\text{decoh}\gtrsim
T_A$.

In practice the condition $T_\text{decoh} \gtrsim T_A$ is much less
restrictive than Eq.~(\ref{bound-decoher-tau}).  In the right panel of
Fig.~\ref{fig:zdecoh-vs-mdelta}, we plotted in black the values of
$z_A=m_\Delta/T_A$ red for different values of
  $\tilde{m}^{\text{eff}}_\Delta$. For example, if
  $\tilde{m}^{\text{eff}}_\Delta=1$~eV, one observes that
  $T_\text{decoh}^\tau\gtrsim T_A$ requires $m_\Delta\lesssim
  10^{9}$~GeV, while if $\tilde{m}_\Delta=100$~eV instead, one
  observes that $T_\text{decoh}^\tau\gtrsim T_A$ requires
  $m_\Delta\lesssim 10^7$~GeV.  Similarly, in the left panel of
Fig.~\ref{fig:zdecoh-vs-mdelta} we added as a function of
$\tilde{m}_\Delta^{\text{eff}}$ the  upper  bound which holds on
$m_\Delta$ if one considers this condition rather than the one in
Eq.~(\ref{bound-decoher-tau}).   The corresponding region are labeled  by ``2(3)-flavor'' following that we require $T^\tau_{\text{decoh}}\gtrsim T_A$ or $T^\mu_{\text{decoh}}\gtrsim T_A$.   One should close this section by saying
again that the use of $T_\text{decoh}$ as a sharp transition
temperature is a reasonable assumption one will make, but it does not
probably take into account the fact that partial decoherence could
already occur at higher temperature.
\subsection{Formal integration of Boltzmann equations}
\label{sec:formal-integration-BEQs}
\begin{figure}
  \centering
  \includegraphics[scale=0.6]{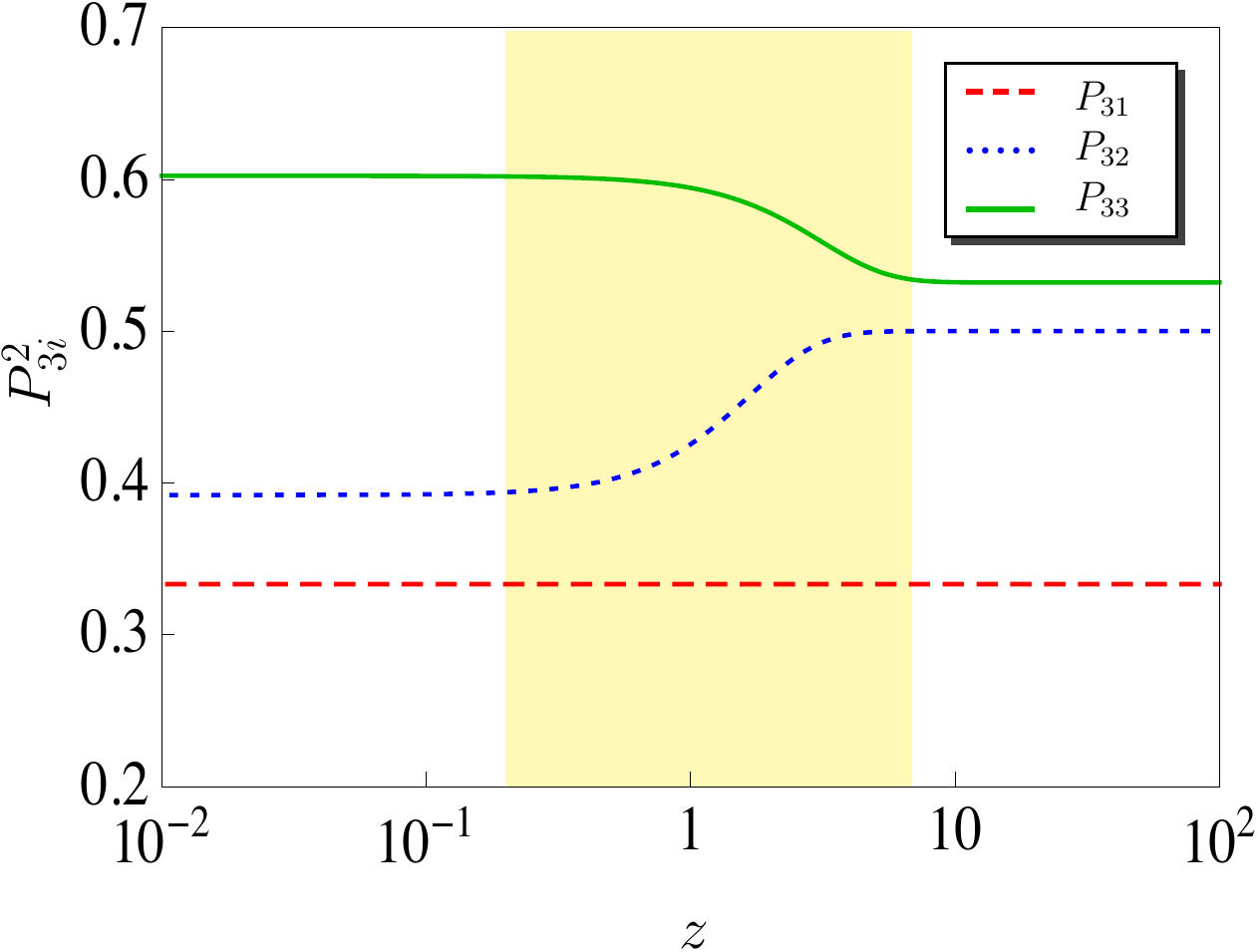}
  \caption{\it ${\cal P}$ eigenvectors-third-component ${\cal P}_{3i}$
    as a function of $z$. The eigenvectors have been evaluated for the
    flavor configuration $B_{\ell_{ii}}=0$ and
    $B_{\ell_{12}}=B_{\ell_{21}}=(1-B_\phi)/2$, with $B_\phi=10^{-4}$.
    We have checked that this result is quite insensitive to changes
    in the flavor configuration. The vertical yellow stripe indicates
    the range where the matrix ${\cal P}$ slightly depends upon $z$.}
  \label{fig:eigenvectors-z-dependence}
\end{figure}

Keeping only leading order terms in Eq.~(\ref{eq:flavored-BEqs3}),
i.e.~dropping third and forth terms, an analytic formal integration of
the equations responsible for the $B-L$ asymmetry can be accomplished,
 basically along the same lines of the type-I
  seesaw case \cite{Antusch:2010ms}.  For definitiveness we will focus on the two flavor
regime, results for the three flavor regime can be readily derived
following the same procedure we will outline. In the two flavor regime
the asymmetry vector introduced in Sec.~\ref{sec:kinetic-eqs} (see
Eq.~(\ref{eq:asymm-vector-full})) is given by
\begin{equation}
  \label{eq:asymmetry-vector}
  \vec Y_\Delta=
  \begin{pmatrix}
    Y_{\Delta_{\Delta_\alpha}}\\
    Y_{\Delta_{B/3-L_a}}\\
    Y_{\Delta_{B/3-L_\tau}}
  \end{pmatrix}\ .
\end{equation}
In terms of this vector, Eqs.~(\ref{eq:flavored-BEqs1}) and
(\ref{eq:flavored-BEqs3}) can be casted in matricial form, namely
\begin{equation}
  \label{eq:matricial-DEqs}
  \frac{d}{dz} \vec{Y}_\Delta(z)
  =-\left(\frac{Y_{\Sigma_\alpha}}{Y_{\Sigma_\alpha}^\text{Eq}}-1\right)
  D(z)\,\vec\varepsilon
  -D(z){\cal M}(z)\vec{Y}_\Delta(z)\ ,
\end{equation}
with
\begin{equation}
  \label{eq:D[z]}
  D(z)=\frac{\gamma_{D_\alpha}(z)}{s(z)\,H(z)\,z}\ ,
\end{equation}
and where the {\it CP-asymmetry-vector} $\vec \varepsilon$ is defined
as
\begin{equation}
  \label{eq:CP-asymmetry-vector}
  \vec \varepsilon=
  \begin{pmatrix}
    0\\
    \epsilon_{\Delta_\alpha}^{\ell_a}\\
    \epsilon_{\Delta_\alpha}^{\ell_\tau}
  \end{pmatrix}
\end{equation}
while the {\it flavor-triplet-coupling-matrix} according to  
\begin{equation}
  \label{eq:flavor-triplet-CM}
  {\cal  M}(z)=
  \begin{pmatrix}
    \frac{1}{Y^\text{Eq}_\Sigma}
    -\frac{\sum_i B_{\ell_i} C^\ell_{i\Delta} - B_\phi C^\phi_\Delta}{Y^\text{Eq}_\ell}
    &-\frac{\sum_i B_{\ell_i} C^\ell_{ia} - B_\phi C^\phi_a}
    {Y^\text{Eq}_\ell}
    &-\frac{\sum_i B_{\ell_i}C^\ell_{i\tau} - B_\phi C^\phi_\tau}
    {Y^\text{Eq}_\ell}
    \\
    -2\sum_j B_{\ell_{aj}}
    \left(
      \frac{1}{Y^\text{Eq}_\Sigma} 
      -
      \frac{1}{2}\frac{C_{aj\Delta}^\ell}{Y^\text{Eq}_\ell}
    \right)
    &\sum_j B_{\ell_{aj}} \frac{C^\ell_{aja}}{Y^\text{Eq}_\ell}
    &\sum_j B_{\ell_{aj}} \frac{C^\ell_{aj\tau}}{Y^\text{Eq}_\ell}
    \\
    -2\sum_j B_{\ell_{\tau j}}
    \left(
      \frac{1}{Y^\text{Eq}_\Sigma} 
      -
      \frac{1}{2}\frac{C_{\tau j\Delta}^\ell}{Y^\text{Eq}_\ell}
    \right)
    &\sum_j B_{\ell_{\tau j}} \frac{C^\ell_{\tau ja}}{Y^\text{Eq}_\ell}
    &\sum_j B_{\ell_{\tau j}} \frac{C^\ell_{\tau j\tau}}{Y^\text{Eq}_\ell}
  \end{pmatrix}\ .
\end{equation} 
In the case ${\cal M}(z)={\cal M}$, the system of equations in
(\ref{eq:matricial-DEqs}) can be decoupled via a rotation of the
asymmetry vector $\vec Y_\Delta$, the matrix accounting for the
rotation being determined by the similarity transformation
\begin{equation}
  \label{eq:similarity-trans}
  {\cal P}^{-1}\,{\cal  M}\,{\cal  P}=\hat {\cal  M}\ ,
\end{equation}
which brings ${\cal M}$ to diagonal form. Strictly speaking ${\cal M}$
does depend on $z$, but it turns out that the $z$ dependence of the
rotation matrix ${\cal P}$ is quite moderate. As can be seen in
Fig.~\ref{fig:eigenvectors-z-dependence}, in the high as well as in
the low temperature regime ${\cal P}(z)={\cal P}$ whereas within the
window $z\subset [0.2,7]$ there is a dependence, which nevertheless is
rather soft.

Thus, taking a $z$ independent change-of-basis-matrix ${\cal P}$ and
rotating the asymmetry vector as
\begin{equation}
  \label{eq:YDelta-rot}
  \vec Y_\Delta'(z)={\cal  P}^{-1}\,\vec Y_\Delta\ ,
\end{equation}
we finally get a decoupled system of differential equations:
\begin{equation}
  \label{eq:Final-BEQs-decoupled}
  \frac{d}{dz}\vec Y_\Delta^\prime(z)=
  -\left(
    \frac{Y_{\Sigma_\alpha}}{Y_{\Sigma_\alpha}^\text{Eq}}-1
  \right)
  D(z)\,\vec{\varepsilon^\prime}
  -D(z)\hat {\cal M}(z)\vec{Y}_\Delta^\prime(z)\ ,
\end{equation}
where the {\it rotated-CP-asymmetry-vector} $\vec{\varepsilon^\prime}$
has been introduced:
\begin{equation}
  \label{eq:rotated-CP-asymm-vec}
  \vec{\varepsilon^\prime}={\cal P}^{-1}\vec\varepsilon\ .
\end{equation}
The decoupled system of equations in (\ref{eq:Final-BEQs-decoupled})
can then be formally integrated through their integrating factor. By
doing so, and assuming vanishing primordial asymmetries, $\vec
Y_\Delta(z_0)=0$ with $z_0\ll 1$, the solution reads
\begin{equation}
  \label{eq:YDelta-final-sol-primed}
  \vec Y_\Delta^\prime(z)=-\int_{z_0}^z\,dz'\,
  \frac{\gamma_{D_\alpha}(z')}
  {\gamma_{D_\alpha}(z')+4\gamma_{A_\alpha}(z')}
  \frac{dY_{\Sigma_\alpha}(z')}{dz'}\,
  e^{-\int_{z'}^{z}\,dz''D(z'')\hat{\cal M}(z'')}\,\vec{\varepsilon^\prime}\ .
\end{equation}
In terms of the ``new'' asymmetries, and due to the diagonal structure
of the matricial damping factor, one can define efficiency functions
$\eta'_i(z)$, which account for the evolution of the primed asymmetries
and their corresponding values at freeze-out ($z\to\infty$), namely
\begin{equation}
  \label{eq:primed-asymm}
  \left[\vec Y^\prime_{\Delta}(z)\right]_i=- \eta^\prime_i(z) \,\varepsilon^\prime_i\,
  Y_{\Sigma_\alpha}^\text{Eq}(z_0)\ ,
\end{equation}
where the efficiency functions can be directly read
from~(\ref{eq:YDelta-final-sol-primed}) by taking into account that, as
usual, they have been normalized to the scalar triplet equilibrium
distribution evaluated at $z_0$. The evolution of these asymmetries,
however, does not describe the evolution of the actual $B/3-L_i$
asymmetries and instead, as can be seen in (\ref{eq:YDelta-rot}), a
superposition which involves the triplet asymmetry as well. A
meaningful description requires switching to the non-primed variables,
which yields\footnote{This result has been derived by using ${\cal
    P}e^{\hat {\cal M}}{\cal P}^{-1}=e^{{\cal M}}$ and
  Eq.~(\ref{eq:flavored-BEqs2}), taking into account that
  $Y_\Sigma(z)$ follows quite closely the equilibrium distribution
  function so $y_\Sigma+1\simeq 2$.}
\begin{equation}
  \label{eq:YDelta-final-sol}
  \vec Y_\Delta(z)=-\int_{z_0}^z\,dz'\,
  \frac{\gamma_{D_\alpha}(z')}
  {\gamma_{D_\alpha}(z')+4\gamma_{A_\alpha}(z')}
  \frac{dY_{\Sigma_\alpha}(z')}{dz'}\,
  e^{-\int_{z'}^{z}\,dz''D(z''){\cal M}(z'')}\,\vec\varepsilon\ .
\end{equation}
In the non-primed basis the matricial damping factor is no longer
diagonal and therefore defining efficiency functions, as it was done
in the primed basis, is no longer possible: both $B/3-L_a$ and
$B/3-L_\tau$ are a superposition of two terms weighted by the
corresponding CP asymmetries $\epsilon_{\Delta_\alpha}^{\ell_a}$ and
$\epsilon_{\Delta_\alpha}^{\ell_\tau}$.  Let us discuss this in more
detail. The $i$-th component of the asymmetry vector in
(\ref{eq:YDelta-final-sol}) can be written as
\begin{equation}
  \label{eq:YDelta-final-sol-ith-comp}
  \left[\vec Y_{\Delta}(z)\right]_i=-\int_{z_0}^z\,dz'\,
  \frac{\gamma_{D_\alpha}(z')}
  {\gamma_{D_\alpha}(z')+4\gamma_{A_\alpha}(z')}
  \frac{dY_{\Sigma_\alpha}(z')}{dz'}\,
  \sum_{k=1,2,3}
  \left[
    e^{-\int_{z'}^{z}\,dz''D(z''){\cal M}(z'')}
  \right]_{ik}
  \,\varepsilon_k\ ,
\end{equation}
thus implying that in the primed basis the flavored asymmetries become
\begin{align}
  \label{eq:flavored-asymm-non-primed-basis}
  Y_{B/3-L_a}(z)&=-
  \left[
   \eta_{aa}(z) \epsilon_{\Delta_\alpha}^{\ell_a}
    +
   \eta_{a\tau}(z) \epsilon_{\Delta_\alpha}^{\ell_\tau}
  \right]Y^\text{Eq}_{\Sigma_\alpha}(z_0)\ ,
  \nonumber\\
  Y_{B/3-L_\tau}(z)&=-
  \left[
    \eta_{\tau a}(z)\epsilon_{\Delta_\alpha}^{\ell_a}
    +
    \eta_{\tau\tau}(z)\epsilon_{\Delta_\alpha}^{\ell_\tau}
  \right]Y^\text{Eq}_{\Sigma_\alpha}(z_0)\ ,
\end{align}
with the flavored efficiency functions defined as:
\begin{equation}
  \label{eq:flavored-efficiency-functions-non-primed-basis}
  \eta_{ik}(z)=
  \frac{1}{Y^\text{Eq}_{\Sigma_\alpha}(z_0)}
  \int_{z_0}^z\,dz'\,
  \frac{\gamma_{D_\alpha}(z')}
  {\gamma_{D_\alpha}(z')+4\gamma_{A_\alpha}(z')}
  \frac{dY_{\Sigma_\alpha}(z')}{dz'}\,
  \left[
  e^{-\int_{z'}^{z}\,dz''D(z''){\cal M}(z'')}
  \right]_{ik}\ .
\end{equation}
So, once lepton flavors are taken into account---in general---the
efficiencies are no longer flavor diagonal. The presence of the flavor
off-diagonal efficiencies is a manifestation of flavor coupling which,
in contrast to the type-I seesaw-based leptogenesis case, persists
even when $C^\ell=\mathbb{I}$, due to the intricate structure of the
{\it flavor-triplet-coupling-matrix}. More precisely this occurs
because, in contrast to the type-I seesaw leptogenesis case, an
asymmetry in the state generating the $B-L$ asymmetry develops
($Y_{\Delta_\Delta}$), and so an additional kinetic equation
accounting for this asymmetry turns out to be mandatory. Due to the
presence of this equation the asymmetries in flavor $a$ and $\tau$ are
indirectly coupled, and such coupling becomes manifest in the
exponential function in
Eq.~(\ref{eq:flavored-efficiency-functions-non-primed-basis}). In
other words, unlike standard leptogenesis, flavor coupling effects are
unavoidable in scalar triplet leptogenesis.

A specific case where flavored efficiency functions, in the same sense
of (\ref{eq:primed-asymm}), can be properly defined corresponds to PFL
scenarios. What actually happens in those cases is that due to the PFL
condition $\sum_i\epsilon_{\Delta_\alpha}^{\ell_i}=0$, which implies  
$\epsilon^\ell_{\Delta_\alpha}\equiv-\epsilon_{\Delta_\alpha}^{\ell_a}
=\epsilon_{\Delta_\alpha}^{\ell_\tau}$, the off-diagonal efficiency
functions can be hidden by suitable redefinitions:
\begin{align}
  \label{eq:flavored-symm-and-efficiencies-PFL-case}
  Y_{B/3-L_a}(z)&=
  \left[\eta_{aa}(z) - \eta_{a\tau}(z)\right]\epsilon^\ell_{\Delta_\alpha}\, 
  Y^\text{Eq}_{\Sigma_\alpha}(z_0)
  \to
  \eta_a(z) \,  \epsilon^\ell_{\Delta_\alpha}\,
  Y^\text{Eq}_{\Sigma_\alpha}(z_0) \ ,
  \nonumber\\
  Y_{B/3-L_\tau}(z)&=
  \left[\eta_{\tau a}(z) - \eta_{\tau\tau}(z)\right] 
  \epsilon^\ell_{\Delta_\alpha}\, Y^\text{Eq}_{\Sigma_\alpha}(z_0)
  \to
  \eta_\tau(z) \, \epsilon^\ell_{\Delta_\alpha}\,  
  Y^\text{Eq}_{\Sigma_\alpha}(z_0)\ ,
\end{align}
and so the total $B-L$ asymmetry can be written as
\begin{equation}
  \label{eq:total-BmL-PFL-case-in-terms-of-eff}
  Y_{\Delta_{B-L}}(z)=
  \left[
    \eta_a(z)
    +
    \eta_\tau(z)
  \right] \epsilon^\ell_{\Delta_\alpha}
\, Y^\text{Eq}_{\Sigma_\alpha}(z_0) \ ,
\end{equation}
with the final value (the value at freeze-out) given by
$Y_{\Delta_{B-L}}=Y_{\Delta_{B-L}}(z\to\infty)$.
\section{Purely flavored triplet leptogenesis}
\label{sec:PFL-scenarios}
For concreteness and in order to analyze as well as to demonstrate the
viability of this scenario, we will fix the triplet mass spectrum to
be hierarchical ($m_{\Delta_\alpha}\ll m_{\Delta_\beta}$ with
$\alpha<\beta$) and assume that the $B-L$ asymmetry is entirely due to
the dynamics of the lightest state  $\Delta_\alpha\equiv\Delta$  (henceforth we drop the
triplet generation index).  We will also consider two-flavored regime
situations where the $B-L$ asymmetry is distributed along the $\tau$
and $a$ lepton flavor directions ($a$ being an admixture of $\mu$ and
$e$ flavors)\footnote{Note that in the regime where all the charged
  lepton SM Yukawa interactions are in thermodynamic equilibrium
  ($T\ll 10^5$~GeV) lepton flavor equilibrating processes would render
  this PFL scenario unviable \cite{AristizabalSierra:2009mq}.}.

As previously argued (see Eq.~(\ref{eq:PFLcondition}) and the
corresponding discussion), when the scalar triplet CP asymmetries
arise from the presence of another scalar triplet, there exists an overall regime
in which the purely flavored CP asymmetries are larger than the lepton
number violating CP asymmetries, thus leading to a natural
realization (to a very good approximation) of a PFL successful
scenario.  Strictly speaking PFL scenarios are defined by the
condition $\sum_i\epsilon^{\ell_i}=0$ \cite{AristizabalSierra:2007ur},
however in a more general fashion whenever the condition
$|\sum_i\epsilon^{\ell_i}_\Delta|< |\epsilon^{\ell_i}_\Delta|$ (for
any given value of $i$) is satisfied a PFL scenario can be defined as
well.  This is actually the condition which is generically satisfied,
as soon as Eq.~(\ref{eq:PFLcondition}) holds, i.e. if one or
both scalar triplets couple substantially less to scalars than they do
to leptons.

The viability of PFL scenarios demands leptogenesis to take place in
the flavored regime, i.e. requires leptogenesis to occur at
$T\leq T_\text{decoh}$ (see
  Sec.~\ref{domain-of-validity}), and furthermore it requires more
than a dominance of the purely flavored CP asymmetries.
Since the
sum of the purely flavored CP asymmetries vanishes (total
lepton number is conserved), if there were only source terms, a net non-vanishing
$B-L$ asymmetry would not develop due to an exact cancellation among
the different $B/3-L_i$ asymmetries.  
This cancellation has to be mandatorily avoided in order that a net
non-vanishing total $B-L$ asymmetry develops.  In type-I seesaw, this
is possible due to the lepton flavor dependence of the washout effect,
which allows the $B/3-L_i$ asymmetries to be washed-out in different
amounts. In other words, the production of a net $B-L$ asymmetry in
the PFL type-I case, which involves $L$-conserving CP asymmetries as
well, is closely related to the action of $L$-violating inverse decay
rates larger than the Hubble Universe expansion rate (fast
$L$-violating inverse decays), so that they reprocess the $B/3-L_i$
asymmetries in different amounts, in such a way that these asymmetries
do not compensate each other anymore.

In the type-II scenario a similar effect is also possible, provided
decay/inverse decay to leptons and to scalars reach thermal
equilibrium at some stage during the production of the $B-L$
asymmetry, so that $L$-violating processes do induce a washout.
Additionally, and this is a new effect which does not exist in the PFL
type-I scenario, this is also possible even if the $L$-breaking
processes present in the heat bath never reach thermal equilibrium.

\begin{figure}
  \centering 
  \includegraphics[width=7.9cm,height=6.5cm]{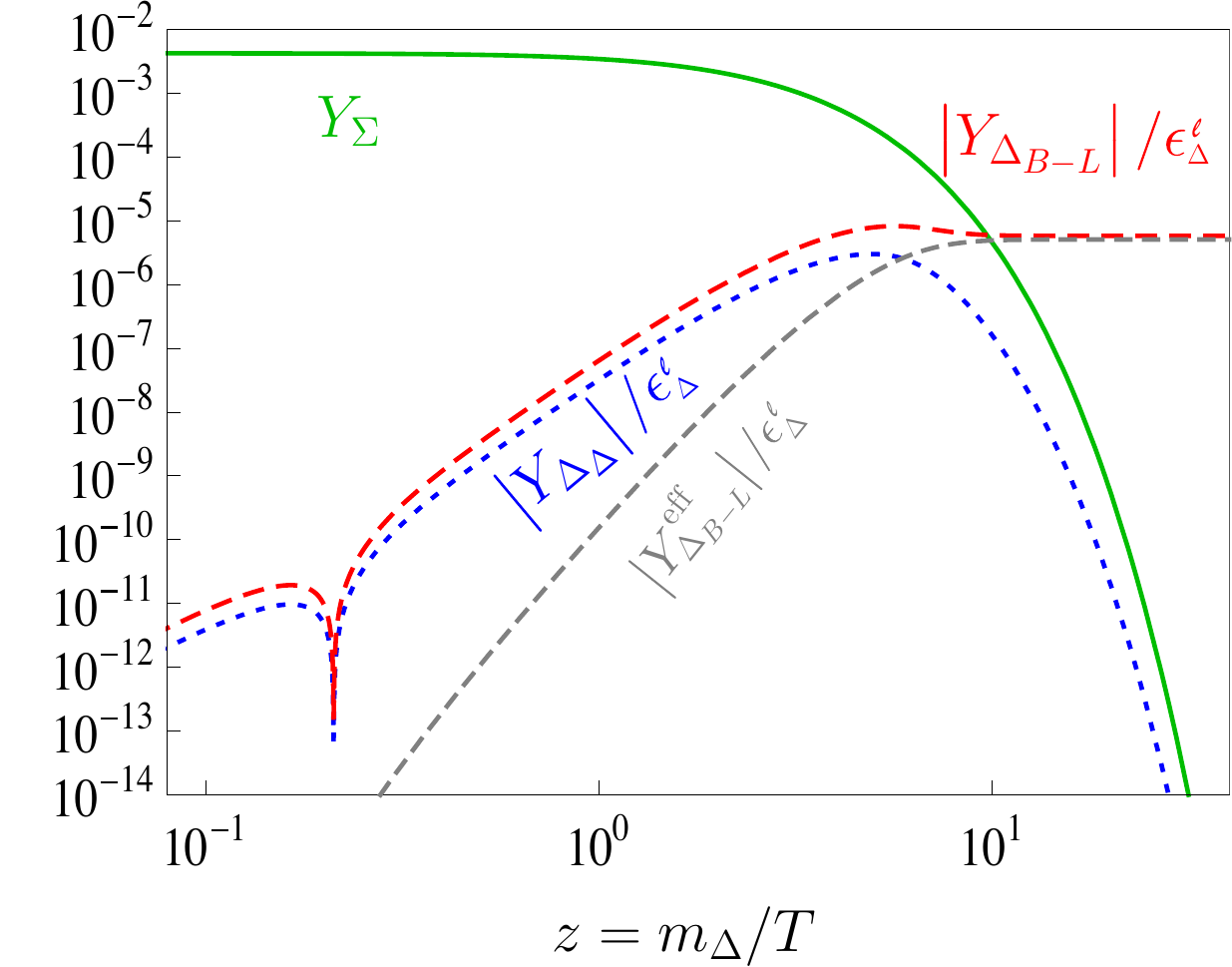}
  \includegraphics[width=7.9cm,height=6.5cm]{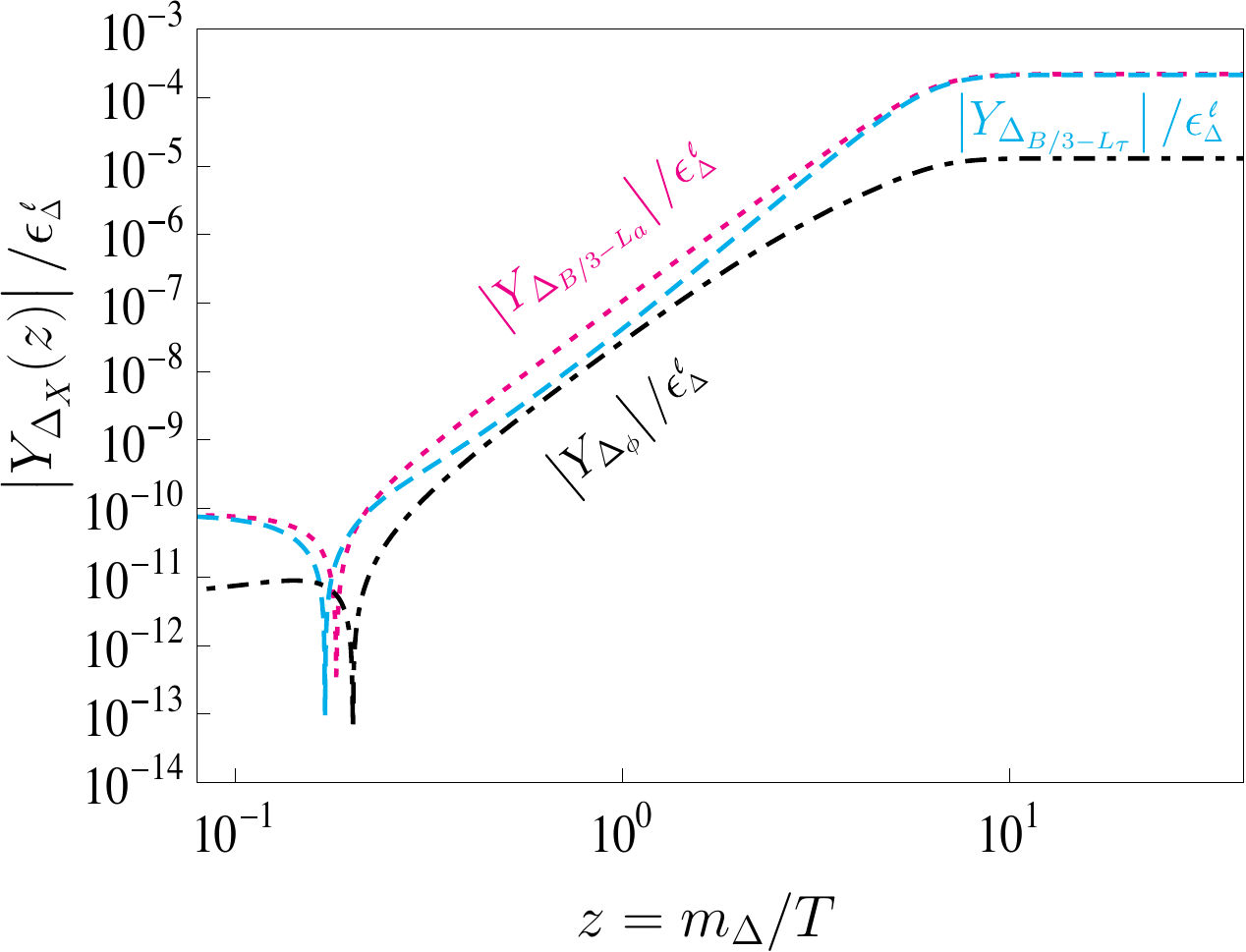}
  \caption{\it Evolution of the different asymmetries $Y_\Sigma$,
    $Y_{\Delta_\Delta}$, $Y_{\Delta_{B-L}}$, $Y_{\Delta_\phi}$ and
    $Y_{\Delta_{B-L}}^{\text{eff}}$ as given by
    Eq.~(\ref{eq:countingatz}), as a function of $z$ for the flavor
    configuration: $B_{\ell_{aa}}=B_{\ell_{a\tau}}=0$,
    $B_{\ell_{\tau\tau}}=(1-B_\phi)$.  The remaining parameters have
    been fixed according to:  
    $m_\Delta=10^9$~GeV,
    $\tilde m_{\Delta_\alpha}= 10^{-2}$~eV and $B_\phi=10^{-4}$.  
    }
  \label{fig:abundances}
\end{figure}

Let us explain already at this point how does this new effect work. To
this end we display in Fig.~\ref{fig:abundances} the evolution of the
different abundances as a function of $z$ for the following parameter
choice\footnote{Using Eq.~(\ref{m-tilde-eff}), this
    choice corresponds to $\tilde{m}_\Delta^{\text{eff}}=1$~eV. From
    Fig.~\ref{fig:zdecoh-vs-mdelta},  it is clear that
      this parameter choice ensures the $B-L$ asymmetry generation
      process to take place in the two-flavor regime where
      Eq.~(\ref{eq:C-ell-PFL-used}) holds.}:  
    $\epsilon_\Delta^{\ell}\equiv\epsilon_\Delta^{\ell_\tau}
    =-\epsilon_\Delta^{\ell_a}=1$, $m_\Delta=10^9$~GeV,
    $\tilde{m}_\Delta= 10^{-2}$~eV, $B_\phi=10^{-4}$,
    $B_{\ell_{aa}}=B_{\ell_{a\tau}}=0$ and
    $B_{\ell_{\tau\tau}}=1-B_\phi$.  As we will discuss further on in
  this section, this $B_{\ell_{ij}}$ flavor configuration maximizes
  the efficiency.

Fig.~\ref{fig:reactionrates} (left-hand side plot)  clearly shows that
for $B_\phi=10^{-4}$  the inverse decays $\phi\phi\to\Delta$ 
have always a rate slower than the Hubble expansion rate.
The fact that for the type-II PFL case, we
do get nevertheless a net non-vanishing $B-L$ asymmetry can then at
first sight appear to be counterintuitive. If for instance only the
channel to leptons does get in thermal equilibrium, as it turns out to
be the case for $B_\phi=10^{-4}$, the scalar triplets   have effectively lepton
number   $L=-2$ and the only active (fast) inverse decays in the
thermal bath, $\Delta \to \bar{\ell}\bar{\ell}$ and $\bar{\Delta} \to
{\ell}{\ell}$,  do not break total lepton number.

However, although the scalar doublet channel never reaches thermal
equilibrium,
 still a portion
of the scalar triplets in the heat bath undergoes decays to scalar
doublets ($\Delta\to\phi\phi$), and these processes do break $L$.  If
the processes $\Delta\to\phi\phi$ and $\bar\Delta\to\bar\phi\bar\phi$
take place at different rates, the thermal bath gets a fraction of
total lepton number each time these reactions occur.  Quantitatively
this means that we can define an effective $B-L$ yield,
$Y_{\Delta_{B-L}}^{\text{eff}}$, determined by the counting of how
many scalar triplets decay times their branching ratio into scalar
doublets, namely
\begin{equation}
  \label{eq:countingatz}
  Y_{\Delta_{B-L}}^{\text{eff}}(z) \approx -2\   \int_{z_0}^{z}\,\frac{dz'}{sHz'}\,
  \frac{ Y_{\Delta_\Delta}}{Y^{\text{Eq}}_{\Sigma}}\, B_\phi \gamma_D \ ,   
\end{equation}
where the factor $2$ comes from the fact that the decay to scalar
doublets violates lepton number by $2$ units.  This effective
quantity holds for the total $B-L$ asymmetry available if one assigns
to $\Delta$ ($\bar{\Delta}$) a lepton number  equal to $-2$
($2$), as we have previously pointed out.  It is related to the
usual $B-L$ yield (where triplets have vanishing lepton number) according to
\begin{equation}
\label{eq:BmL-relations}
Y_{\Delta_{B-L}}(z)=-2 Y_{\Delta_\Delta}(z) + Y_{\Delta_{B-L}}^{\text{eff}} (z)\,.
\end{equation} 
Since ultimately all triplets decay  (their density vanishes), the
final $B-L$ asymmetry simply reads
\begin{equation}
  \label{eq:counting}
  Y_{\Delta_{B-L}}=Y_{\Delta_{B-L}}^{\text{eff}}(z\to\infty)\approx 
  -2\ \int_{z_0}^{\infty}\, \frac{dz'}{sHz'}\, 
  \frac{ Y_{\Delta_\Delta}}{Y^{\text{Eq}}_{\Sigma}}\, B_\phi \gamma_D \ .   
\end{equation}
In order to prove that this formula reproduces the correct $B-L$
asymmetry yield at freeze-out, we have inserted in
Eq.~(\ref{eq:countingatz}) the $Y_{\Delta_\Delta}$ asymmetry obtained
by solving numerically the set of Boltzmann equations. The result is
shown in Fig.~\ref{fig:abundances} (left-hand
side plot)  by the dashed gray curve. It
clearly shows that Eq.~(\ref{eq:counting}) reproduces very well the
numerical result (red dashed curve) for the $B-L$ asymmetry yield at
freeze-out, up to a small deviation of order $~30\%$.  This deviation
can be fully traced back to the effect of the inverse decay processes,
$\phi\phi\to \Delta$ and $\bar\phi\bar\phi\to\bar\Delta$, i.e.~of the
term $\propto B_\phi^\alpha$ in Eq.~(\ref{eq:flavored-BEqs1}). These
scalar inverse decays are not as numerous as scalar decays but not
negligible either.

\begin{figure}
  \centering 
  \includegraphics[width=0.75\linewidth]{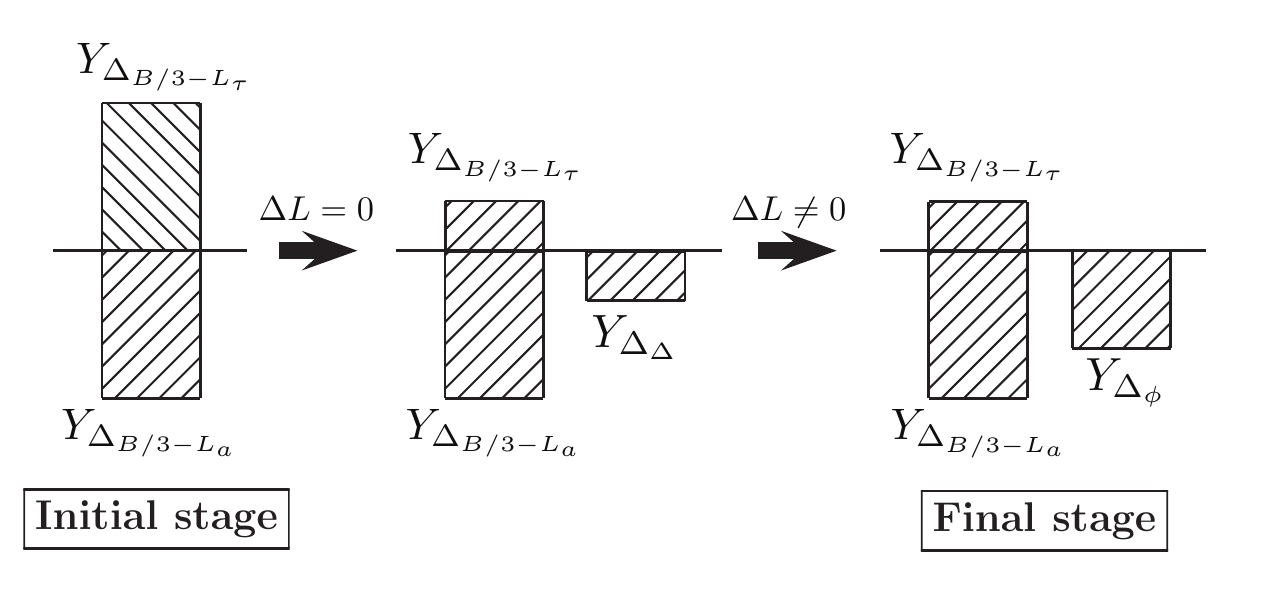}  
  \caption{\it Sketch of the type-II PFL mechanism. See text for further details.} 
  \label{fig:PFLmechanism}
\end{figure}

The generation of a baryon asymmetry, through decays rather than
through inverse decay washout effects, is thus closely related to the
possibility of creating a scalar triplet asymmetry (something
obviously not possible for a right-handed neutrino due to its Majorana
nature). The role of flavor effects is in fact to generate such a
triplet asymmetry.  To see that, it is useful to write down the
relevant terms in Eq.~(\ref{eq:flavored-BEqs1}),
\begin{equation}
\label{eq:deltadeltagenerator}
\dot Y_{\Delta_{\Delta}}\supset
\sum_k
\sum_i B_{\ell_i}C^\ell_{ik}
\frac{Y_{\Delta_k}}{Y^\text{Eq}_\ell}
\gamma_{D}\ .
\end{equation}
This expression clearly shows that a triplet asymmetry can be
generated by two kinds of flavor effects:
\begin{itemize}
\item[-] The first possibility arises if the $C^\ell_{ij}$ have a
  flavor structure. For instance, if the $\tau$ Yukawa is in
  equilibrium, once a lepton doublet $\ell_\tau$ is produced, it has
  the time to interact through the Yukawa coupling and a fraction of
  the $\tau$ flavor is transferred from $\ell_\tau$ lepton doublets to
  $e_\tau$ lepton singlets, while this is not the case for flavor $a$.
  These transferred fractions are just given by the $C^\ell$ matrices
  which are dictated by the chemical potential equilibrium equations,
  see Eq.~(\ref{eq:flavor-coupling-matrices}). This means that there
  are less $\ell_\tau$ than $\ell_a$ lepton doublets available for
  inverse decays to scalar triplets. So, even if there is no flavor
  structure in the branching ratios (i.e. $B_{\ell_a}=B_{\ell_\tau}$)
  and even if, at the onset, $Y_{\Delta_{L_\tau}}=-Y_{\Delta_{L_a}}$,
  the number of $\Delta$ produced is different from the number of
  $\bar{\Delta}$ produced because their production rate is
  proportional to $Y_{\overline{\ell}_\tau }+Y_{\overline{\ell}_a }$
  and $Y_{\ell_\tau}+Y_{\ell_a}$ respectively, which are
  unequal\footnote{It is worth noting that we have the same
    reprocessing concerning the $\phi$ asymmetry created from the slow
    $\Delta$ decays.  This latter asymmetry is partly reprocessed
    through $L$-conserving SM Yukawa interactions into chiral
    asymmetries for charged leptons, which modifies back the $\Delta$
    asymmetry, hence the number of $\Delta$ decaying into SM scalars,
    hence the $B-L$ asymmetry. This effect is nevertheless mild.}.
\item[-] The second possibility arises from the flavor structure of
  scalar triplet decays, i.e. the $B_{\ell_i}$. If $B_{\ell_a}\neq
  B_{\ell_\tau}$, a triplet asymmetry can be produced even if the
  $C^\ell$ coefficients do not distinguish the $a$ and $\tau$ flavors.
  In this case, even if at the onset,
  $Y_{\Delta_{L_\tau}}=-Y_{\Delta_{L_a}}$, with for example
  $B_{\tau}\gg B_{a}$ and $Y_{L_\tau}>0$, inverse decays involving the
  $\tau$ flavor are much more frequent than those involving the $a$
  flavor and inverse decays  
  $  \ell _\tau   \ell _{a,\tau}\to \bar \Delta $ 
  occur more frequently than $
   \ell  _a\ell _{a,\tau}\to\bar\Delta $ 
  inverse decays, resulting in the generation of a $Y_{\Delta_\Delta}$
  asymmetry (of negative sign in this case).
 \end{itemize}
 In other words, in the PFL case there is no $L$-violating CP
 asymmetry. The fact that a final $B-L$ asymmetry can be generated in
 this case, even without $L$-violating processes attaining thermal
 equilibrium, i.e.~$B_\phi\ll B_\ell$, can be understood as a three
 step process, summarized in Fig.~\ref{fig:PFLmechanism}.  Firstly, an
 asymmetry $Y_{\Delta_{L_\tau}}=|Y_{\Delta_{L_a}}|\neq 0$ is created
 from the source term in Eq.~(\ref{eq:flavored-BEqs3}). Secondly, thanks 
 to flavor effects, this asymmetry induces a triplet asymmetry via
 Eq.~(\ref{eq:deltadeltagenerator}), due to the flavor structure
 encoded in $C^\ell_{ij}$ and/or due to the flavor structure encoded
 in the $B_{\ell_i}$. And finally, once a scalar triplet asymmetry is
 created, a $B-L$ asymmetry develops in turn because each time a
 triplet (anti-triplet) decays to scalars, a pair less of anti-leptons
 (leptons) is produced back from the decay of a triplet
 (anti-triplet). The more $Y_{\Delta_\Delta}$ asymmetry is produced,
 the bigger the efficiency.  This PFL production mechanism, based on
 the chain of processes $\ell_i\ell_j\leftrightarrow \bar\Delta \to
 \bar\phi\bar\phi$ and $\bar\ell_i\bar\ell_j\leftrightarrow \Delta \to
 \phi \phi$, is therefore very different from the PFL type-I
 scenario. It stems from the fact that in the type-II scenario, a
 seesaw state asymmetry develops, and in its last step this asymmetry
 generates a final $B-L$ asymmetry from a production mechanism which
 is due to out-of-thermal equilibrium decays, i.e.~from the $\Delta
 \to \phi\phi$ and $\bar\Delta \to \bar\phi\bar\phi$ $L$-violating
 processes ($B_\phi\ll B_\ell$)\footnote{This production mechanism
   driven by a tiny coupling is in many ways similar to the dark
   matter freeze-in production mechanism, as Eq.~(\ref{eq:counting})
   shows.  However there are important differences.  Firstly, this
   equation involves as a source term an asymmetry,
   $Y_{\Delta_\Delta}$, and not the symmetric component of a particle
   species as in the freeze-in scenario. Secondly, since we are
   dealing with decay rates much larger that the one of the dark
   matter freeze-in, still a small amount of inverse decays occurs,
   as we have pointed out.}.

 Let us emphasize once again that this $B_\phi\ll 1$ case is the
 situation which leads naturally to PFL, since this condition leads to
 a natural dominance of the purely flavored CP asymmetries. It must be
 noted that PFL could nevertheless work for larger values of $B_\phi$
 too, in a way more similar to the more involved PFL scenarios in the
 type-I context, see Sec.~\ref{Bphidependance} below.
  
 In the following section we will analyze, along these lines, the
 efficiency dependence upon the relevant parameters.  We will discuss
 in particular the flavor configurations which minimize, or maximize,
 the production of $Y_{\Delta_\Delta}$.  We will then discuss the
 flavored CP asymmetry parameter dependence and show how the
 configurations that maximize the efficiency minimize the flavored CP
 asymmetry. 
 The production of the $B-L$ asymmetry,
 which is given by the product of the flavored CP asymmetry and the
 efficiency, results therefore from the balance of both effects.

\subsection{PFL scenario efficiency}

The problem of quantifying the efficiency is---in principle---an eight
parameters problem: $\epsilon^{\ell_{\tau,a}}_\Delta$, $m_\Delta$,
$\tilde m_\Delta$, $B_\phi$, $B_{\ell_{aa}}$, $B_{\ell_{\tau\tau}}$
and $B_{\ell_{a\tau}}$, which reduces to six parameters due to the
constraints $B_\ell+B_\phi=1$ and
$\epsilon^\ell_\Delta\equiv\epsilon^{\ell_\tau}_\Delta =-
\epsilon^{\ell_a}_\Delta$. Since the efficiency does not depend on
$\epsilon^\ell_\Delta$---see
Eq.~(\ref{eq:total-BmL-PFL-case-in-terms-of-eff})---we will analyze
the dependence of the efficiency upon the 5 remaining parameters:
$m_\Delta$, $\tilde m_\Delta$, $B_\phi$, $B_{\ell_{aa}}$ and
$B_{\ell_{\tau\tau}}$.

We start by analyzing the dependence upon $B_{\ell_{ij}}$ for fixed
$B_\phi$, $m_\Delta$ and $\tilde m_\Delta$. We will see that different
flavor configurations ($B_{\ell_{ij}}$ configurations) will produce a
minimal or maximal efficiency. However, as we will latter show in
Sec.~\ref{Maximal-and-minimal-B-L-asymmetry}, the configurations that
maximize the efficiency do not necessarily maximize the final $B-L$
asymmetry.  We then proceed by analyzing the dependence of the
efficiency with $\tilde m_\Delta$ for fixed $B_\phi$, $m_\Delta$ and
$B_{\ell_{ij}}$, and finally the dependence of the efficiency with
$B_\phi$ for fixed $\tilde m_\Delta$, $m_\Delta$ and $B_{\ell_{ij}}$.
This will allow us to understand and distinguish the main features of
the type-II seesaw PFL scenario.
\subsubsection{Efficiency: $B_{\ell_{ij}}$ dependence}
\label{Bell-vs-efficiency}
\begin{figure}
  \centering
  \includegraphics[scale=0.7]{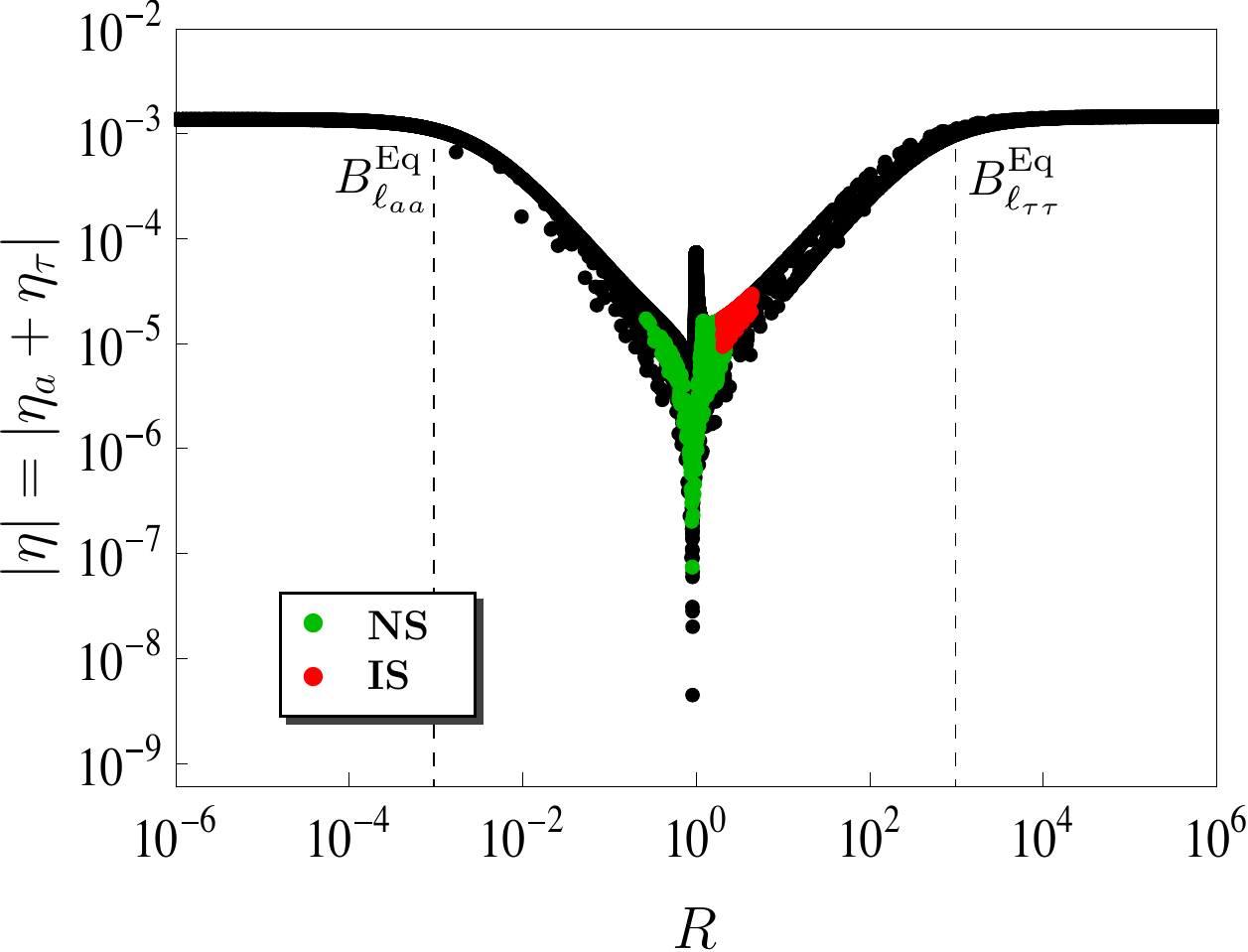}
  \caption{\it Efficiency as a function of the $R$ parameter for  
    $m_\Delta=10^9$~GeV, 
    $\tilde{m}_\Delta=10^{-2}$~eV and   $B_\phi=10^{-4}$. The green (red) dots indicate  the
    allowed range for the efficiency, as required by neutrino data
    ($3\sigma$ level \cite{Tortola:2012te}) for the inverted (normal)
    hierarchical light neutrino mass spectrum. In addition to the
    constraints on $R$, we also took into account   the constraints
    imposed by data on the different $B_{\ell_{ij}}$ elements (see
    Fig.~\ref{fig:minmax-values-for-R}). We stress that these
    constraints apply only if the neutrino mass matrix is entirely
    dominated by the lightest scalar triplet contribution.}
\label{fig:efficiency-vs-R}
\end{figure}

In order to proceed, we first solve numerically the system of kinetic
equations in (\ref{eq:flavored-BEqs1})-(\ref{eq:flavored-BEqs3}) for
different flavor configurations. We then provide some physical
arguments supporting the special flavor configurations that
maximize/minimize the efficiency. For concreteness, we fix three out
of the five relevant parameters as follows: 
\begin{equation}
  \label{eq:fix-parameter-space-point}
  m_\Delta=10^9\,\mbox{GeV}  \ ,\quad
  \tilde m_\Delta= 10^{-2}\,\mbox{eV}
  \ ,\quad  B_\phi= 10^{-4}  
  \ .
\end{equation} 
Once these parameters are fixed, the efficiency is entirely dictated
by the flavor configurations determined by the values of the
$B_{\ell_{ij}}$ parameters. It turns out that the flavor dependence is
well described by the quantity:
\begin{equation}
  \label{eq:R-definition-twoFlavored-regime}
  R\equiv\frac{B_{\ell_{a}} }{B_{\ell_{\tau}}}=
  \frac{B_{\ell_{aa}}+B_{\ell_{a\tau}}}
  {B_{\ell_{\tau a}}+B_{\ell_{\tau\tau}}}\ ,
  \end{equation}
  which represents the ratio of triplet decay branching ratios to
  different lepton-flavor final states.  The importance of this
  quantity can be understood from Eq.~(\ref{eq:deltadeltagenerator}),
  where we see that it is precisely through the $B_{\ell_i}$ that a
  triplet asymmetry is generated.  We plot in
  Fig.~\ref{fig:efficiency-vs-R} the efficiency as a function of this
  parameter $R$ for the parameters fixed according to
  Eq.~(\ref{eq:fix-parameter-space-point}).

\begin{figure}
  \centering
  \includegraphics[width=14cm,height=8cm]{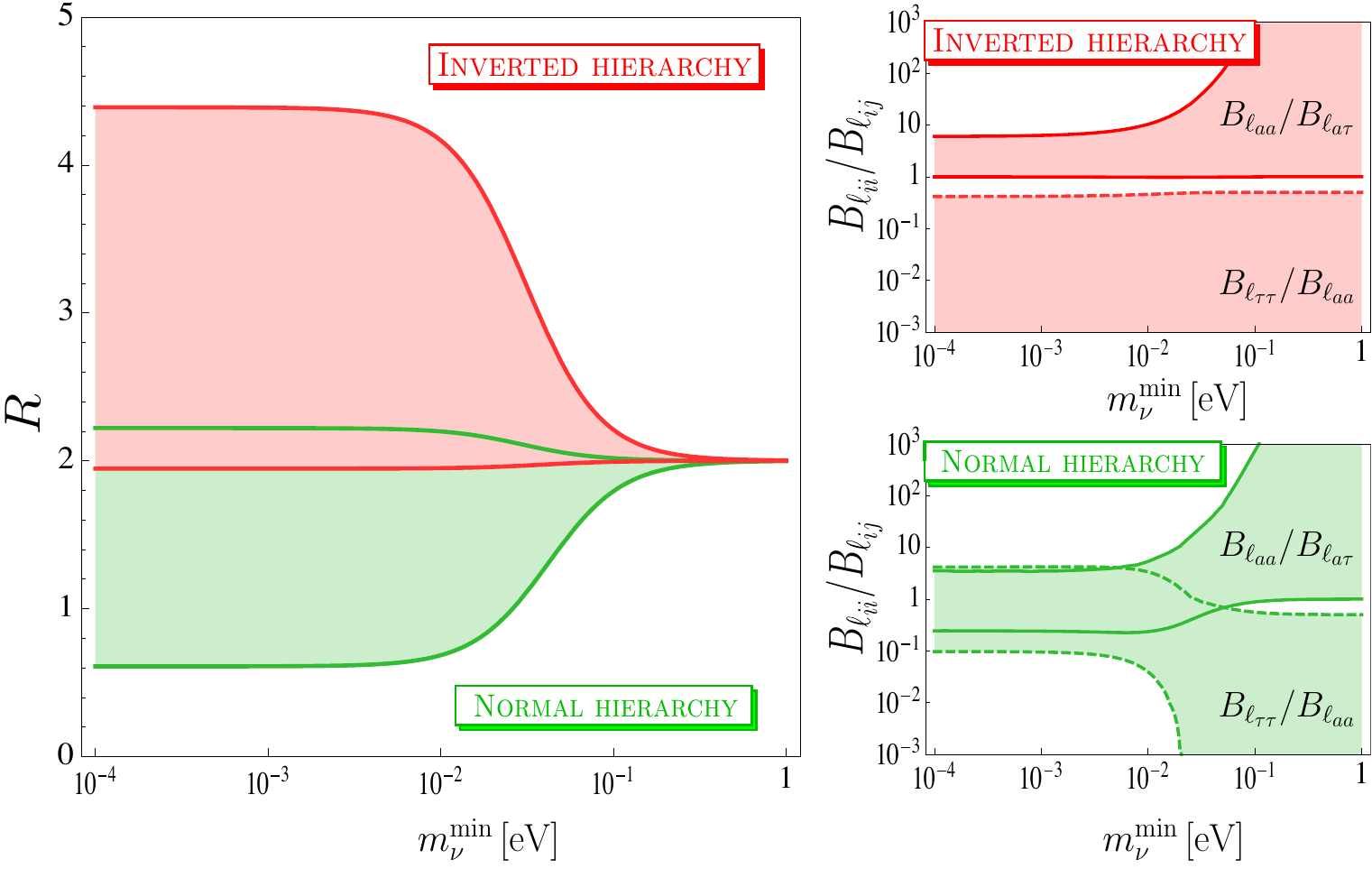}
  \caption{\it Allowed ranges for $R$ (left plot) and the ratios of
    branching ratios $B_{\ell_{ii}}/B_{\ell_{ij}}$ (right plots) as a
    function of the lightest neutrino mass for both normal (green) and
    inverted mass spectrum (red).  The results have been derived by
    varying the neutrino oscillation parameters in the $3\sigma$ range
    according to \cite{Tortola:2012te}.}
  \label{fig:minmax-values-for-R}
\end{figure}

A viable scalar triplet leptogenesis setup requires---of course---consistency with neutrino data
\cite{Tortola:2012te,Fogli:2012ua,GonzalezGarcia:2012sz}. If the most
relevant contribution to the neutrino mass matrix in
Eq.~(\ref{eq:neutrino-mass-matrix-pure}) is given by the lightest
triplet, which can be regarded as a quite reasonable possibility
(assumption), the determination of the available flavor configurations
can be done directly via neutrino oscillation data.  We present in
Fig.~\ref{fig:minmax-values-for-R} the constraints on $R$ (left panel)
and on the ratios of branching ratios $B_{\ell_{ii}}/B_{\ell_{ij}}$
(right panel) as a function of the lightest neutrino mass, for the
normal (green) and inverted (red) light neutrino mass spectrum. We
fixed the neutrino oscillation parameters according to their upper and
lower $3\sigma$ limits \cite{Tortola:2012te}.  It can be seen that the
$R$ configuration leading to a vanishing $B-L$ asymmetry, although
showing up at the $3\sigma$ level in the normal spectrum case, can be
readily evaded, thus showing the viability of the PFL scenario even in
its most constrained form.

Fig.~\ref{fig:efficiency-vs-R} clearly shows that the efficiency
exhibits four special configurations, namely ($i$, $ii$) two global
maxima at $R\ll 1$ and $R\gg 1$, ($iii$) one local maximum and ($iv$)
one global minimum near $R\sim 1$. We now aim to understand the
physical reasons behind these special configurations.

{\bf Configurations ($i$) and ($ii$):} following
Eq.~(\ref{eq:R-definition-twoFlavored-regime}), these global maxima
correspond to the flavor alignments $B_{\ell_{a}}\ll B_{\ell_{\tau}}$
and $B_{\ell_{a}}\gg B_{\ell_{\tau}}$. The effect seems entirely
driven by the $B_{\ell_i}$, so we will not consider the possible
effects of the $C^\ell_{ij}$ and $C^\phi_k$ elements in the analysis.
More precisely, these maxima are reached whenever the inverse decays
involving the $a$ or $\tau$ flavor never enter in thermal equilibrium,
i.e. for $B_{\ell_{ii}}<B^{\text{Eq}}_{\ell_{ii}}$ where
$B^{\text{Eq}}_{\ell_{ii}}$ is determined by: 
 \begin{equation}
 \label{Bell-equilibrium}
 B^{\text{Eq}}_{\ell_{ii}}\ \frac{Y_\Sigma^{\text{Eq}}}{Y_\ell^{\text{Eq}}}\ \frac{ \gamma_D }{H n_\gamma}\Big|_{\text{max}}= 1 
 \quad 
 \text{which gives}\quad B^{\text{Eq}}_{\ell_{ii}}\approx 10^{-3}\ ,
 \end{equation} 
 where we used in the last equality the parameter values given in
 Eq.~(\ref{eq:fix-parameter-space-point}). This value is in good
 agreement with the numerical results shown in
 Fig.~\ref{fig:efficiency-vs-R}, where the two maxima are reached for
 $R\lesssim 10^{-3}$ and $R\gtrsim 10^{3}$.  For these configurations,
 only the asymmetry produced in one flavor is transferred through
 inverse decays $\ell_i\ell_i\to\bar\Delta$ to a triplet asymmetry,
 which is therefore maximal since the two flavor asymmetries have
 opposite signs. As a consequence, one asymmetry is depleted through
 the chain $\ell_i\ell_i\leftrightarrow\bar\Delta\to
 \bar\phi\bar\phi$, while the other flavor asymmetry remains
 unaffected, clearly leading to a maximal efficiency.

 {\bf Configuration ($iii$):} this local maximum is in fact reached for
 $R\approx 1$ when $B_{\ell_{ii}}\ll B_{\ell_{a\tau}}$. In this
 configuration, only the inverse decays $\ell_a\ell_\tau\to\bar\Delta$
 reach thermal equilibrium, and one expects no production of a triplet
 asymmetry, and therefore no production of a final $B-L$ asymmetry,
 since the flavor asymmetries are depleted by the same
 amount. However, this is not the case because the $C^\ell_{ij}$
 elements have a flavor structure, which plays a crucial role. The
 point is that when inverse decays are in thermal equilibrium, the
 combination of processes $\ell_a\ell_\tau\leftrightarrow\bar\Delta\to
 \bar\phi\bar\phi$ and $\bar\ell_a\bar\ell_\tau\leftrightarrow
 \Delta\to\phi\phi$ tends to equilibrate the flavor asymmetries in
 lepton doublets $Y_{\Delta_{\ell_\tau}}\approx -Y_{\Delta_{\ell_a}}
 $, while in the meantime decreasing the separated asymmetries by a
 small amount\footnote{Indeed, if $Y_{\ell_a}\cdot Y_{\ell_\tau}>
   Y_{\bar{\ell_a}}\cdot Y_{\bar{\ell_\tau}}$, that is if
   $Y_{\Delta_{\ell_\tau}} +Y_{\Delta_{\ell_a}} > 0$, there will be
   more $\ell_a\ell_\tau\leftrightarrow\bar\Delta\to \bar\phi\bar\phi$
   processes than $\bar\ell_a\bar\ell_\tau\leftrightarrow
   \Delta\to\phi\phi$ processes, so that statistically
   $Y_{\Delta_{\ell_\tau}}+Y_{\Delta_{\ell_a}}$ will decrease, as well
   as the separated asymmetries $ Y_{\Delta_{\ell_\tau}}$ and
   $|Y_{\Delta_{\ell_a}}|$. This lasts until
   $Y_{\Delta_{\ell_\tau}}\approx -Y_{\Delta_{\ell_a}} $, and from
   that moment no more triplet asymmetry can be generated and the
   asymmetries $ Y_{\Delta_{\ell_\tau}}$ and $|Y_{\Delta_{\ell_a}}|$
   are left invariant.}.  But due to the chemical equilibrium
 conditions, the total lepton flavors asymmetries
 $Y_{\Delta_{B/3-L_i}}$ are in general different. Indeed, using
 Eq.~(\ref{eq:lepton-doublets-scalar-doublet-BmL-asymm}), the total
 $B-L$ asymmetry at freeze-out is related to the lepton flavor doublet
 asymmetries through:
 \begin{equation}
   Y_{\Delta_{B-L}}= Y_{\Delta_a}+  Y_{\Delta_\tau}=
   -\frac{Y_{\Delta_{\ell_a}}\left(C^\ell_{\tau\tau}- C^\ell_{\tau a}\right)
     + Y_{\Delta_{\ell_\tau}}\left(C^\ell_{aa}- C^\ell_{a \tau}\right)}
   {C^\ell_{\tau\tau} C^\ell_{aa} -C^\ell_{a \tau}C^\ell_{  \tau a}} \ .
 \end{equation}
 In the PFL regime, in the case where the final  lepton doublet
 asymmetries are equal and opposite, $Y_{\Delta_{\ell_\tau}}\approx -
 Y_{\Delta_{\ell_a}} $ (as for the case $B_{\ell_{ii}}=0$), a final
 $B-L$ asymmetry can be produced only if the $C^\ell_{ij}$ elements
 have a flavor structure.  This $B-L$ asymmetry can be quite large
 because the flavor asymmetries in lepton doublets
 $Y_{\Delta_{\ell_i}}$ decrease  only slightly for this special
 configuration.

 Any significant deviation from this special configuration, e.g.
 $B_{\ell_{ii}}>B^{\text{Eq}}_{\ell_{ii}} $, would not only tend to
 equilibrate the flavor asymmetries in the lepton doublets, but also
 the $Y_{\Delta_{\ell_i}}$ separately through the chain
 $\bar\ell_i\bar\ell_i\leftrightarrow \Delta\to\phi\phi$. All in all,
 the efficiency has in consequence a local maximum for
 $B_{\ell_{a\tau}}\approx (1-B_\phi)/2$.

 {\bf Configuration ($iv$):} shifted to the left of the maximum defining
 configuration ($iii$), a minimal efficiency (almost vanishing
 efficiency) can be seen, it lies at about $R\approx 3/4$. In order to
 understand the reason for this configuration to show up, we can look
 in a first step if analytically the efficiency may vanish for some
 value of the flavor parameters $B_{\ell_{ij}}$.  Using
 Eq.~(\ref{eq:total-BmL-PFL-case-in-terms-of-eff}), we see that a
 vanishing efficiency is obtained whenever $\eta_\tau(z)=-\eta_a(z)$
 for all $z$, which means through
 Eq.~(\ref{eq:flavored-efficiency-functions-non-primed-basis}):
 \begin{equation}
  \label{eq:vanishing-BmL-combination-PFL}
  \sum_{i,k=2,3}\left(e^{-\int_{z_0}^{z}\,dz'D(z'){\cal M}(z')}\right)_{ik}(-1)^{1+k}=0
  \quad \forall z \ ,
\end{equation}
which is satisfied as long as all the coefficients of the exponential
power series expansion vanish, i.e.
\begin{equation}
  \label{eq:vanishing-BmL-combination-expansion-PFL}
  \sum_{i,k=2,3} \int_{z_0}^{z}\,dz'D(z')(-1)^{1+k}
  \left[
    {\cal  M}_{ik}(z')
    - \frac{1}{2}\int_{z_0}^{z}\,dz''
    D(z'')\sum_{j=1}^3{\cal  M}_{ij}(z'){\cal  M}_{jk}(z'')+\dots
  \right]=0\ .
\end{equation} 
We have found this turns out to be the case if the {\it
  flavor-triplet-coupling-matrix} entries satisfy the following two
conditions
\begin{equation}
  \label{eq:vanishing-BmL-cond}
 {\cal  M}_{12}={\cal  M}_{13}
  \quad\mbox{and}\quad
   \sum_{i,k=2,3}(-1)^{1+k}{\cal  M}_{ik}=0 \ ,
\end{equation}
where the corresponding elements must not depend on $z$, which is
indeed our case---see Eq. (\ref{eq:flavor-triplet-CM}).  This result
in turn can be understood using Eqs.~(\ref{eq:flavored-BEqs1}) and
(\ref{eq:flavored-BEqs3}). In the two flavor PFL scenario, since the
source terms for both flavors are equal and opposite, a vanishing
efficiency will be generated if the washouts of the two flavors are
also equal and opposite, which is nothing but the conditions in
Eq.~(\ref{eq:vanishing-BmL-cond}).
 
More precisely, for this to be achieved, we need that
$Y_{\Delta_{\tau}}=-Y_{\Delta_{a}}$ remains valid at any time. As
Eq.~(\ref{eq:flavored-BEqs3}) shows, this requires: (a)
$Y_{\Delta_\Delta}=0$ and (b) $\sum_{i,j,k} C^\ell_{ijk}B_{\ell_{ij}}
Y_{\Delta_k}=0$ at any time. These two relations hold simultaneously
if both conditions in Eq.~(\ref{eq:vanishing-BmL-cond}) are
fulfilled. Indeed, if relation (a) holds, (b) can be rewritten as the
second condition in Eq.~(\ref{eq:vanishing-BmL-cond}). On the other
hand, if relation (b) holds, (a) can be rewritten using
Eq.~(\ref{eq:flavored-BEqs3}) as $\sum_{i,k} (C^\ell_{ik} B_{\ell_{i}}
-B_\phi C^\phi_k)Y_{\Delta_k} =0$, which is nothing but the first
condition in Eq.~(\ref{eq:vanishing-BmL-cond}).

Using Eq.~(\ref{eq:flavor-triplet-CM}), these conditions can be
simultaneously fulfilled only in the limit $B_\phi\to 0$, in which
case the triplet flavor configuration must satisfy the simple
relation: 
\begin{equation}
  \label{eq:R-minimum-twoFlavored-regime}
  R=\frac{B_{\ell_{aa}}+B_{\ell_{a\tau}}}{B_{\ell_{\tau a}}+B_{\ell_{\tau\tau}}}
  =\frac{C^{\ell}_{\tau\tau} - C^\ell_{\tau a}}{C^\ell_{aa} - C^\ell_{a\tau}} 
  \approx 0.74\ .
\end{equation} 
Strictly speaking, since $B_\phi\neq 0$, the efficiency is not
vanishing for any value of $B_{\ell_{ij}}$. However, for small
$B_\phi$, the efficiency does not vanish exactly anymore but shows now
a minimum for $R\approx 3/4$, which is in good agreement with the
numerical results shown in Fig.~\ref{fig:efficiency-vs-R}.

\subsubsection{Efficiency: $\tilde{m}_\Delta$ dependence}
\label{sec:mtilde-dependence}
\begin{figure}
  \centering 
  \includegraphics[width=7.7cm,height=6.5cm]{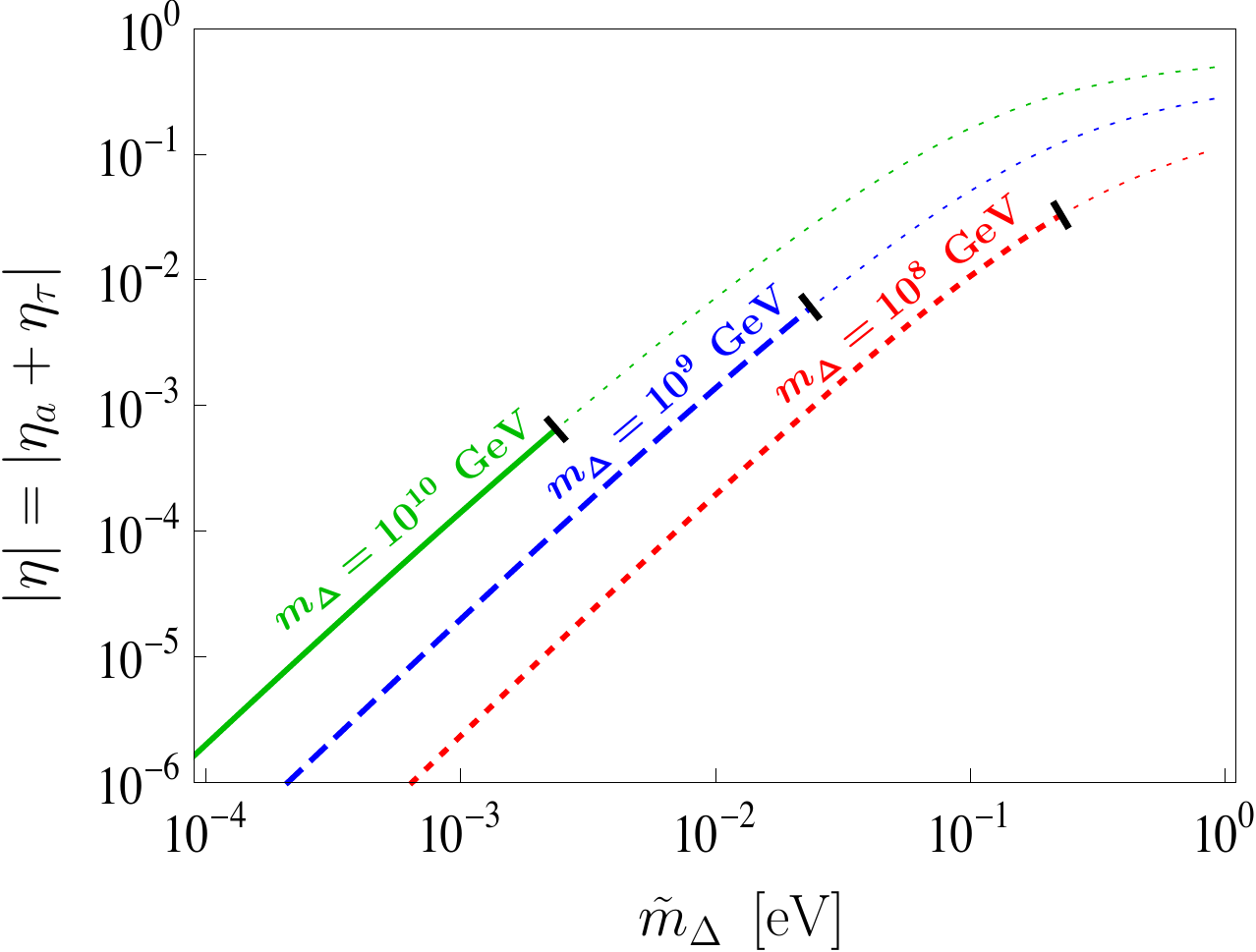}
  \caption{\it Efficiency as a function of $\tilde m_\Delta$ for
    several values of the scalar triplet mass. The parameters have
    been fixed according to $B_\phi=10^{-4}$ and
      $B_{\ell_{aa}}=B_{\ell_{a\tau}}=0$. The lines are cut
      whenever the 2-flavor regime condition
     ceases to be fulfilled (see
      Sec.~\ref{domain-of-validity}).} 
  \label{fig:YDBmL-v-mtildeDelta}
\end{figure}

By fixing  $B_\phi=10^{-4}$  as in the previous section, and taking as an
example $B_{\ell_{aa}}=B_{\ell_{a\tau}}=0$, and
$B_{\ell_{\tau\tau}}=1-B_\phi$, we display in
Fig.~\ref{fig:YDBmL-v-mtildeDelta} the dependence of the efficiency
with $\tilde m_\Delta$, for the three benchmark triplet masses
 $m_\Delta=10^8,10^9,10^{10}$ GeV.  It can be seen that
irrespective of the triplet mass, the smaller $\tilde m_\Delta$ the
smaller the resulting efficiency. The reason for this behavior follows
directly from the relative strength of gauge and Yukawa induced
reactions: the larger $\tilde m_\Delta$ the most likely triplets will
decay rather than scatter, thus implying a larger efficiency.

On the other hand, we see that the efficiency decreases with
$m_\Delta$. This is also due to gauge reactions: the smaller
$m_\Delta$ the most likely the triplet will scatter rather than decay,
thus implying a smaller efficiency.  More
precisely, as for the unflavored case, when gauge scatterings are
faster than decays they suppress $Y_\Sigma-Y_\Sigma^{Eq}$ in
Eq.~(\ref{eq:flavored-BEqs2}) by a factor $\gamma_D/\gamma_A$, which
implies an equal suppression of the source term in
Eq.~(\ref{eq:flavored-BEqs3}).

In the unflavored case one can distinguish two regimes
\cite{Hambye:2005tk}, the gauge and Yukawa regimes, depending on the
values of $m_\Delta$ and $\tilde m_\Delta$.  While in the unflavored
case a maximum efficiency is obtained at the transition between both
regimes, this is in general not anymore the case in the flavored
leptogenesis scenario. Depending on the flavor configuration, a maximum
efficiency can be obtained far in the Yukawa regime because of flavor
effects, see Sec.~\ref{sec:single-triplet-scenarios} for a more
detailed explanation.

\subsubsection{Efficiency: $B_\phi$ dependence}
\label{Bphidependance}
\begin{figure}
  \centering
  \includegraphics[width=7.7cm,height=6.5cm]{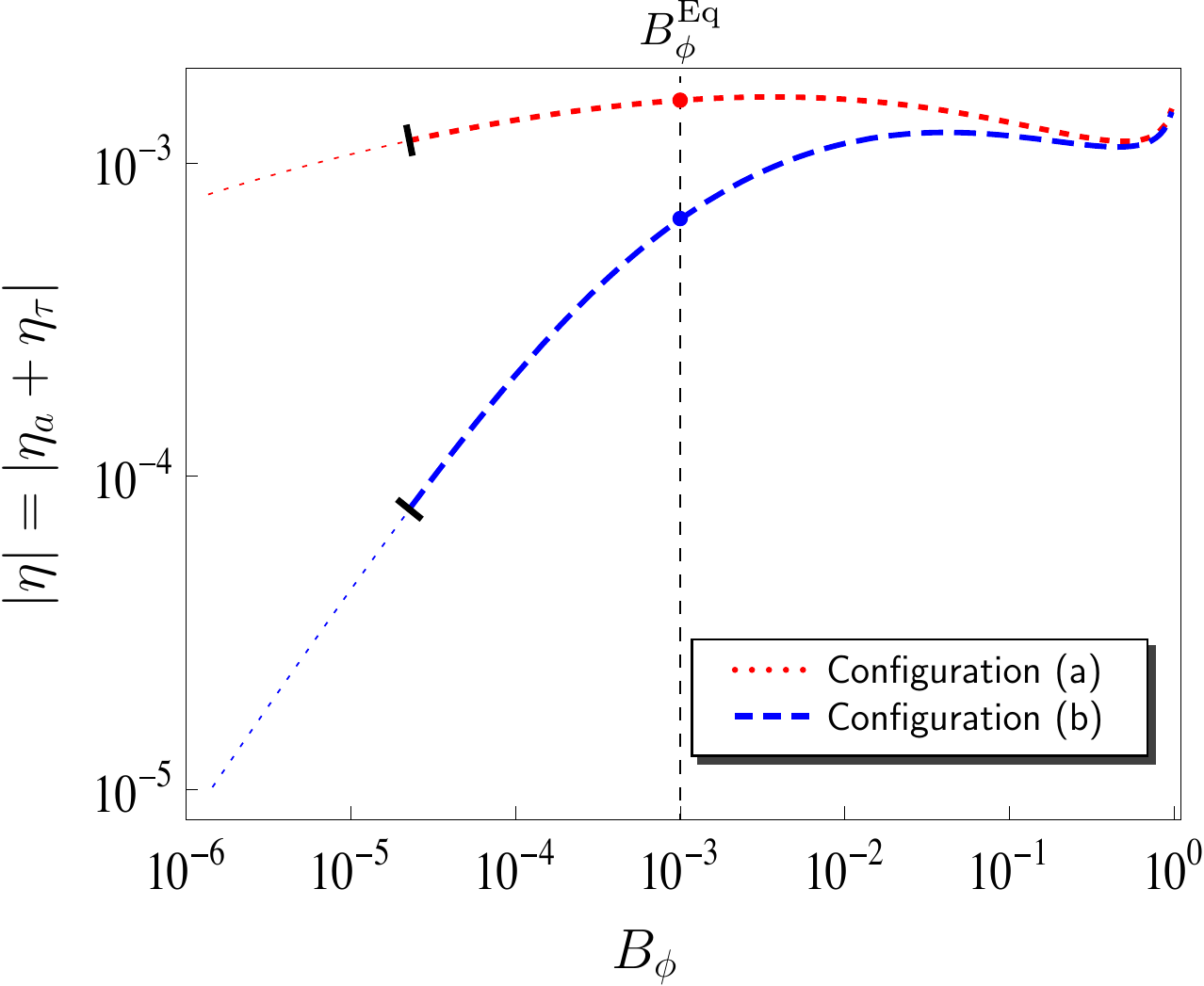} 
  \caption{\it  Efficiency as a function of $B_\phi$
      for $m_\Delta=10^{9}$~GeV and $\tilde{m}_\Delta=10^{-2}$ eV.
      Configuration (a) corresponds to
      $B_{\ell_{aa}}=B_{\ell_{a\tau}}= 0$ (i.e.~$R=0$) while
      configuration (b) corresponds to
      $B_{\ell_{aa}}=B_{\ell_{\tau\tau}}/99=(1-B_\phi)/100$
      (i.e.~$R\simeq 10^{-2}$). The lines are cut when the
      2-flavor regime condition  
      ceases to be fulfilled (see Sec.~\ref{domain-of-validity}).} 
  \label{fig:YDBmL-v-bphi}
\end{figure}

We present in Fig.~\ref{fig:YDBmL-v-bphi} the dependence of the
efficiency upon $B_\phi$ in the range $[10^{-6},1]$ 
for fixed values of $m_\Delta$ and $\tilde m_\Delta$.  We considered
two particular flavor configurations for $B_{\ell_{ij}}$. The red
curve (configurations (a)) corresponds to one of the two flavor
configurations that maximize the efficiency (see
Sec.~\ref{Bell-vs-efficiency}).  The blue curve (configurations (b))
corresponds instead to the configuration  
$B_{\ell_{aa}}=B_{\ell_{\tau\tau}}/99=(1-B_\phi)/100$ and
$B_{\ell_{a\tau}}=0$.
These two
configurations show two different behaviors, that are in fact
representative of any other flavor configuration.

For $B_\phi\leq 10^{-1}$, which is the interesting region for this PFL
scenario, we can distinguish two distinct regimes. They are separated
by $B^{\text{Eq}}_\phi$, the value at which the inverse decays
$\phi\phi\to\Delta$ become active, determined by the condition 
  \begin{equation}
 \label{Bphi-equilibrium}
 B^{\text{Eq}}_{\phi}\ \frac{Y_\Sigma^{\text{Eq}}}{Y_\ell^{\text{Eq}}}\  \frac{ \gamma_D }{H n_\gamma}\Big|_{\text{max}}= 1 
 \quad \text{which gives}\quad 
 B^{\text{Eq}}_{\phi}    \approx  10^{-3}\ ,
 \end{equation} 
 where we used in the last equality the parameter value
domain-of-validity  $\tilde{m}_\Delta=10^{-2}$ eV.  The way the efficiency scales with
 $B_\phi$ depends on the flavor configurations. For $B_\phi\lesssim
 B_\phi^{\text{Eq}}$ the efficiency always increases with $B_\phi$ as
 a result of the fact that the larger $B_\phi$ the faster the decay to
 SM scalars, as can be seen in Eq.~(\ref{eq:counting}), but the exact
 scaling actually also depends on the interplay of the
 $Y_{\Delta_\Delta}$ and $Y_{\Delta_{B/3-L_i}}$ asymmetries.

 Now, as soon as $B_\phi\gtrsim B_\phi^{\text{Eq}}$, inverse decays
 $\phi\phi\to\Delta$ become efficient, implying that lepton number is
 broken by processes in thermal equilibrium (fast processes). This
 brings a new $\sqrt{B_\phi}$ suppression in the efficiency, resulting
 in an efficiency increasing less with $B_\phi$ or even decreasing,
 depending on the flavor configuration, see
 Fig.~\ref{fig:YDBmL-v-bphi}.

 To conclude, we see that for the flavor configuration that maximizes
 the efficiency, the value of $B_\phi$ which gives the maximal
 efficiency is obtained for $B_\phi \sim B_\phi^{\text{Eq}}$, that is
 to say for the value of $B_\phi$ at which the $\phi\phi\rightarrow
 \Delta$ inverse decays are about to be active. In this case, the
 efficiency can be as large as unity for values of $m_\Delta\gtrsim
 10^{12}$~GeV, or less for smaller values of $m_\Delta$ (due to the
 gauge scattering thermalization effect).  For other configurations
 that lead to smaller efficiencies, the maximum efficiency is obtained
 for much larger values of $B_\phi\sim 1$.
  
\subsection{Minimal and maximal  $B-L$ asymmetry}
\label{Maximal-and-minimal-B-L-asymmetry}
As stressed above, a PFL scenario is naturally favored as soon
as $\epsilon^{\ell_i (\not{F})}_{\Delta_\alpha}$ dominates the CP
asymmetry, which naturally holds if Yukawa couplings are larger than
scalar couplings, i.e.~when Eq.~(\ref{eq:PFLcondition}) holds. This
equation can also be recasted in terms of the triplet branching ratios
to scalar and lepton final states
\begin{equation}
  \label{eq:PFLcondition-v2}
  \sqrt{\frac{B^\alpha_\phi B^\beta_\phi}{B^\alpha_\ell B^\beta_\ell}}\ll
  \frac{m_{\Delta_\alpha}}{m_{\Delta_\beta}}
  \frac{\mbox{Tr}[\mathcal{M}_{\alpha}^\nu\mathcal{M}_{\beta}^{\nu\dagger}]}
  {\tilde{m}_{\Delta_\alpha}\tilde{m}_{\Delta_\beta}}
  \leq \frac{m_{\Delta_\alpha}}{m_{\Delta_\beta}} \ ,
\end{equation}
where the last inequality comes from the Cauchy-Schwarz inequality: 
\begin{equation}
  \label{eq:trace-inequality}
  \left|\text{Tr}\left[A B \right]\right|
  \leq 
  \sqrt{\text{Tr}\left[AA^\dagger\right]}
  \sqrt{\text{Tr}\left[	BB^\dagger\right]}\ .
\end{equation}
As an example, taking a smooth triplet mass hierarchy
$m_{\Delta_\alpha}/m_{\Delta_\beta}\sim 10^{-1} \ (10^{-2})$ and
assuming the upper bound
$\mbox{Tr}[\mathcal{M}_{\alpha}^\nu\mathcal{M}_{\beta}^{\nu\dagger}]\approx
\tilde{m}_{\Delta_\alpha}\tilde{m}_{\Delta_\beta}$, a PFL scenario
will be naturally dominant as soon as $B^{\alpha,\beta}_\phi\ll
10^{-1}\ ( 10^{-2})$.

We have seen in the previous sections that the efficiency strongly
depends on the flavor parameters $B_{\ell_{ij}}$. Explicitly, we have
shown that the efficiency has a minimum at $R\approx 3/4$, global
maxima at $B_{\ell_{aa}}=B_{\ell_{a\tau}}\approx 0$ and
$B_{\ell_{\tau\tau}}=B_{\ell_{a\tau}}\approx 0$, and a local maximum
at $B_{\ell_{aa}}=B_{\ell_{\tau\tau}}\approx 0$.  However, a maximal
efficiency does not imply a maximal $B-L$ asymmetry.  Indeed, using
Eq.~(\ref{eq:CPpuretype2-LNC-2}), we can actually compute a general
upper bound for the purely flavored CP asymmetry:
\begin{align}
  \label{eq:CPpuretype2-LNC-2-scale}
  |\epsilon^{\ell(\not{F})}_{\Delta_\alpha}| &\leq 
  \frac{1}{2\pi}\, g (m^2_{\Delta_\alpha}/m^2_{\Delta_\beta} )\, 
  \left[
    \sqrt{B^\ell_{aa}B^\ell_{\tau\tau}} 
    + 
    \sqrt{B^\ell_{a\tau}\left(B^\ell_{aa}+B^\ell_{\tau\tau}\right)} 
  \right]\ , 
  \end{align} 
  where we assumed perturbative Yukawa couplings for the second
  triplet generation, i.e. $|Y_\beta|\leq 1$.  This expression shows
  clearly that the three configurations that maximize the efficiency
  give vanishing CP asymmetries! This can be understood easily from
  the fact these configurations involve a Yukawa coupling only for one
  flavor.  We see also that the upper bound on the CP asymmetry is
  directly related to the hierarchy between the different triplet
  masses, which is compatible with the requirement in
  Eq.~(\ref{eq:PFLcondition-v2}), i.e.~a smooth triplet mass hierarchy
  favors PFL scenario and allows for a large CP asymmetry.

\begin{figure}
  \centering
    \includegraphics[height=7.cm]{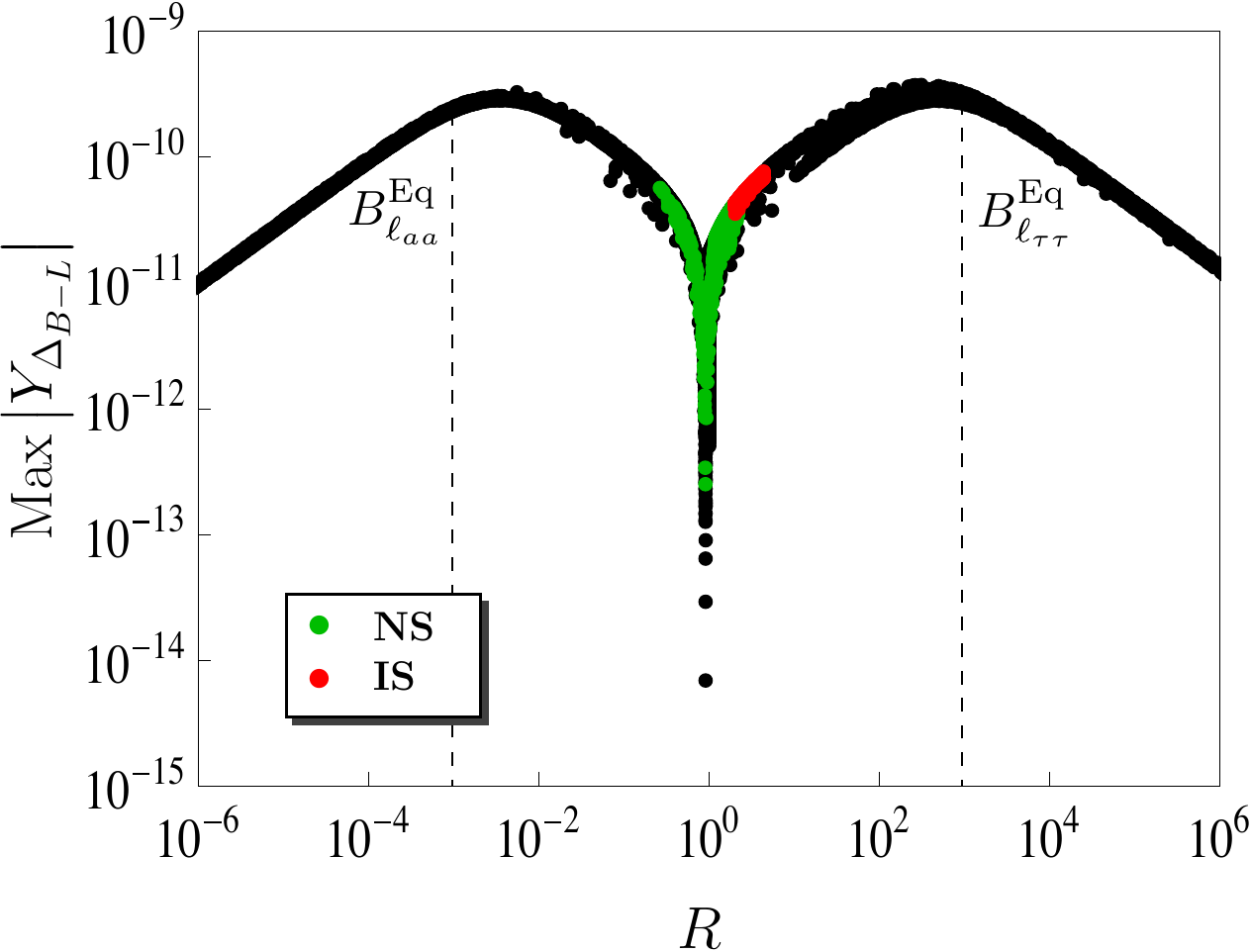}
    \caption{\it Maximum attainable final $B-L$ asymmetry as a
      function of the $R$ parameter, for 
      $m_{\Delta_\alpha}=10^9$~GeV, $m_{\Delta_\beta}=10^{10}$~GeV,
      $\tilde{m}_\Delta=10^{-2}$~eV and $B_\phi=10^{-4}$. Neutrino data constraints have been
      imposed as in Fig.~(\ref{fig:efficiency-vs-R}).}
  \label{fig:Max-Final-Asymmetry}
\end{figure}

  We plot in Fig.~\ref{fig:Max-Final-Asymmetry} the resulting maximal
   $B-L$ final asymmetry that can be achieved, as a function of the flavor
  parameter $R$, for $m_{\Delta_\alpha}/m_{\Delta_\beta}=10^{-1}$.  To
  this end we have considered the same parameter configuration used in
  Fig.~\ref{fig:efficiency-vs-R}.  It can be seen that the maximal
  $B-L$ asymmetry that can be achieved can account for the observed
  baryon asymmetry of the Universe for a large range of $R$ values,
  except at $R\approx 3/4$. We also point out that, if the neutrino
  mass matrix is dominated by the light scalar triplet, the
  constraints coming from neutrino data are compatible with successful
  PFL scenario.  One   realizes as well that two of the $B-L$
  asymmetry global maxima are shifted with respect to the efficiency
  maxima, and are now located around the points at which
  $\ell_i\ell_i\to\bar\Delta$ inverse decay rates are of the order of
  the Universe Hubble expansion rate, where
  $B_{\ell_{ii}}=B_{\ell_{ii}}^{\text{Eq}}$, see
  Eq.~(\ref{Bell-equilibrium}). As a final remark, it is worth noting
  that the local maximum at $R\approx 1$ has gone away.

  This result has to be compared with the unflavored case, where the
  CP asymmetry is very suppressed for $B_\phi\ll B_\ell$ or $B_\phi\gg
  B_\ell$, since the CP asymmetry is proportional to $\sqrt{B_\phi
    B_\ell}$---see Eqs.~(\ref{eq:CPpuretype2-LNC-1}) and
  (\ref{eq:CPpuretype2-bis}). This is no more the case in PFL
  leptogenesis, since the lepton number conserving and flavor
  violating CP asymmetries depend only on Yukawa couplings.

\section{General triplet flavored leptogenesis}
\label{sec:single-triplet-scenarios}
Having discussed the viability of the PFL scenario in pure type-II
seesaw models, we are now in a position to analyze the impact that
flavor effects may have in general triplet flavored leptogenesis
models. Here, as already defined in the introduction, by ``general models'' we
refer to models where the lepton number violating CP asymmetries are
relevant or even dominate over the lepton number conserving CP
asymmetries which drive PFL. Accordingly, if the extra degrees of
freedom enabling a non-vanishing CP asymmetry are additional triplets,
a general model will be defined by
Eq.~(\ref{eq:flavored-CP-asymmetry-type-II-scenario}), while if the
extra degrees of freedom are RH neutrinos---as will be the case in
models featuring interplay between type-I and type-II seesaws---the CP
asymmetry in Eq.~(\ref{eq:CPmixedtype1+2}), being lepton number
violating, will always define a ``general model''. In what follows we
will assume the asymmetry is entirely generated via the decays of the
lightest triplet, something that can be achieved by taking a heavy
mass spectrum obeying the following hierarchy: $m_\Delta\ll
M_{\Delta_\alpha, N_\alpha}$.

In ``general'' scenarios, since the CP asymmetries are lepton number
breaking, successful leptogenesis is possible in the absence of lepton
flavor effects, in contrast to PFL where flavor effects are
mandatory. In what follows we will quantify the enhancement
that the inclusion of flavor effects may have in the final $B-L$
asymmetry, and in order to do that and to put the discussion in
context we will start by briefly reviewing some general well known
results of the unflavored regime, which have been derived from kinetic
equations in which none of the SM reactions were taken into account (see
Sec.~\ref{sec:chemical-equilibrium}) \cite{Hambye:2012fh}. 

\begin{figure}[t!]
  \centering
  \includegraphics[width=7.2cm,height=4.5cm]
  {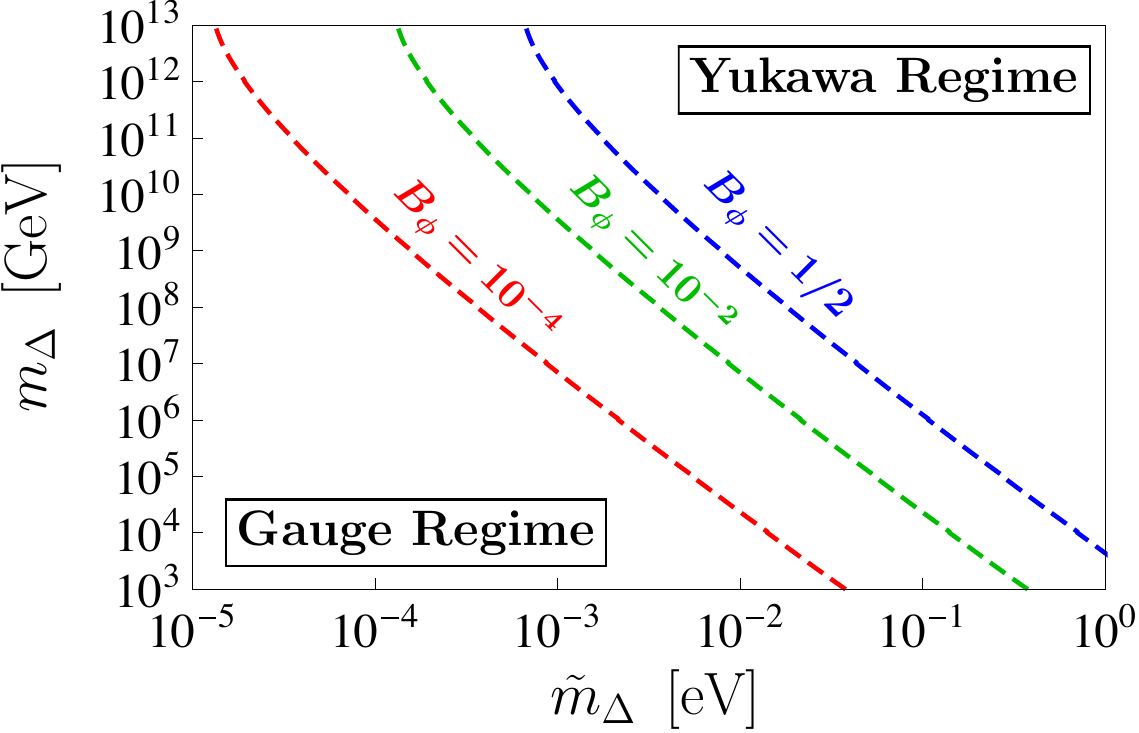}
  \hspace{0.5cm}
  \includegraphics[width=7.2cm,height=4.5cm]
  {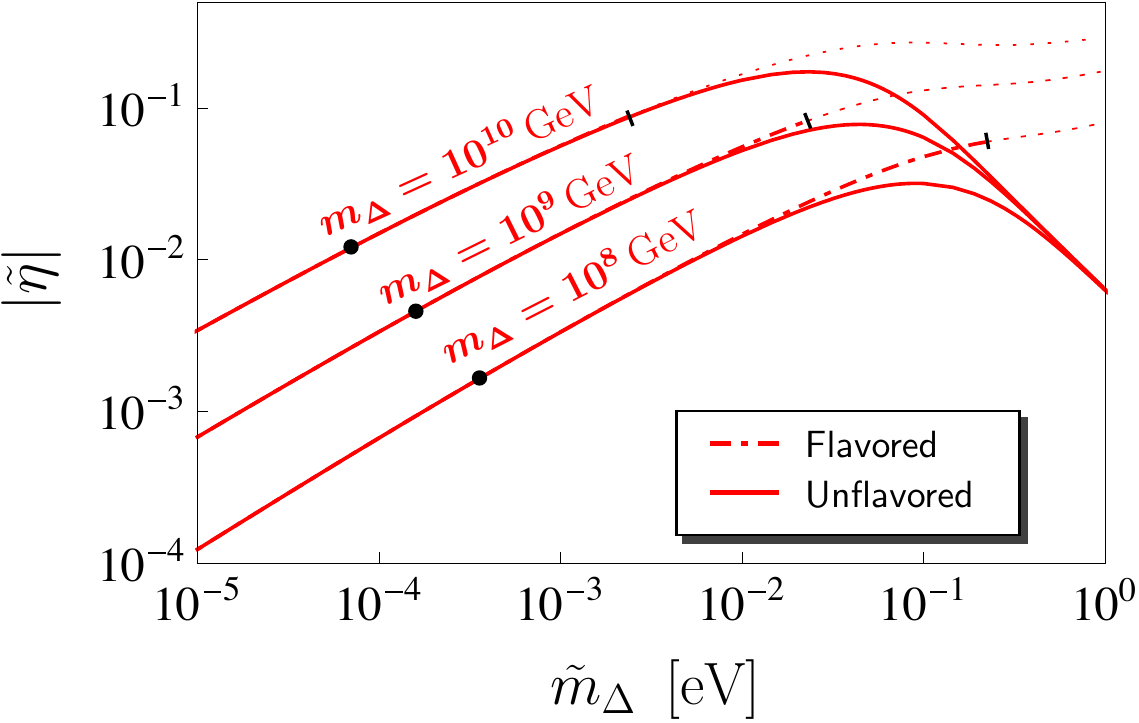}\vspace{0.4cm}\\
  \includegraphics[width=7.2cm,height=4.5cm]
  {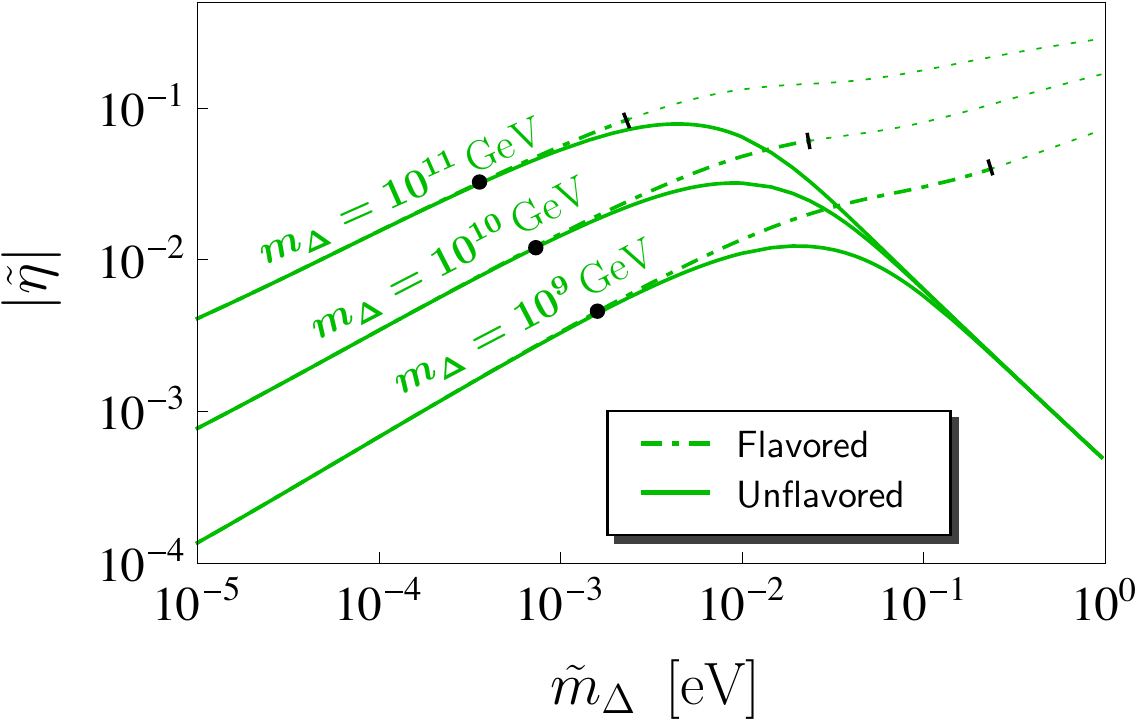}
  \hspace{0.5cm}
  \includegraphics[width=7.2cm,height=4.5cm]
  {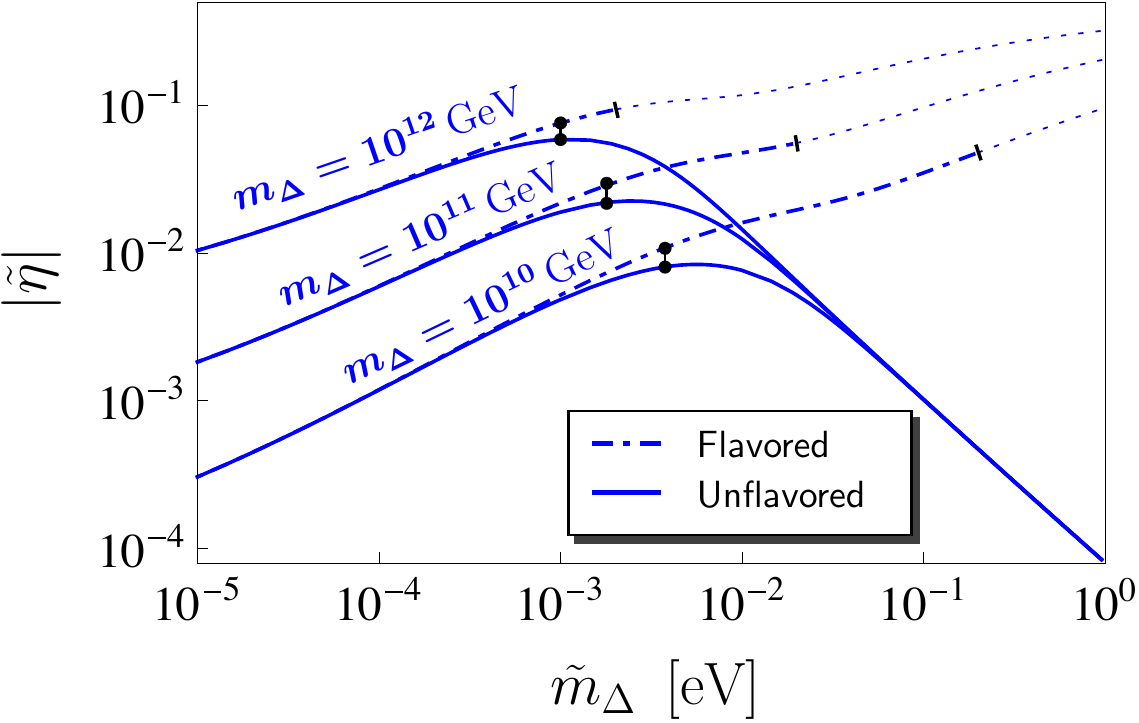}
  \caption{\it Upper left-hand side plot: Yukawa and gauge regimes for
    different values of $B_\phi$ in the plane $m_\Delta-\tilde
    m_\Delta$. The remaining plots show the dependence of the
    efficiency-like parameter $\tilde\eta$ (for the unflavored case
    $\tilde\eta$ refers to the efficiency) with the effective mass
    parameter $\tilde m_\Delta$,  for $B_\phi=10^{-4}$ (upper
    right-hand side plot), $B_\phi=10^{-2}$ (lower left-hand side
    plot) and $B_\phi=1/2$  (lower right-hand side
    plot). We fixed $\bar\epsilon_\Delta=0$ and the flavor
    configuration according to $B_{\ell_{aa}}=B_{\ell_{a\tau}}=0$ and
    $B_{\ell_{\tau\tau}}= 1-B_\phi $. Black dots correspond to
    $\tilde{m}_\Delta=\tilde{m}^*_\Delta$.  The lines are cut when the
      2-flavor regime condition 
      ceases to be fulfilled (see Sec.~\ref{domain-of-validity}).} 
  \label{fig:Yukawa-gauge-regime-Fl-UnFl-asymm-vs-mtilde}
\end{figure}
In the unflavored case, an efficiency function accounting for the $z$
(temperature) evolution of the unflavored $B-L$ asymmetry can be
defined\footnote{With the procedure followed in
  Sec.~\ref{sec:formal-integration-BEQs}, but using instead the system
  of equations in (\ref{eq:network-aligned-Triplet}) and
  (\ref{eq:network-aligned-BmL}), an explicit expression for $\eta(z)$
  can be derived for the unflavored regime.}:
\begin{equation}
  \label{eq:eff-unflavored-sec5}
  \eta(z)=-\frac{Y_{\Delta_{B-L}}(z)}
  {\epsilon_\Delta\,Y^{\text{Eq}}_\Sigma(z_0)}\ ,
\end{equation}
where at freeze out $\eta\equiv\eta(z\to \infty)$. As in fermion
triplet leptogenesis, in this case one can also define a gauge and a
Yukawa regime, which boundaries in the $\tilde m_\Delta-m_\Delta$
parameter space plane are determined by the values of $B_\phi$, as
displayed in
Fig.~\ref{fig:Yukawa-gauge-regime-Fl-UnFl-asymm-vs-mtilde} (upper
left-hand side plot). While in the gauge regime triplet dynamics is
dominated by gauge-mediated triplet annihilation, in the Yukawa regime
the dynamics is driven by Yukawa-induced reactions, and so it is in
the latter where flavor effects can have striking implications. For a
fixed triplet mass, the transition between both regimes becomes
determined by a ``critical'' $\tilde m_\Delta$, that we denote by
$\tilde m_\Delta^\star$\footnote{In practice, for a given value of
  $m_\Delta$, $\tilde m_\Delta^\star$ is defined as the value of
  $\tilde m_\Delta$ above (below) which the inverse decays are (not)
  in thermal equilibrium once the gauge scatterings cease to dominate
  the whole process (i.e. it leads to
  $\gamma_D/{n_\Delta^{\text{Eq}}H}=1$ when $\gamma_A$ goes below
  $\gamma_D$ at a temperature $z=z_A$).  Note that
    $\tilde m_\Delta^\star$ and $\tilde{m}_\Delta^{\text{eff}}$,
    defined in Eq.~(\ref{m-tilde-eff}), are unrelated parameters.}.

The behavior of the efficiency (i.e.~of the $B-L$ asymmetry) is to a
large extent determined by the regime where leptogenesis takes place
(gauge or Yukawa), or in other words by the location of the boundary
in the $\tilde m_\Delta-m_\Delta$ plane, determined in turn by the
value of $B_\phi$. For the parameter space points shown in
Fig.~\ref{fig:Yukawa-gauge-regime-Fl-UnFl-asymm-vs-mtilde} (upper
left-hand side plot), the behavior of the $B-L$ asymmetry goes along
the following lines:
\begin{itemize}
\item {\it The $B_\phi=1/2$ case}:\\
  In the gauge (Yukawa) regime the asymmetry increases (decreases)
  with $\tilde{m}_\Delta$. In the gauge regime this is due to the fact
  that there is no substantial production of the asymmetry until $z$
  approaches the value $z=z_A$ where $\gamma_A/\gamma_D$ goes below
  unity, $z_A\sim 3$ in the right-hand side plot in
  Fig.~\ref{fig:reactionrates}. The generation of the $B-L$ asymmetry
  can then be understood as proceeding in two stages determined by two
  $z$ (temperature) windows: $z<z_A$ and $z>z_A$. For $z<z_A$ the
  $B-L$ asymmetry is suppressed by a factor $\gamma_D/\gamma_A\propto
  m_\Delta\tilde{m}_\Delta/g^4$, so for a fixed scalar triplet mass
  the smaller $\tilde{m}_\Delta$, the smaller the ratio
  $\gamma_D/\gamma_A$, and so the asymmetry produced.  In this range
  most of the production occurs when $z\lesssim z_A$. Before,
  $\gamma_D/\gamma_A$ is exponentially suppressed.  For $z>z_A$
  instead, $\gamma_A$ becomes irrelevant and, since in the gauge
  regime there is no suppression effect from inverse decays, the
  asymmetry produced within this $z$ range is simply equal to the
  number of triplets left times the CP asymmetry,
  $Y_{\Delta_{B-L}}\simeq \epsilon _\Delta Y_\Delta^{\text{Eq}}(z\sim
  z_A)$. The relevance of this contribution is determined by
  $\tilde{m}_\Delta$: large values of this parameter imply small
  values for $z_A$, which then in turn imply a less Boltzmann
  suppressed $Y_\Delta^{\text{Eq}}(z_A)$.  Thus, the $B-L$ asymmetry
  generated consist of two contributions, one generated at $z<z_A$ and
  a second produced at $z> z_A$, namely
  \begin{equation}
    \label{eq:upperbound}
    Y_{\Delta_{B-L}}\simeq 
    \epsilon_\Delta \int_{z_{0}}^{z_A} \frac{dY_\Sigma^{\text{Eq}}}{dz}
    \frac{\gamma_D}{4 \gamma_A} dz 
    +
    \epsilon_\Delta Y_\Sigma^{\text{Eq}}(z_A)\simeq \epsilon_\Delta^\ell
    Y_\Sigma^{\text{Eq}}(z_A) (z_A/4+1)\ .
  \end{equation}
  In the Yukawa regime instead, the efficiency decreases with
  $\tilde{m}_\Delta$ because in this case still there is no
  substantial asymmetry produced until $z$ approaches $z_A$, and
  because the asymmetry produced afterwards is further washed-out by
  the inverse decay whose magnitude increases with $\tilde{m}_\Delta$.
\item {\it The small $B_\phi$ case  ($B_\phi=10^{-2}$ or
    $B_\phi=10^{-4}$)}:\\
  As can be seen in the upper right-hand and lower left-hand plots in
  Fig.~\ref{fig:Yukawa-gauge-regime-Fl-UnFl-asymm-vs-mtilde} the
  efficiency goes on to increase with $\tilde{m}_\Delta$ well inside
  the Yukawa regime. This can be understood from the fact that in this
  case, even if the total decay rate is well in thermal equilibrium 
  and faster than the gauge scattering rate, the decay to a pair of
  scalars remains out-of-equilibrium (see
  Fig.~\ref{fig:reactionrates}, left-hand side plot), which implies
  that lepton number is not broken by inverse decay, resulting in no
  washout from these processes. The efficiency is suppressed only by
  gauge-mediated triplet scatterings which, as pointed out in the
  previous item, precludes any substantial production of $B-L$
  asymmetry until $z$ reaches $\sim z_A$. All in all, despite standing
  in the Yukawa regime, the $B-L$ asymmetry is only suppressed by
  gauge scatterings in the way stressed in the previous item, and so
  the total $B-L$ asymmetry is again given by
  Eq.~(\ref{eq:upperbound}).  For large values of $\tilde{m}_\Delta$, the
  gauge suppression is nevertheless faint because $z_A$ is not much
  larger than unity. This results in very large efficiency for
  $B_\phi\ll 1/2$.  Only when $\tilde{m}_\Delta\gg
  \tilde{m}_\Delta^\star$ is lepton number effectively broken by
  scalar doublet-triplet inverse decays and the efficiency decreases
  with $\tilde{m}_\Delta$.  Note that, even if large efficiencies can
  be obtained in this way for $B_\phi\ll 1/2$, since lepton number is
  unbroken in the $B_\phi\rightarrow 0$ limit, these efficiency
  enhancements are accompanied by a suppression of the CP asymmetry,
  so that still the maximum $B-L$ asymmetry is obtained for values of
  $B_\phi$ not far from its maximum value 1/2.
\end{itemize}
The picture described in the items above is expected to change as soon
as one hits the flavor regime, if the parameters are
such that triplet dynamics takes place in the Yukawa regime. In order
to discuss the impact that flavor effects may have, it is convenient
to introduce an efficiency-like parameter. Let us discuss this in some
more detail.  Flavor coupling does not allow a conventional definition
of an efficiency, however a parameter resembling the efficiency of the
unflavored case can be defined:
\begin{equation}
  \label{eq:asymmetry-efficiency-general}
  \tilde\eta=
  - \frac{Y_{\Delta_{B-L}}(z\to \infty)}
  {\epsilon_\Delta Y^\text{Eq}_{\Sigma}(z_0)}  \ ,
\end{equation}
with $\tilde\eta$ given by
\begin{equation}
  \tilde\eta\equiv\frac{1}{2}
  \left[
    \left(
      \eta_{aa}+\eta_{a\tau}+\eta_{\tau a}+\eta_{\tau\tau}
    \right)
    + \bar\epsilon_\Delta \left(\eta_a+\eta_\tau\right)
  \right]\ ,
\end{equation}
where the flavored efficiency functions have been defined in
Eq.~(\ref{eq:flavored-efficiency-functions-non-primed-basis}) and with
\begin{equation}
  \label{eq:epsilon-bar}
  \epsilon_\Delta=  \epsilon^{\ell_a}_\Delta + \epsilon^{\ell_\tau}_\Delta
  \qquad \text{and}\qquad 
  \overline \epsilon_\Delta =
  \frac{\epsilon_\Delta^{\ell_a}-\epsilon_\Delta^{\ell_\tau}}
  {\epsilon_\Delta}\ .
\end{equation}
Note that the definition of $\tilde \eta$ is such that when taking the
limit $\epsilon_\Delta^{\ell_a}\to \epsilon_\Delta^{\ell_\tau}$, one
recovers the usual definition of the efficiency.  This parameter
proves to be useful in particular when comparing the results obtained
in the flavored regime with those arising from the unflavored limit.
Instead, the parameter $\overline\epsilon_\Delta$, introduced in the
definition of $\tilde\eta$, has a two-fold utility: first of all it
``measures'' the deviation from the PFL ($\bar \epsilon_\Delta\gg 1$)
and the general scenarios ($\bar \epsilon_\Delta\ll 1$); secondly, it
``measures'' the flavor misalignment of the source terms in the
evolution equations of the $B/3-L_i$ charges.

In order to quantify the impact that flavor effects have on the $B-L$
asymmetry, it is useful to consider first a case where both CP
flavored asymmetries are equal, i.e. $\overline\epsilon_\Delta=0$,
that is to say in a way the extreme opposite to the PFL case.  This
will allow to discuss flavor effects that are different from the ones
we discussed in the previous section for the PFL case.  For this case,
we show in Fig.~\ref{fig:Yukawa-gauge-regime-Fl-UnFl-asymm-vs-mtilde}
the efficiency-like parameter $\tilde{\eta}$ as a function of $\tilde
m_\Delta$ for different values of ($m_\Delta$, $B_\phi$), overlapped
with the results we got for the unflavored case.

Some comments are in order regarding these results. Either in the
gauge or in the Yukawa regime (for $\tilde m_\Delta\sim \tilde
m^\star_\Delta$), gauge scatterings preclude any substantial creation
of a $B-L$ asymmetry as long as $\gamma_A/\gamma_D\gg 1$, that is to
say as long as $z$ is below $\sim z_A$.  The $B/3-L_i$ asymmetry
production is anyway suppressed by a $\gamma_D/\gamma_A$ factor as in
Eq.~(\ref{eq:upperbound}). Gauge scatterings, being flavor ``blind'',
are insensitive to lepton flavor effects and so the suppressions
they induce cannot be overcome.  This means that, as long as we consider values of
parameters which in the unflavored case gives
Eq.~(\ref{eq:upperbound}) i.e. the maximum efficiency allowed by gauge
scattering (in the gauge regime, or in the Yukawa regime for
$B_\phi<1/2$ and not too large values of $\tilde{m}$), flavor effects
cannot further enhance the efficiency.  However, in the Yukawa regime,
for large values of $\tilde{m}_\Delta$, since inverse decay washouts
are flavor sensitive, flavor effects allow to largely avoid this
effect, so that the efficiency goes on to increase also there, as
Fig.~\ref{fig:Yukawa-gauge-regime-Fl-UnFl-asymm-vs-mtilde} shows. As a
result in this case too, one is left only with the unavoidable gauge
scattering suppression. This suppression is nevertheless very mild for
large values of $\tilde{m}$ (i.e.~small values of $z_A$). Hence, large
enhancement of the efficiency can be obtained from flavor effects,
especially for large values of $B_\phi$.  In other words, deep inside
the Yukawa region ($\tilde m_\Delta\gg \tilde m^\star_\Delta$) where
gauge scattering suppression is faint, flavor effects start showing up
and become even striking as $\tilde m_\Delta$ increases and $B_\phi$
approaches 1/2.  Summarizing, in this equal flavored CP asymmetries
case we consider here ($\overline\epsilon_\Delta=0$),
Eq.~(\ref{eq:upperbound}) can still be used as an approximate upper
bound of the $B-L$ asymmetry one can reach in all regimes, even deep
in the Yukawa regime. We have checked that this upper bound can be
saturated in all regimes up to a factor~$\sim 2$.

\begin{figure}
  \centering
  \includegraphics[scale=0.6]
  {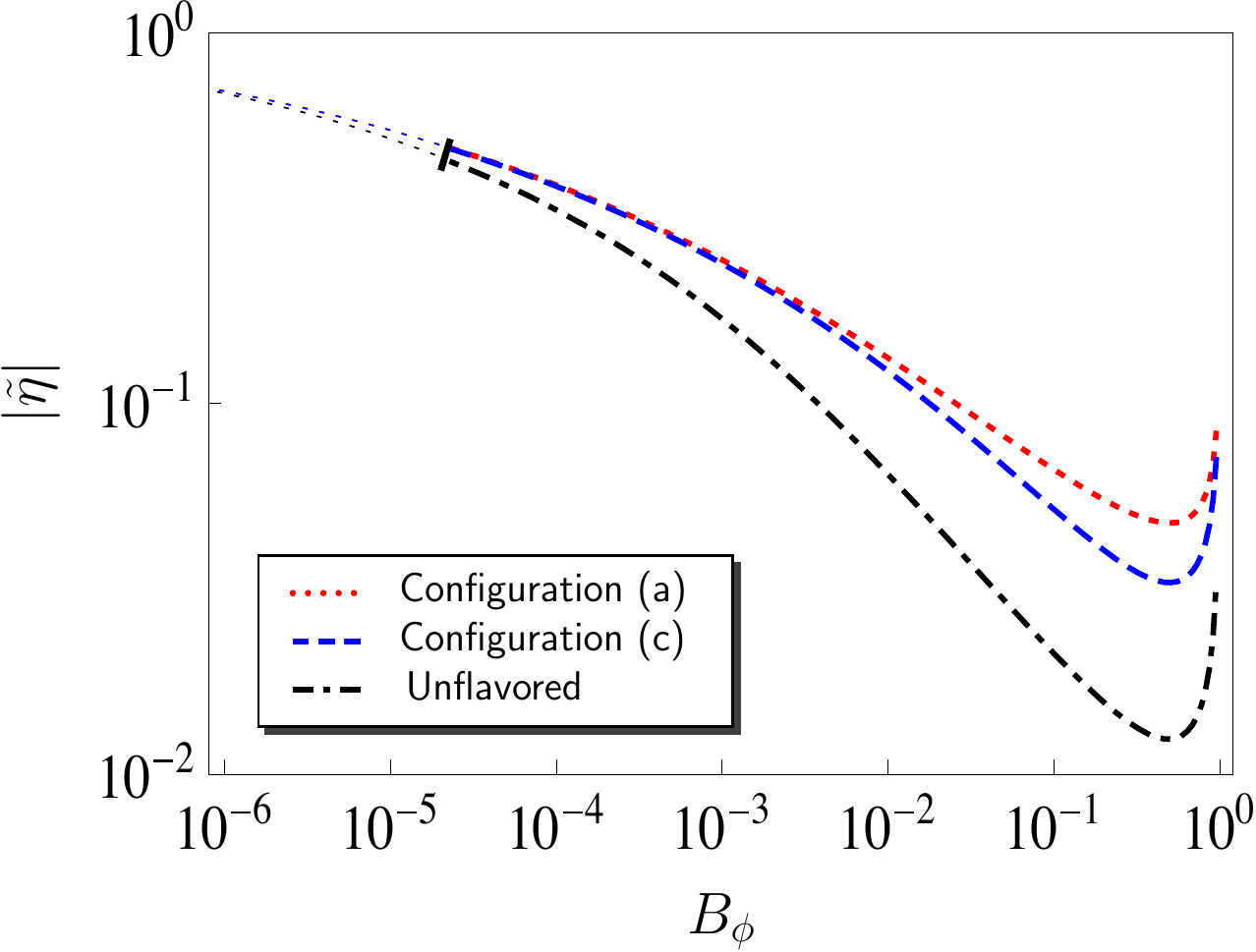}
  \includegraphics[scale=0.6]
  {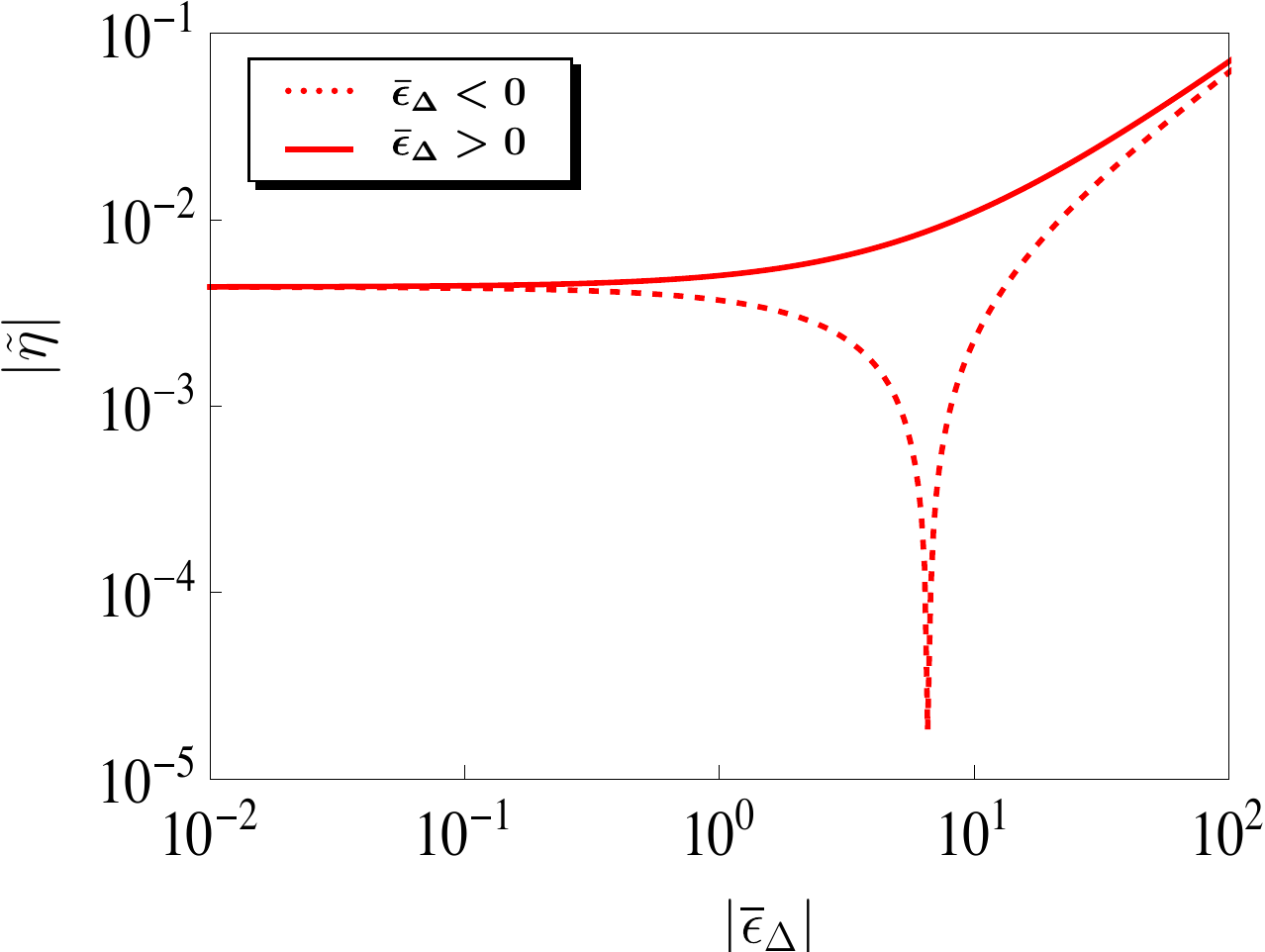}
  \caption{\it Left-hand plot: Efficiency-like parameter
    $\tilde{\eta}$ as a function of $B_\phi$, for
    $\epsilon_\Delta^{\ell_a}=\epsilon_\Delta^{\ell_\tau}$.  The flavor
    configuration (a) corresponds to the one used in Fig.~\ref{fig:YDBmL-v-bphi}, while (c) corresponds to $B_{\ell{aa}}=B_{\ell{\tau\tau}}=(1-B_\phi)/2$  (for
    the unflavored case $\tilde\eta$ refers to the efficiency).
    Right-hand plot: Efficiency-like parameter $\tilde{\eta}$ as a
    function of $|\overline \epsilon_\Delta|$ for $B_\phi=1/2$ and
    flavor configuration (a). In both plots we fixed 
    $m_\Delta=10^{11}$°GeV and $\tilde{m}_\Delta=10^{-2}$~eV. The lines are cut
      whenever the 2-flavor regime condition
       ceases to be fulfilled (see
      Sec.~\ref{domain-of-validity}).}
\label{fig:YDBmL-vs-epsiloni-over-epsilon}
\end{figure}

To further emphasize the effects of flavor in the
$\bar\epsilon_\Delta=0$ case, we have calculated the efficiency-like
parameter $\tilde\eta$ as a function of $B_\phi$. The calculation has
been done for fixed parameters
$m_\Delta$ and $\tilde{m}_\Delta$, and  for two flavor configurations (a)---the one already
used in Fig.~\ref{fig:YDBmL-v-bphi}---and (c) which corresponds to $B_{\ell{aa}}=B_{\ell{\tau\tau}}=(1-B_\phi)/2$,  i.e. without any flavor structure.   The results are displayed in
Fig.~\ref{fig:YDBmL-vs-epsiloni-over-epsilon} (left-hand side plot),
where the flavored and unflavored (as e.g. in
Refs.~\cite{Hambye:2012fh, Hambye:2005tk}) outputs are compared.  It
can be seen that considering only the effects of the SM interactions
(i.e.~configuration  (c)), one can get an enhancement of order 2 with
respect to the unflavored case, whereas for the flavor configuration
(a) one can get a further one-order of magnitude enhancement, as can
be seen in particular for $B_\phi=B_\ell=1/2$.

Finally, let us discuss what happens very qualitatively in cases other
than the pure PFL case,
$\epsilon_\Delta^{\ell_\tau}=-\epsilon_\Delta^{\ell_a}$ and the
``opposite'' case,
$\epsilon_\Delta^{\ell_\tau}=\epsilon_\Delta^{\ell_a}$. In these
``intermediate'' cases the ``efficiency'' as defined in
Eq.~(\ref{eq:asymmetry-efficiency-general}) cannot be considered as an
efficiency anymore, because it can be larger than one. For instance in
the pure PFL case it is infinity since $\epsilon_\Delta=0$. As a
result it is difficult to span the range of possibilities in simple
terms for these cases. To get a reliable idea of the behavior of the
efficiency-like parameter, and thus of the $B-L$ asymmetry, in a
specific case, the most efficient procedure is probably to integrate
first the full set of Boltzmann equations in a ``blind'' way and see
what the result looks like before trying to understand it by simple
means. But the basic picture qualitatively remains clear. As long as
$z<z_A$ any flavor asymmetry production is suppressed by a factor of
$\gamma_D/\gamma_A$, and afterwards the $B-L$ asymmetry that can be
produced can anyway not be larger than the number of triplets
remaining at $z\sim z_A$ times the sum of the absolute values of the
flavor asymmetries. The important flavor effects stressed above, from
the $L$-violating inverse decays as well as from the $L$-violating
decays, will be operative in a way which may depend non trivially on
basically all parameters, the flavor CP asymmetries, the
$C^{\ell,\phi}$ constants, the total decay rate and the various
branching ratios.

As an illustration of the efficiency dependence on the mismatch
between the flavored CP asymmetries, parameterized by
$\bar\epsilon_\Delta$, on the right-hand side plot in
Fig.~\ref{fig:YDBmL-vs-epsiloni-over-epsilon} we show the dependence
of $\tilde{\eta}$ with $\bar\epsilon_\Delta$ for $B_\phi=1/2$ for the
flavor configuration (a).  In the region where $\overline
\epsilon_\Delta\ll 1$ is small ($\epsilon_\Delta^{\ell_a}\sim
\epsilon_\Delta^{\ell_\tau}$), as previously stressed, any possible
mismatch between the asymmetries in flavor $a$ and $\tau$ can only be
due to the flavor dependence of the washout terms. As $\overline
\epsilon_\Delta$ increases, the source terms start having a flavor
dependence as well, and so an imbalance between production in flavor
$a$ and $\tau$ appears. The flavor dependence of both production and
washout at large $\overline\epsilon_\Delta$, yields larger values for
$\tilde{\eta}$.  In other words, flavor effects are diminished in
those regions of parameter space where $\overline \epsilon_\Delta\ll
1$ and become more remarkable in regions where $\overline
\epsilon_\Delta\gg 1$.  Accordingly, in the various flavor regimes,
enhancements of the efficiency-like parameter $\tilde \eta$ with
respect to the unflavored case are a consequence of combined effects:
the mismatch between the different flavored CP asymmetries
$\epsilon^{\ell_i}_\Delta$, the SM interactions through the $C^\ell$
and $C^\phi$ matrices, and the flavor configurations encoded in
$B_{\ell_{ij}}$.
\section{Conclusions}
\label{sec:concl}
We have considered scalar triplet leptogenesis scenarios where the
states enabling successful production of the cosmic baryon asymmetry
are either extra triplets or RH neutrinos.  We have derived for the
first time the complete set of flavored classical Boltzmann equations
governing the evolution of the different relevant asymmetries,
including the effects of those SM reactions which in the leptogenesis
era may be fast: charged lepton and quark Yukawa reactions as well as
QCD and electroweak sphaleron processes. The resulting network of
kinetic equations combined with the different {\it asymmetry coupling
  matrices}, which follow from the chemical equilibrium conditions
enforced by the fast SM processes, provide the tools for studying
triplet scalar leptogenesis in full generality. 
 Furthermore, by requiring that the decoherence rate to be faster than the leptonic inverse decay rate during the leptogenesis era, we determined the domain of validity of the various flavor regimes.

In scenarios involving an additional triplet (purely type-II seesaw
scenarios), we have identified a novel class of models where the
flavored CP asymmetries, consisting of lepton number violating and
lepton number conserving contributions, become dominated by the lepton
number conserving piece. Such a dominance naturally shows up as soon
as the couplings of at least one triplet (i.e. not necessarily of all
seesaw states as for PFL type-I seesaw scenarios) approximately
conserve $L$, in practice simply that it couples more to leptons than
to scalars.  The purely flavored CP asymmetries have no reasons to be
suppressed by the smallness of the light neutrino masses since, in
contrast to the lepton-number-violating CP asymmetries, they only
involve $L$-conserving couplings.

With the aid of the derived flavored Boltzmann equations and {\it
  asymmetry coupling matrices}, we have carried out a throughout study
of the PFL scenario in the two flavor regime, for definitiveness.  The
way this PFL scenario works is totally novel (for small values of
$B_\phi$ which gives natural dominance of the purely flavored CP
asymmetries): in this case there is no $L$-violating process in
thermal equilibrium at any epoch but yet flavor effects do allow the
creation of a $B-L$ asymmetry from the $L$-violating slow decay of the
triplet to SM scalars.  We have proved its viability by calculating
the $B-L$ yield, finding that, for utterly reasonable and wide ranges
of parameter values, a baryon asymmetry consistent with observation
can always be achieved. By exploring the $B-L$ asymmetry parameter
space dependence, we have determined the lepton flavor configuration
that maximizes the efficiency, finding that the same structure renders
the flavored CP asymmetry minimal. Our findings show that maximal
$B-L$ yield is achieved for intermediate lepton flavor configurations.

Finally, we discussed general scenarios, which we have defined by the
condition of the CP asymmetry involving lepton number violation.
These scenarios can arise either in models with extra triplets or with
RH neutrinos (models exhibiting interplay between type-I and type-II
seesaw). We discussed the impact that lepton flavor effects may have
in the final $B-L$ asymmetry, showing that relevant flavor effects can
only be achieved in the Yukawa regime, being more striking as deeper
one moves into that regime, and depending on the parameter flavor
configuration. Our results show that for certain flavor
structures---once lying in the Yukawa regime---the asymmetry may be
enhanced by several orders of magnitude.  In both regimes, the $B-L$
asymmetry production is suppressed as long as the gauge scattering
rate is faster than the decay rate (i.e.~for $z\ll z_A$), and the
asymmetry produced afterwards is proportional to the number of triplet
remaining afterwards. 
The latter being  Boltzmann suppressed if $z_A>1$, the suppression is more pronounced for smaller values of $m_\Delta$, given that as $m_\Delta$ decreases $z_A$ increases.  
Deep in the Yukawa regime, however, the decay rate becomes faster than the gauge scattering rate at very early epochs, $z_A\lesssim 1$ and this Boltzmann suppression goes away. 
In this way, deep in the Yukawa
regime, one can basically avoid all efficiency suppressions, from
gauge scattering as well as, through flavor effects, from
$L$-violating inverse decays.
\section{Acknowledgements}
We want to thank Enrico Nardi for many stimulating discussions and
valuable comments.  DAS wants also to thanks Juan Racker for useful
comments. DAS is supported by the Belgian FNRS agency through a
``Charg\'e de Recherches'' contract and will like to thank the
``Service de Physique Th\'eorique'' of the ``Universit\'e Libre de
Bruxelles'' for the warm hospitality during the completion of this
work. The work of MD and TH is supported by the FNRS, the IISN and by
the Belgian Science Policy, IAP VII/37.
\appendix
\section{Conventions and definitions}
\label{ThermoDef}
Here in this appendix we collect all the relevant formul{\ae} we used
throughout the paper.  We stress we have used Maxwell-Boltzmann
distributions, so for the number of relativistic degrees of freedom we
have used $g_\star=\sum_{i=\text{All species}}g_i=118$ ($T\gg 300\,$
GeV) while for the entropy density
\begin{equation}
  \label{eq:entropy-density}
  s(z)=\frac{4 m_\Delta^3g_\star}{z^3\pi^2}\ ,
\end{equation}
with $z=m_\Delta/T$. For the expansion rate of the Universe we used
\begin{equation}
  \label{eq:expansion-rate-Universe}
  H(z)=\sqrt{\frac{8 g_\star}{\pi}}\frac{m_\Delta^2}{M_\text{Planck}}\frac{1}{z^2}\ .
\end{equation}
Decay and scattering $1\leftrightarrow 2$ and $2\leftrightarrow 2$
reaction densities are given by:
\begin{align}
  \label{eq:reaction-densities-decay}
  \gamma_D&=\frac{K_1(z)}{K_2(z)}\,n^\text{Eq}_\Sigma\,\Gamma^\text{Tot}_\Delta\ ,
  \\
  \label{eq:reaction-densities-scattering}
  \gamma_S&=\frac{m_\Delta^4}{64\,\pi^4}\int_{x_\text{min}}^\infty
  \,dx\sqrt{x}\,\frac{K_1(z\sqrt{x})\;\widehat\sigma_S}{z}\ .
\end{align}
Here $n^\text{Eq}_\Sigma$ is the $\Sigma=\Delta+\Delta^\dagger$ number
density (number density for a non-relativistic species),
$x=s/m_\Delta^2$ ($s$ being the center-of-mass energy),
$\Gamma_\Delta^\text{Tot}$ denotes the triplet total decay width,
given in Eq.~(\ref{eq:Gammatot}), whereas $\widehat \sigma_S$ the
reduced cross section. The integration upper and lower limits are
determined by the kinematics of the corresponding scattering process:
for gauge boson mediated processes $x_\text{min}=4$, for Yukawa (or
scalar) induced reactions $x_\text{min}=0$.

Denoting $\delta=\Gamma_\Delta^\text{Tot}/m_\Delta$, we have found
that the reduced cross sections for the $s$ and $t$ channel $\Delta
L=2$ processes can be written as:
\begin{align}
  \label{eq:DeltaLEq2-reactions-reduced-cross-sections}
  \widehat \sigma^{\phi\phi}_{\ell_i \ell_j}&=
  64\pi\,B_\phi\,B_{\ell_{ij}}
  \,\delta^2\,
  \frac{x}{(x-1)^2+\delta^2}\ ,
  \nonumber\\
  \widehat \sigma^{\phi\ell_j}_{\phi \ell_i}&=
  64\pi\,B_\phi\,B_{\ell_{ij}}\,\delta^2\,
  \frac{1}{x}
  \left[
    \ln(1+x)
    -
    \frac{x}{1+x}
  \right]\ .
\end{align}
The reduced cross sections for the $s$ and $t$ channel flavor
violating reactions, instead, can be written according to:
\begin{align}
  \label{eq:flavor-violating-reduced-cross-sections}
  \widehat \sigma^{\ell_n\ell_m}_{\ell_i\ell_j}&=
  64\pi\,B_{\ell_{nm}}\,B_{\ell_{ij}}\,\delta^2\,
  \frac{x^2}{(1-x)^2+\delta^2}\ ,
  \nonumber\\
  \widehat \sigma^{\ell_j\ell_m}_{\ell_i\ell_n}&=
  64\pi\,B_{\ell_{nm}}\,B_{\ell_{ij}}\,\delta^2\,
  \left[
    \frac{x+2}{x+1}
    -
    \ln(1+x)
  \right]\ .
\end{align}
Finally, the reduced cross section for gauge induced processes reads
\cite{Hambye:2012fh,Hambye:2005tk,Cirelli:2007xd}
\begin{align}
  \label{eq:reduced-cross-section-gauge}
  \widehat \sigma_A=&
  \frac{2}{72\pi}
  \left\{
    \left(
      15 C_1 - 3C_2 
    \right)r
    +
    \left(
      5 C_2 - 11C_1 
    \right)r^3
  \right.
  \nonumber\\
  &\left. 
    + 3\left(r^2-1\right)
    \left[
      2 C_1 +  C_2\left(r^2-1\right) 
    \right]
    \ln\left(\frac{1+r}{1-r}\right)
  \right\}
  + \left(\frac{50 g^4 + 41 g^{\prime 4}}{48 \pi}\right)\,r^{3/2}\ ,
\end{align}
where the following notation has been adopted: $r=\sqrt{1-4/x}$ and
$C_1=12g^4+3g_Y^4+12g^2g^2_Y$ and $C_2=6g^4+3g_Y^4+12g^2g_Y^2$ (with
$g$ and $g_Y$ the $SU(2)$ and $U(1)$ SM gauge coupling
constants).

The reaction densities with a resonant intermediate state subtracted
can be calculated from Eqs.~(\ref{eq:reaction-densities-decay}),
(\ref{eq:reaction-densities-scattering}),
(\ref{eq:DeltaLEq2-reactions-reduced-cross-sections}) and
(\ref{eq:flavor-violating-reduced-cross-sections}) as follows:
\begin{align}
  \gamma^{\prime\phi\phi}_{\ell_i\ell_j}&=
  \gamma^{\phi\phi}_{\ell_i\ell_j}-B_{\ell_{ij}}\,B_\phi\,\gamma_D\ ,
  \nonumber\\
  \gamma^{\prime\ell_n\ell_m}_{\ell_i \ell_j}&=
  \gamma^{\ell_n\ell_m}_{\ell_i \ell_j}-B_{\ell_{ij}}B_{\ell_{nm}}\gamma_D\ .
\end{align}
Rates for the different SM reactions are approximately
given by \cite{Moore:1997im,Bento:2003jv,Campbell:1992jd,Cline:1993bd}:
\begin{alignat}{6}
  \label{eq:QCD-instantons}
  \mbox{QCD instantons:}&\qquad &\gamma_\text{QCD}(T)&\simeq 312\,\alpha_\text{S}\,T^4\ ,\\
  \label{eq:EW-sphalerons}
  \mbox{Electroweak sphalerons:}&\qquad &\gamma_\text{EW}(T)&\simeq 26\,\alpha_\text{EW}\,T^4\ ,\\
  \label{eq:Yuk-reactions}
  \mbox{Yukawa reactions:}&\qquad &\gamma_{f_i}(T)
  &\simeq 5\times 10^{-3}\,h_{f_i}^2\,T\,n^\text{Eq}_{f_i}
  =5\times 10^{-4}\,h_{f_i}^2\,T^4\ ,
\end{alignat}
where $h_{f_i}$ denotes the Yukawa coupling of fermion $f_i$.

\section{Summary of the different $C^\ell$ and $C^\phi$ matrices}
\label{Other-Cell}

\subsection{$C^\ell$ matrices in all possible regimes}
\label{Cell-other-regime}
\begin{table}[h!]
\centering
\begin{tabular}{| l | l | c | l |}
\hline\hline
$T$ (GeV) &In equilibrium &Flavor(s)&  Global symmetries  of the effective $\mathcal{L}$\\
\hline
\hline   
 &&&\\
$\gtrsim 10^{15}$   & Hyp.  &1  &    $  U(1)_Y \times U(1)_B \times U(1)_{E_R}\times U(1)_{PQ}\times$\\
& & & $  SU(3)_Q \times  SU(3)_u\times SU(3)_d \times SU(3)_e$  \\ [6pt]
\hline  
 &&&\\
$ [10^{12},10^{15}]$  & Hyp., $t$&1 & $ U(1)_Y \times U(1)_B  \times U(1)_{E_R}\times$\\
& &&$SU(2)_Q \times  SU(2)_u\times SU(3)_d \times SU(3)_e$       \\  [6pt]
\hline
 &&&\\
   $ [10^{9}, 10^{12}]$ :&&&\\ [8pt]
\quad$[T^\tau_\text{decoh},10^{12}]$& Hyp., Sphal.,  &1  &$ U(1)_Y\times U(1)_Q \times  U(1)_u\times SU(2)_d \times$\\
&  $t$,$b$,$c$  & & $ SU(3)_e   $      \\ [5pt]
\quad$[10^{9},T^\tau_\text{decoh}]$& Hyp., Sphal.,    &2& $U(1)_Y\times U(1)_Q \times  U(1)_u\times SU(2)_d \times $\\
&  $t$,$b$,$c$,$\tau $  &&$SU(2)_e   $      \\ [6pt]
\hline
 &&&\\
$[10^{5}, 10^{9}] $  :  &&&\\ [8pt]
\quad$[T^{\tau}_\text{decoh},10^{9}]$& Hyp., Sphal.,   &1& $U(1)_Y\times U(1)_Q\times  U(1)_u\times U(1)_d  \times$        \\
& $t$,$b$,$c$,$s$  && $ SU(3)_e $\\[5pt]
\quad$[T^{\mu}_\text{decoh},T^{\tau}_\text{decoh}]$  & Hyp., Sphal.,  & 2 &$U(1)_Y\times U(1)_Q\times  U(1)_u\times U(1)_d  \times$        \\
& $t$,$b$,$c$,$s$,$\tau$  && $ SU(2)_e $\\[5pt]
\quad$[10^5,T^{\mu}_\text{decoh}]$   & Hyp., Sphal.,   &3 & $U(1)_Y\times U(1)_Q\times  U(1)_u\times U(1)_d  \times $        \\
& $t$,$b$,$c$,$s$,$\tau$,$\mu$  && $U(1)_e $\\  [6pt]
\hline
 &&&\\
$ \lesssim 10^5 $  :  &&&\\[8pt]
\quad$[T^{\tau}_\text{decoh},10^{5}]$& Hyp., Sphal.,   &1& $  U(1)_Y\times SU(3)_e$        \\
& $t$,$b$,$c$,$s$,$u$,$d$  &&  \\ [5pt]
\quad$[T^{\mu}_\text{decoh},T^{\tau}_\text{decoh}]$  & Hyp., Sphal.,   &2& $ U(1)_Y\times SU(2)_e$        \\
& $t$,$b$,$c$,$s$,$u$,$d$,$\tau$  &&  \\ [5pt]
\quad$[T^{e}_\text{decoh},T^{\mu}_\text{decoh}]$   & Hyp., Sphal.,   &3& $ U(1)_Y\times  U(1)_e $        \\
& $t$,$b$,$c$,$s$,$u$,$d$,$\tau$,$\mu$  &&  \\ [5pt]
\quad$\lesssim T^{e}_\text{decoh}$   & Hyp., Sphal.,   &3& $ U(1)_Y$     \\
& $t$,$b$,$c$,$s$,$u$,$d$,$\tau$,$\mu$,$e$  && \\ [6pt]
\hline
\end{tabular} 
\caption{\it Temperature ranges and the corresponding 
  reactions which are in thermal equilibrium. In the third column we 
  show the number of flavor(s) that has (have) to be considered in 
  the kinetic equations, and in the fourth column are the global 
  symmetries of the early Universe effective Lagrangian are displayed.}
\label{table-chemical-pot-summary} 
\end{table}
\begin{table}[t!]
\centering
\begin{tabular}{|l|c| c | c|}
\hline\hline
$T$ (GeV) & Flavor(s)    &$C_\ell$&  $C_\phi$  \\
\hline  &&&\\
$\gtrsim 10^{15}$   &$L$&   $\begin{pmatrix}0&\frac{1}{2}\end{pmatrix}$  &$\begin{pmatrix}3&\frac{1}{2}\end{pmatrix}$     \\  [8pt]
\hline &&&\\
$ [10^{12},10^{15}]$  & $L$&$\begin{pmatrix}0&\frac{1}{2}\end{pmatrix}$  &$\begin{pmatrix}2&\frac{1}{3}\end{pmatrix}$    \\[8pt]\hline  &&&\\
$ [10^{9}, 10^{12}]$ : &&&\\ [8pt]
 \quad$[T^\tau_\text{decoh},10^{12}]$&$B-L$&  $\begin{pmatrix}0&\frac{3}{10}\end{pmatrix}$  &$\begin{pmatrix}\frac{3}{4}&\frac{1}{8}\end{pmatrix}$    \\  [7pt]
\quad$[10^{9},T^\tau_\text{decoh}]$&$B/3-L_{\tau,a}$&  $\begin{pmatrix}-\frac{6}{359}&\frac{307}{718}&-\frac{18}{359}\\ \frac{39}{359}&-\frac{21}{718}& \frac{117}{359}\end{pmatrix}$  &$\begin{pmatrix}\frac{258}{359}&\frac{41}{359}&\frac{56}{359}\end{pmatrix}$    \\  [15pt]
\hline  &&&\\
$ [10^{5},10^{9}]$ : &&& \\ [8pt]
\quad$[T^{\tau}_\text{decoh},10^{9}]$&$B-L$& $\begin{pmatrix}0&\frac{3}{10}\end{pmatrix}$  &$\begin{pmatrix}\frac{3}{4}&\frac{1}{8}\end{pmatrix}$   \\  [7pt]
\quad$[T^{\mu}_\text{decoh},T^{\tau}_\text{decoh}]$ &$B/3-L_{\tau,a}$&$\begin{pmatrix}-\frac{6}{359}&\frac{307}{718}&-\frac{18}{359}\\ \frac{39}{359}&-\frac{21}{718}& \frac{117}{359}\end{pmatrix}$  &$\begin{pmatrix}\frac{258}{359}&\frac{41}{359}&\frac{56}{359}\end{pmatrix}$  \\ [13pt]
\quad$[10^5,T^{\mu}_\text{decoh}]$   &$B/3-L_{\tau,\mu,a}$&$\begin{pmatrix}-\frac{6}{179}&\frac{151}{358}&-\frac{10}{179}&-\frac{10}{179}\\ \frac{33}{358}&-\frac{25}{716}& \frac{172}{537}&-\frac{7}{537}\\\frac{33}{358}&-\frac{25}{716}&-\frac{7}{537}& \frac{172}{537}\end{pmatrix}$  &$\begin{pmatrix}\frac{123}{179}&\frac{37}{358}&\frac{26}{179}&\frac{26}{179}\end{pmatrix}$  \\ [23pt]
\hline  &&&\\
$\lesssim 10^{5}$ : &&&\\ [8pt]
\quad$[T^{\tau}_\text{decoh},10^{5}]$   &$B/3-L $&   $\begin{pmatrix}0&\frac{3}{10}\end{pmatrix}$  &$\begin{pmatrix}\frac{6}{11}&\frac{1}{11}\end{pmatrix}$     \\  [7pt]
\quad$[T^{\mu}_\text{decoh},T^{\tau}_\text{decoh}]$  &$B/3-L_{\tau,a}$&   $\begin{pmatrix}-\frac{3}{244}&\frac{209}{488}&-\frac{3}{61}\\ \frac{39}{488}&-\frac{33}{976}& \frac{39}{122}\end{pmatrix}$  &$\begin{pmatrix}\frac{519}{976}&\frac{199}{1952}&\frac{31}{244}\end{pmatrix}$    \\ [13pt]
\quad$[T^{e}_\text{decoh},T^{\mu}_\text{decoh}]$  &$B/3-L_{\tau,\mu,a}$&   $\begin{pmatrix}-\frac{12}{481}&\frac{11}{26}&-\frac{2}{37}&-\frac{2}{37}\\ \frac{33}{481}&-\frac{1}{26}& \frac{35}{111}&-\frac{2}{111}\\\frac{33}{481}&-\frac{1}{26}&-\frac{2}{111}& \frac{35}{111}\end{pmatrix}$  &$\begin{pmatrix}\frac{256}{481}&\frac{1}{13}&\frac{4}{37}&\frac{4}{37}\end{pmatrix}$      \\ [20pt]
\quad$\lesssim T^{e}_\text{decoh}$    &$B/3-L_{\tau,\mu,e}$&   $\begin{pmatrix} \frac{9}{158}&\frac{221}{711}&-\frac{16}{711}&-\frac{16}{711}\\ \frac{9}{158}&-\frac{16}{711}& \frac{221}{711}&-\frac{16}{711}\\\frac{9}{158}&-\frac{16}{711}&-\frac{16}{711}& \frac{221}{711}\end{pmatrix}$  &$\begin{pmatrix}\frac{39}{79}&\frac{8}{79}&\frac{8}{79}&\frac{8}{79}\end{pmatrix}$      \\ [23pt]
\hline
\end{tabular} 
\caption{\it Temperature ranges, as in Tab.~\ref{table-chemical-pot-summary}. 
  In the second column, we show the asymmetries for which kinetic 
  equations have to be written. In the third and fourth columns the different
  $C^\ell$ and $C^\phi$ matrices holding in each regime. Note that these 
  matrices reduce to those found in the type-I seesaw case when removing 
  their first column.
}
\label{table-cell-summary} 
\end{table}
 As it has been discussed in
  Sec. \ref{domain-of-validity}, in scalar triplet flavored
  leptogenesis there are parameter space configurations for which
  lepton flavor coherence is not lost when the SM tau Yukawa reaction
  (or any other SM lepton Yukawa interaction) becomes fast. In those
  cases, the $C^\ell$ and $C^\phi$ matrices 
  certainly differ from those derived in
  Sec.~\ref{sec:chemical-equilibrium}, which hold when lepton flavor
  decoherence takes place at the same temperature at which the
  corresponding SM Yukawa coupling becomes fast. Although this lepton
  flavor decoherence ``delay'' is not inherent to scalar triplet
  flavored letogenesis, and it is rather a consequence of parameter
  choices, here we summarize all possible $C^\ell$ and $C^\phi$
  matrices including as well those cases. The list presented here thus
  encompasses all the scenarios one can consider when tracking the
  $B-L$ asymmetry in triplet scalar flavored leptogensis scenarios. 


Tab.~\ref{table-chemical-pot-summary} displays the different possible
temperature regimes, the corresponding reactions which are faster than
the Hubble expansion rate, the lepton flavor regimes (one-, two- or
three-flavor regimes) and the corresponding global symmetries of the
early Universe effective Lagrangian. In Tab.~\ref{table-cell-summary},
instead, we specify for the different temperature regimes the
asymmetry charges for which kinetic evolution equations have to be
written and the corresponding $C^\ell$ and $C^\phi$ matrices valid in
each case. We remind that $T_\text{decoh}^{f_i}$, as defined in
Sec.~\ref{domain-of-validity}, refers to the temperature at which the
lepton-related triplet inverse decay becomes smaller than the SM
lepton $f_i$ Yukawa interaction.

\subsection{$C^\ell$ and $C^\phi$ matrices in the lepton one-flavor limit}
\label{Cell-one-flavor-limit}

\begin{itemize}
\item {\it QCD instantons, electroweak sphalerons, bottom, charm and
    tau Yukawa-related reactions in
    thermal equilibrium, $T\subset [10^{9},10^{12}]$~GeV}:\\
  Sticking to the one lepton flavor approximation means first choosing
  a direction in the $\tau-a$ flavor space. Taking either
  $B_{\ell_{aa}}=1-B_\phi$ or $B_{\ell_{\tau\tau}}=1-B_\phi$, both
  cases are governed by the system of equations in
  (\ref{eq:flavored-BEqs2}), (\ref{eq:network-aligned-Triplet}) and
  (\ref{eq:network-aligned-BmL}) with the structure of the
  $C^{\ell,\phi}$ matrices determined by the corresponding choice. If
  $B_{\ell_{aa}}=1-B_\phi$, the asymmetry is entirely projected along
  the $a$ flavor direction and so the one lepton flavor approximation
  $C^{\ell,\phi}$ matrices are given by
  \begin{equation}
    \label{eq:unflavored-Cell-Cphi-R2-BaaEq1}
    C^\ell=
    \begin{pmatrix}
      - 6/359 & 307/718
    \end{pmatrix}\ ,
    \qquad
    C^\phi=
    \begin{pmatrix}
       258/359 & 41/359
    \end{pmatrix}\ .
  \end{equation}
  If instead the asymmetry is projected along the $\tau$ flavor the
  matrices read:
  \begin{equation}
    \label{eq:unflavored-Cell-Cphi-R2-BtautauEq1}
    C^\ell=
    \begin{pmatrix}
        39/359&  117/359
    \end{pmatrix}\ ,
    \qquad
    C^\phi=
    \begin{pmatrix}
      258/359 & 56/359
    \end{pmatrix}\ .
  \end{equation}
\item {\it  Strange and muon Yukawa interactions in thermal 
    equilibrium, $T\subset [10^5,10^{9}]$~GeV:}\\
  Since in this regime the flavor basis is completely defined, there
  are several flavor projections which render flavor alignment. The
  corresponding $C^{\ell,\phi}$ matrices for the alignments
  $B_{\ell_{ee}}=1-B_\phi$ and $B_{\ell_{\tau\tau}}=1-B_\phi$ (the
  results for $B_{\ell_{\mu\mu}}=1-B_\phi$ match those of
  $B_{\ell_{\tau\tau}}=1-B_\phi$) read:
  \begin{alignat}{6}
    B_{\ell_{ee}}&=1-B_\phi:&\quad 
    C^\ell&=
    \begin{pmatrix}
      -6/179 & 151/358\\
    \end{pmatrix}\ ,
    &\quad
    C^\phi&=
    \begin{pmatrix}
      123/179 & 37/358
    \end{pmatrix}\\
    B_{\ell_{\tau\tau}}&=1-B_\phi:&\quad 
    C^\ell&=
    \begin{pmatrix}
      33/358 & 172/537\\
    \end{pmatrix}\ ,
    &\quad
    C^\phi&=
    \begin{pmatrix}
      123/179 & 26/179
    \end{pmatrix}
  \end{alignat}
 \item {\it  All SM reactions in thermal equilibrium, $T\lesssim
    10^5$~GeV:}\\
    In this case as well one can define a one-flavor approximation by
  fixing an alignment in flavor space. For $B_{\ell_{ii}}=1-B_\phi$
  ($i=e, \mu, \tau$) the $C^{\ell,\phi}$ matrices are given by
  \begin{alignat}{6}
    B_{\ell_{ii}}&=1-B_\phi:&\quad 
    C^\ell&=
    \begin{pmatrix}
      9/158 & 221/711\\
    \end{pmatrix}\ ,
    &\quad
    C^\phi&=
    \begin{pmatrix}
      39/79 & 8/79
    \end{pmatrix}\ .
  \end{alignat}
\end{itemize}



\begin{thebibliography}{99}
\bibitem{Tortola:2012te} 
  D.~V.~Forero, M.~Tortola and J.~W.~F.~Valle,
  Phys.\ Rev.\ D {\bf 86}, 073012 (2012)
  [arXiv:1205.4018 [hep-ph]].

\bibitem{GonzalezGarcia:2012sz} 
  M.~C.~Gonzalez-Garcia, M.~Maltoni, J.~Salvado and T.~Schwetz,
  JHEP {\bf 1212}, 123 (2012)
  [arXiv:1209.3023 [hep-ph]].

\bibitem{Fogli:2012ua} 
  G.~L.~Fogli, E.~Lisi, A.~Marrone, D.~Montanino, A.~Palazzo and A.~M.~Rotunno,
  Phys.\ Rev.\ D {\bf 86}, 013012 (2012)
  [arXiv:1205.5254 [hep-ph]].

\bibitem{Hinshaw:2012aka} 
  G.~Hinshaw {\it et al.}  [WMAP Collaboration],
  arXiv:1212.5226 [astro-ph.CO].

\bibitem{Ade:2013zuv} 
  P.~A.~R.~Ade {\it et al.}  [Planck Collaboration],
  arXiv:1303.5076 [astro-ph.CO].

\bibitem{seesaw} P. Minkowski, {\it Phys. Lett.} B {\bf 67} 421
  (1977); T.  Yanagida, in {\it Proc. of Workshop on Unified Theory
    and Baryon number in the Universe}, eds. O. Sawada and
  A. Sugamoto, KEK, Tsukuba, (1979) p.95; M. Gell-Mann, P. Ramond and
  R. Slansky, in {\it Supergravity}, eds P.  van Niewenhuizen and
  D. Z. Freedman (North Holland, Amsterdam 1980) p.315; P.  Ramond,
  {\it Sanibel talk}, retroprinted as hep-ph/9809459; S. L. Glashow,
  in{\it Quarks and Leptons}, Carg\`ese lectures, eds M. L\'evy,
  (Plenum, 1980, New York) p. 707; R. N. Mohapatra and
  G. Senjanovi\'c, {\it Phys. Rev.  Lett.} {\bf 44}, 912 (1980);
  J.~Schechter and J.~W.~F.~Valle,
  Phys.\ Rev.\  D {\bf 22} (1980) 2227;
  Phys.\ Rev.\  D {\bf 25} (1982) 774.

\bibitem{Schechter:1980gr}
  J.~Schechter and J.~W.~F.~Valle,
  Phys.\ Rev.\  D {\bf 22}, 2227 (1980);
  G.~Lazarides, Q.~Shafi and C.~Wetterich,
  Nucl.\ Phys.\  B {\bf 181}, 287 (1981);
  R.~N.~Mohapatra and G.~Senjanovic,
  Phys.\ Rev.\  D {\bf 23}, 165 (1981);
  C.~Wetterich,
  Nucl.\ Phys.\  B {\bf 187}, 343 (1981);

\bibitem{Foot:1988aq}
  R.~Foot, H.~Lew, X.~G.~He and G.~C.~Joshi,
  Z.\ Phys.\  C {\bf 44}, 441 (1989).

\bibitem{Davidson:2008bu} 
  S.~Davidson, E.~Nardi and Y.~Nir,
  Phys.\ Rept.\  {\bf 466}, 105 (2008)
  [arXiv:0802.2962 [hep-ph]].

\bibitem{Fong:2013wr} 
  C.~S.~Fong, E.~Nardi and A.~Riotto,
  Adv.\ High Energy Phys.\  {\bf 2012}, 158303 (2012)
  [arXiv:1301.3062 [hep-ph]].

\bibitem{Hambye:2012fh} 
  T.~Hambye,
  New J.\ Phys.\  {\bf 14}, 125014 (2012)
  [arXiv:1212.2888 [hep-ph]].

\bibitem{Ma:1998dx} 
  E.~Ma and U.~Sarkar,
  Phys.\ Rev.\ Lett.\  {\bf 80}, 5716 (1998)
  [hep-ph/9802445].

\bibitem{Hambye:2000ui}
  T.~Hambye, E.~Ma and U.~Sarkar,
  Nucl.\ Phys.\ B {\bf 602} (2001) 23
  [hep-ph/0011192].

\bibitem{Hambye:2003ka} 
  T.~Hambye and G.~Senjanovic,
  Phys.\ Lett.\ B {\bf 582}, 73 (2004)
  [hep-ph/0307237].

\bibitem{Hambye:2005tk} 
  T.~Hambye, M.~Raidal and A.~Strumia,
  Phys.\ Lett.\ B {\bf 632}, 667 (2006)
  [hep-ph/0510008].


\bibitem{Hambye:2003rt} 
  T.~Hambye, Y.~Lin, A.~Notari, M.~Papucci and A.~Strumia,
  Nucl.\ Phys.\ B {\bf 695}, 169 (2004)
  [hep-ph/0312203].

\bibitem{AristizabalSierra:2010mv} 
  D.~Aristizabal Sierra, J.~F.~Kamenik and M.~Nemevsek,
  JHEP {\bf 1010}, 036 (2010)
  [arXiv:1007.1907 [hep-ph]].

\bibitem{AristizabalSierra:2011ab} 
  D.~Aristizabal Sierra, F.~Bazzocchi and I.~de Medeiros Varzielas,
  Nucl.\ Phys.\ B {\bf 858}, 196 (2012)
  [arXiv:1112.1843 [hep-ph]].

\bibitem{Franceschini:2008pz} 
  R.~Franceschini, T.~Hambye and A.~Strumia,
  Phys.\ Rev.\ D {\bf 78}, 033002 (2008)
  [arXiv:0805.1613 [hep-ph]].

\bibitem{delAguila:2008cj} 
  F.~del Aguila and J.~A.~Aguilar-Saavedra,
  Nucl.\ Phys.\ B {\bf 813}, 22 (2009)
  [arXiv:0808.2468 [hep-ph]].

\bibitem{Strumia:2008cf} 
  A.~Strumia,
  Nucl.\ Phys.\ B {\bf 809}, 308 (2009)
  [arXiv:0806.1630 [hep-ph]].

\bibitem{Branco:2011zb}
  G.~C.~Branco, R.~G.~Felipe and F.~R.~Joaquim,
  Rev.\ Mod.\ Phys.\  {\bf 84} (2012) 515
  [arXiv:1111.5332 [hep-ph]].

\bibitem{Felipe:2013kk} 
  R.~G.~Felipe, F.~R.~Joaquim and H.~Serodio,
  arXiv:1301.0288 [hep-ph].

\bibitem{AristizabalSierra:2007ur} 
  D.~Aristizabal Sierra, M.~Losada and E.~Nardi,
  Phys.\ Lett.\ B {\bf 659}, 328 (2008)
  [arXiv:0705.1489 [hep-ph]].

\bibitem{AristizabalSierra:2009bh} 
  D.~Aristizabal Sierra, L.~A.~Munoz and E.~Nardi,
  Phys.\ Rev.\ D {\bf 80}, 016007 (2009)
  [arXiv:0904.3043 [hep-ph]].

\bibitem{AristizabalSierra:2009mq}
  D.~Aristizabal Sierra, M.~Losada and E.~Nardi,
  JCAP {\bf 0912}, 015 (2009)
  [arXiv:0905.0662 [hep-ph]].

\bibitem{GonzalezGarcia:2009qd} 
  M.~C.~Gonzalez-Garcia, J.~Racker and N.~Rius,
  JHEP {\bf 0911}, 079 (2009)
  [arXiv:0909.3518 [hep-ph]].

\bibitem{Kolb:1979qa} 
  E.~W.~Kolb and S.~Wolfram,
  Nucl.\ Phys.\ B {\bf 172}, 224 (1980)
  [Erratum-ibid.\ B {\bf 195}, 542 (1982)].

\bibitem{Antusch:2004xy} 
  S.~Antusch and S.~F.~King,
  Phys.\ Lett.\ B {\bf 597}, 199 (2004)
  [hep-ph/0405093].

\bibitem{Abada:2008gs} 
  A.~Abada, P.~Hosteins, F.~-X.~Josse-Michaux and S.~Lavignac,
  Nucl.\ Phys.\ B {\bf 809}, 183 (2009)
  [arXiv:0808.2058 [hep-ph]].

\bibitem{Fong:2010bv} 
  C.~S.~Fong, M.~C.~Gonzalez-Garcia and E.~Nardi,
  JCAP {\bf 1102}, 032 (2011)
  [arXiv:1012.1597 [hep-ph]].

\bibitem{McDonald:2001vt}
  J.~McDonald,
  Phys.\ Rev.\ Lett.\  {\bf 88}, 091304 (2002)
  [hep-ph/0106249];
  L.~J.~Hall, K.~Jedamzik, J.~March-Russell and S.~M.~West,
  JHEP {\bf 1003}, 080 (2010)
  [arXiv:0911.1120 [hep-ph]];
  C.~Cheung, G.~Elor, L.~J.~Hall and P.~Kumar,
  JHEP {\bf 1103}, 042 (2011)
  [arXiv:1010.0022 [hep-ph]];
  C.~E.~Yaguna,
  JHEP {\bf 1108}, 060 (2011)
  [arXiv:1105.1654 [hep-ph]];
  M.~Frigerio, T.~Hambye and E.~Masso,
  Phys.\ Rev.\ X {\bf 1}, 021026 (2011)
  [arXiv:1107.4564 [hep-ph]];
  X.~Chu, T.~Hambye and M.~H.~G.~Tytgat,
  JCAP {\bf 1205}, 034 (2012)
  [arXiv:1112.0493 [hep-ph]].

\bibitem{Buchmuller:2001sr} 
  W.~Buchmuller and M.~Plumacher,
  Phys.\ Lett.\ B {\bf 511}, 74 (2001)
  [hep-ph/0104189].

\bibitem{Nardi:2005hs} 
  E.~Nardi, Y.~Nir, J.~Racker and E.~Roulet,
  JHEP {\bf 0601}, 068 (2006)
  [hep-ph/0512052].

\bibitem{Nardi:2007jp} 
  E.~Nardi, J.~Racker and E.~Roulet,
  JHEP {\bf 0709}, 090 (2007)
  [arXiv:0707.0378 [hep-ph]].

\bibitem{Barbieri:1999ma} 
  R.~Barbieri, P.~Creminelli, A.~Strumia and N.~Tetradis,
  Nucl.\ Phys.\ B {\bf 575}, 61 (2000)
  [hep-ph/9911315].

\bibitem{Harvey:1990qw} 
  J.~A.~Harvey and M.~S.~Turner,
  Phys.\ Rev.\ D {\bf 42}, 3344 (1990).

\bibitem{Moore:1997im} 
  G.~D.~Moore,
  Phys.\ Lett.\ B {\bf 412}, 359 (1997)
  [hep-ph/9705248].

\bibitem{Bento:2003jv} 
  L.~Bento,
  JCAP {\bf 0311}, 002 (2003)
  [hep-ph/0304263].

\bibitem{Nardi:2006fx} 
  E.~Nardi, Y.~Nir, E.~Roulet and J.~Racker,
  JHEP {\bf 0601}, 164 (2006)
  [hep-ph/0601084].

\bibitem{Blanchet:2006ch}
  S.~Blanchet, P.~Di Bari and G.~G.~Raffelt,
  JCAP {\bf 0703} (2007) 012
  [hep-ph/0611337].

\bibitem{Abada:2006fw} 
  A.~Abada, S.~Davidson, F.~-X.~Josse-Michaux, M.~Losada and A.~Riotto,
  JCAP {\bf 0604}, 004 (2006)
  [hep-ph/0601083].

\bibitem{Antusch:2010ms} 
  S.~Antusch, P.~Di Bari, D.~A.~Jones and S.~F.~King,
  Nucl.\ Phys.\ B {\bf 856}, 180 (2012)
  [arXiv:1003.5132 [hep-ph]].

\bibitem{Cirelli:2007xd} 
  M.~Cirelli, A.~Strumia and M.~Tamburini,
  Nucl.\ Phys.\ B {\bf 787}, 152 (2007)
  [arXiv:0706.4071 [hep-ph]].

\bibitem{Campbell:1992jd} 
  B.~A.~Campbell, S.~Davidson, J.~R.~Ellis and K.~A.~Olive,
  Phys.\ Lett.\ B {\bf 297}, 118 (1992)
  [hep-ph/9302221].

\bibitem{Cline:1993bd} 
  J.~M.~Cline, K.~Kainulainen and K.~A.~Olive,
  Phys.\ Rev.\ D {\bf 49}, 6394 (1994)
  [hep-ph/9401208].
\end{thebibliography}
\end{document}